%% file: main.tex
\pdfoutput=1
\newcommand*{\ATLASLATEXPATH}{atlaslatex/latex/}
\documentclass[cernpreprint, UKenglish, texlive=2016]{\ATLASLATEXPATH atlasdoc}

\usepackage[subcaption,backend=biber]{\ATLASLATEXPATH atlaspackage}
\usepackage{\ATLASLATEXPATH atlasbiblatex}

\usepackage[process]{\ATLASLATEXPATH atlasphysics}

\graphicspath{{atlaslatex/logos/}{figures/}}

\usepackage{color}
\usepackage{acronym}
\usepackage{main-defs}
\usepackage{rotating}
\usepackage{multirow}
\usepackage{bm}
\usepackage{bigfoot}

\input{main-metadata}
\hypersetup{pdftitle={ATLAS document},pdfauthor={The ATLAS Collaboration}}

\begin{document}
\LEcontact{Richard Keeler rkeeler@uvic.ca}
\maketitle

\tableofcontents

\input{sections/introduction}
\input{sections/atlas_detector}
\input{sections/mc_modelling}
\input{sections/object_selection}

\input{sections/event_selection}
\input{sections/background_estimation}

\input{sections/ttbar_bkg_fit}
\input{sections/unfolding}
\input{sections/systematics}
\input{sections/results}
\input{sections/summary}

\section*{Acknowledgements}

\input{atlaslatex/acknowledgements/Acknowledgements}




\printbibliography

\clearpage
\input{atlas_authlist}


\end{document}

%% file: main-metadata.tex
\AtlasTitle{Measurements of inclusive and differential fiducial cross-sections
of $t\bar{t}$ production with additional heavy-flavour jets in proton--proton
collisions at $\sqrt{s} = 13$~\TeV\ with the ATLAS detector}

\AtlasAbstract{This paper presents measurements of $t\bar{t}$ production in
  association with additional $b$-jets in $pp$ collisions at the LHC at a
  centre-of-mass energy of 13~\TeV. The data were recorded with the ATLAS
  detector and correspond to an integrated luminosity of \lumi. 
  Fiducial cross-section measurements are
  performed in the dilepton and lepton-plus-jets $t\bar{t}$ decay channels.
  Results are presented at particle level in the form of inclusive
  cross-sections of $t\bar{t}$ final states with three and four $b$-jets as well
  as differential cross-sections as a function of global event properties and
  properties of $b$-jet pairs. The measured inclusive fiducial cross-sections
  generally exceed the $t\bar{t}b\bar{b}$ predictions from various
  next-to-leading-order matrix element calculations matched to a parton shower
  but are compatible within the total uncertainties. The experimental
  uncertainties are smaller than the uncertainties in the predictions. Comparisons of
  state-of-the-art theoretical predictions with the differential measurements
  are shown and good agreement with data is found for most of them.}

\author{The ATLAS Collaboration}

\AtlasRefCode{TOPQ-2017-12}

\PreprintIdNumber{CERN-EP-2018-276}

\AtlasJournal{JHEP}
\AtlasJournalRef{JHEP 04 (2019) 046}
\AtlasDOI{DOI:10.1007/JHEP04(2019)046}

\AtlasCoverSupportingNote{Dilepton note}{https://cds.cern.ch/record/2270265}
\AtlasCoverSupportingNote{\ljets note}{https://cds.cern.ch/record/2270273}
\AtlasCoverCommentsDeadline{11th October 2018}

\AtlasCoverAnalysisTeam{Georges Aad, Henri Bachacou, Rafa\l\ Bielski, Nihal
Brahimi, Yasiel Delabat, Frederic Deliot, Christoph Eckardt, Paul Glaysher,
Stefan Guindon,
Mahsana Haleem, Vivek Jain, Judith Katzy, Thorsten Kuhl, Tom Neep, Yvonne
Peters, Matthieu Robin, Stewart Swift, Timoth\'{e}e Theveneaux-Pelzer, Laurent
Vacavant, Akanksha Vishwakarma, Peter Weber}

\AtlasCoverEdBoardMember{Jiri~Kvita~(chair), Michele Pinamonti, Jie~Yu}

\AtlasCoverEgroupEditors{atlas-topq-2017-12-editors@cern.ch}

\AtlasCoverEgroupEdBoard{atlas-topq-2017-12-editorial-board@cern.ch}

%% file: sections/introduction.tex
\section{Introduction}
\label{sec:introduction}

\begin{figure}
  \subcaptionbox{\label{fig:ttbb}}{
    \includegraphics[width=0.30\textwidth]{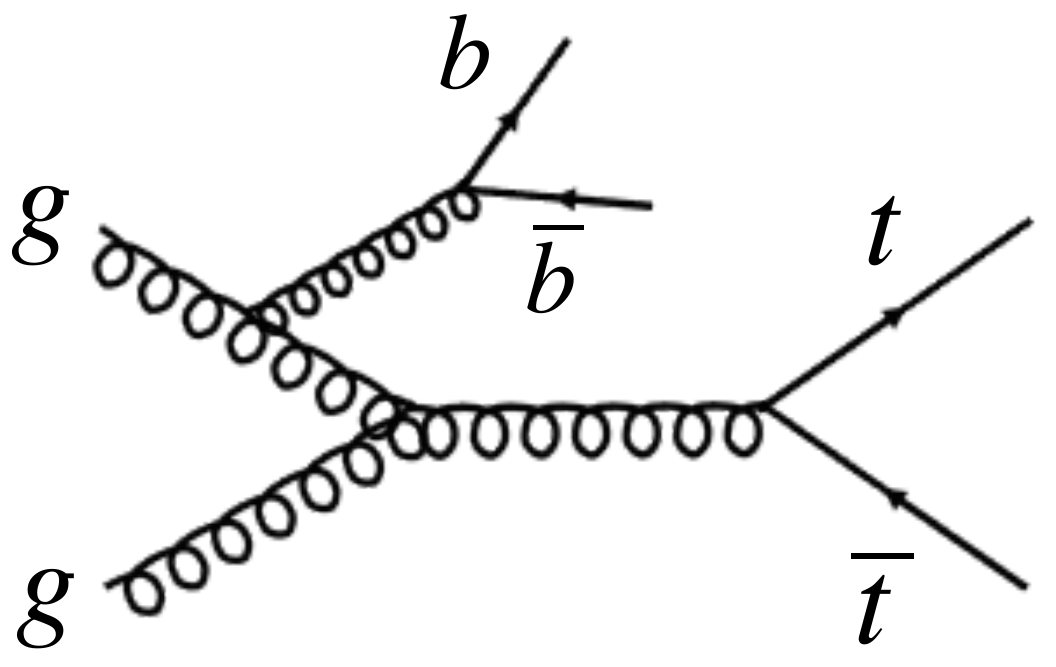}
  }
  \subcaptionbox{\label{fig:ttHbb}}{
    \includegraphics[width=0.30\textwidth]{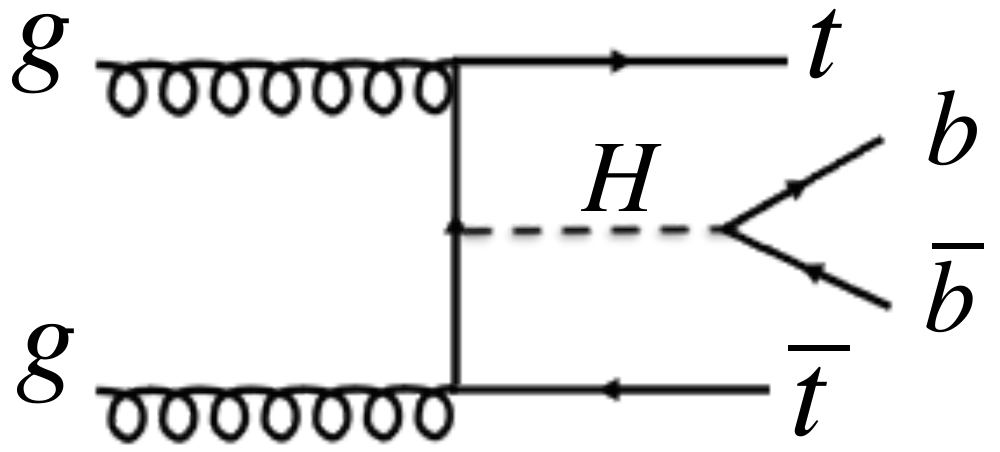}
  }
  \subcaptionbox{\label{fig:ttZ}}{
    \includegraphics[width=0.30\textwidth]{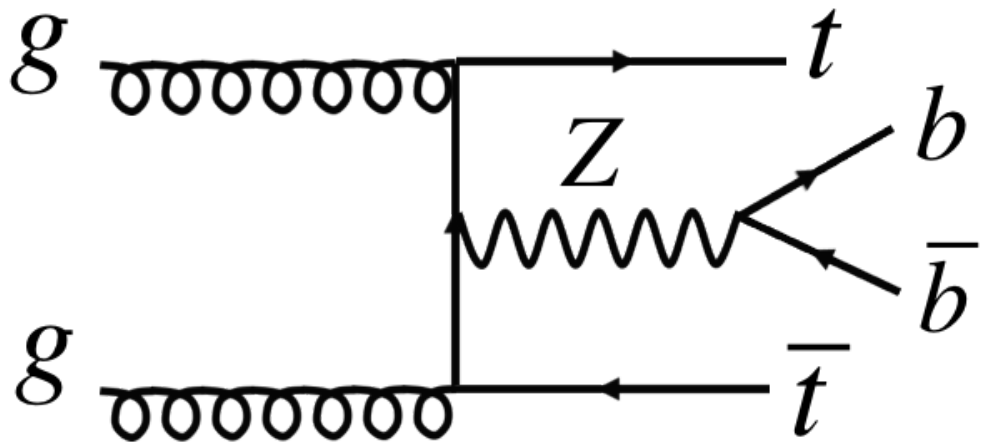}
  }
  \caption{Example Feynman diagrams of processes leading to a \ttbb final
    state, including \protect\subref{fig:ttbb} QCD \ttbb production,
    \protect\subref{fig:ttHbb} $\ttbar H (H \to b\bar{b})$, and
    \protect\subref{fig:ttZ} $\ttbar Z (Z \to b\bar{b})$.}
    \label{fig:feynmanDia}
\end{figure}

Measurements of the production cross-section of top-antitop quark pairs
(\ttbar) with additional jets provide important tests of quantum chromodynamics
(QCD) predictions. Among these, the process of \ttbar produced in association
with jets originating from $b$-quarks ($b$-jets) is particularly important to measure, as there are many uncertainties in the calculation of the process. 
For example, calculating the amplitude for the process shown in Figure~\ref{fig:ttbb} is a challenge 
due to the  non-negligible mass of the $b$-quark.
It is therefore important to compare the predictions with both inclusive and
differential experimental cross-section measurements of \ttbar production with
additional \bjets. State-of-the-art QCD calculations give
predictions for the \ttbar production cross-section with up to two additional
massless partons at \ac{NLO} in perturbation theory matched to a
parton shower~\cite{Hoeche:2014qda}, and the QCD production of \ttbb is
calculated at NLO 
matched to a parton shower~\cite{Cascioli:2013era,Garzelli:2014aba,Bevilacqua:2017cru,Jezo:2018yaf}.

Moreover, since the discovery of the Higgs
boson~\cite{HIGG-2012-27,CMS-HIG-12-028}, the determination of the Higgs
coupling to the heaviest elementary particle, the top quark, is a crucial test
of the Standard Model (SM). Direct measurements of the top-quark Yukawa
coupling are performed in events where a Higgs boson is produced in association
with a top-quark pair (\ttH)~\cite{HIGG-2018-13,CMS-HIG-17-035}. 
The Higgs branching ratios are dominated by the $H \to b\bar{b}$
decay~\cite{CMS-HIG-18-016,HIGG-2018-04}, and therefore the \ttH
process can be measured with the best statistical precision using
events where the Higgs boson decays in this manner, leading to a \ttbb
final state as shown in Figure~\ref{fig:ttHbb}.  However, this channel
suffers from a large background from QCD \ttbb production indicated in
Figure~\ref{fig:ttbb}~\cite{HIGG-2017-03-FIXED,CMS-HIG-17-026}.

Measurements of $\ttbar H (H \to b\bar{b})$ would benefit from a better understanding of the  QCD production of
\ttbb as predicted by the SM and, in particular, improved \ac{MC} modelling. The measurements presented in this paper 
were chosen in order to provide data needed 
to improve the QCD \ac{MC} modelling of the \ttbb process.
The differential observables are particularly interesting as they are sensitive to the relative contribution of events from \ttbar-associated Higgs production (\ttH) with $H
\rightarrow b\bar{b}$ decays to QCD-produced \ttbb events in various phase space regions.
Even though the aim is to improve the modelling of QCD production of additional
$b$-jets in \ttbar events, this analysis measures their production without separating the different production channels such as \ttH or \ttbar
in association with a vector boson (\ttV), for example the $\ttbar Z$ process
shown in Figure~\ref{fig:ttZ}.

In this paper, measurements of fiducial cross-sections are presented using
data recorded by the ATLAS detector during 2015 and 2016 in proton--proton ($pp$)
collisions at a centre-of-mass energy $\sqrt{s} = 13$~\TeV, corresponding to a
total integrated luminosity of \lumi. In addition, differential measurements at this centre-of-mass energy are
presented as a function of various observables. 
Previous measurements of \ttbar production with additional heavy-flavour jets have been
reported by ATLAS at $\sqrt{s} = 7$~\TeV~\cite{TOPQ-2012-16} and both 
CMS and ATLAS at $\sqrt{s} = 8$~\TeV~\cite{TOPQ-2014-10,CMS-TOP-12-041,CMS-TOP-13-010}. 
CMS has also reported a measurement of the inclusive \ttbb cross-section using 2.3~\ifb\
at $\sqrt{s} = 13$~\TeV~\cite{CMS-TOP-16-010}. 

Since the top quark decays into a $b$-quark and $W$~boson nearly 100\% of the time,
\ttbar events are typically classified according to how the two $W$~bosons decay. In
this analysis, two channels are considered: the $e\mu$ channel, in which both
$W$~bosons decay leptonically, one into a muon and muon neutrino and the other into an
electron and electron neutrino,   and the
lepton-plus-jets channel (\ljets), in which one $W$~boson decays into an
isolated charged lepton (an electron or muon) and corresponding neutrino and the
other $W$~boson decays into a pair of quarks.
Electrons and muons produced either directly in the decay of the $W$~boson or
via an intermediate $\tau$-lepton are included in both channels.

The decay of a top-quark pair results in two $b$-quarks and therefore a final
state which includes the production of two additional $b$-quarks may contain up
to four \bjets. The inclusive fiducial cross-sections are presented for events
with at least three $b$-jets and for events with at least four $b$-jets. The
differential cross-sections are presented for events with at least three \bjets
in the $e\mu$ channel and with at least four \bjets in the \ljets channel. The
results are obtained as a function of the transverse momentum
(\pt)\footnote{ATLAS uses a right-handed coordinate system with its origin at
  the nominal interaction point (IP) in the centre of the detector and the
  $z$-axis along the beam pipe. The $x$-axis points from the IP to the centre of
  the LHC ring, and the $y$-axis points upward. Cylindrical coordinates
  $(r,\phi)$ are used in the transverse plane, $\phi$ being the azimuthal angle
  around the $z$-axis. The pseudorapidity is defined in terms of the polar angle
  $\theta$ as $\eta=-\ln\tan(\theta/2)$. The angular separation between two
  points in $\eta$ and $\phi$ is defined as $\Delta R = \sqrt{(\Delta\eta)^2 +
    (\Delta\phi)^2}$.} of each of the \bjets, the scalar sum of the \pt of the
lepton(s) and jets in the events ($H_{\mathrm{T}}$) and of only jets in the
events ($H^{\mathrm{had}}_{\mathrm{T}}$) and as a function of the $b$-jet
multiplicity ($N_{b\mathrm{\textrm{-}jets}}$).

This analysis does not attempt to identify the origin of the $b$-jets, i.e.\ it does not
distinguish between additional \bjets 
and \bjets that come from the top-quark decays. This is to avoid using
simulation-based information to attribute $b$-jets to a particular production
process, which would lead to significant modelling uncertainties. Instead, differential
cross-sections are measured as a function of kinematic distributions of pairs of
\bjets. The reported distributions could be used to distinguish the contribution of specific production
mechanisms: the pair made from the two \bjets closest in angular distance is expected to be formed by \bjets from gluon splitting and the pair made from the two
highest-\pt \bjets is expected to be dominated by top-pair production. For each of these
pairs, the distributions are measured for the angular separation between the $b$-jets ($\Delta
R(b,b)$), the invariant mass ($m_{bb}$) and transverse momentum
($p_{\mathrm{T},bb}$).  It should be noted that for  events with at least three \bjets, it is
likely that one of the two closest \bjets originates from the top quark.
Hence the simple picture that the two closest \bjets are usually from gluon
splitting may not apply. However, 
$\Delta R$, $m_{bb}$ and 
$p_{\mathrm{T},bb}$ 
are used for reconstruction
of the final state in analyses with multiple $b$-jets and therefore probing the
modelling of these observables is important.

 The cross-sections are obtained by subtracting the estimated number of
 non-\ttbar background events from the data distributions. At detector level,
 jets are identified as containing $b$-hadrons (``$b$-tagging'') by a multivariate
 algorithm~\cite{PERF-2016-05}. The \ttbar background resulting from additional light-flavour and
 charm-quark jets wrongly identified as \bjets is evaluated using a template
 fit, in which the templates are constructed from the output discriminant of the
 $b$-tagging algorithm. The background-subtracted distributions are corrected
 for acceptance and detector effects using an unfolding technique that includes
corrections for the \ttbar-related backgrounds.

 This paper is laid out as follows. The experimental set-up for the collected data is
 described in Section~\ref{sec:atlas_detector}. Details of the simulation used
 in this analysis are provided in Section~\ref{sec:mc_modelling}. The
 reconstruction and identification of leptons and jets, the $b$-tagging of jets at
 detector level, and the definitions of objects at particle level are described in
 Section~\ref{sec:object_selection}. The selection of reconstructed events and
 the definition of the fiducial phase space are given in
 Section~\ref{sec:event_selection_fiducial_def}. Estimation of the background from
 non-\ttbar processes is described in Section~\ref{sec:background_estimation}.
 The method to estimate the \ttbar background with additional jets
 misidentified as \bjets and the unfolding procedure to correct the data to
 particle level for fiducial cross-section measurements are explained in
 Section~\ref{sec:fiducial_xsec}. Sources of systematic uncertainties and their
 propagation to the measured cross-sections are described in
 Section~\ref{sec:systematics}. The measured inclusive and normalised
 differential fiducial cross-sections and the comparison with various theoretical
 predictions are presented in Section~\ref{sec:results}. Finally, the results are
 summarised in Section~\ref{sec:summary}.

%% file: sections/atlas_detector.tex
\section{ATLAS detector}
\label{sec:atlas_detector}

The ATLAS detector~\cite{PERF-2007-01} at the LHC covers nearly the entire solid
angle around the collision point. It consists of an inner-tracking detector
surrounded by a thin superconducting solenoid, electromagnetic and hadronic
calorimeters, and a muon spectrometer incorporating three large superconducting
toroidal magnets. 

The \ac{ID} system is immersed in a \SI{2}{\tesla}
axial magnetic field and provides charged-particle tracking in the
pseudorapidity range $|\eta| < 2.5$. The ID is composed of silicon detectors and the transition radiation tracker.
The high-granularity silicon pixel detector covers the interaction region and is
followed by the silicon microstrip tracker. The innermost silicon pixel layer, added 
to the inner detector before the start of Run-2 data
taking~\cite{Capeans:2010jnh,Abbott:2018ikt}, improves the identification of $b$-jets. The
tracking capabilities of the silicon detectors are augmented by the transition radiation tracker, which is located at a larger radius and
enables track reconstruction  up to $|\eta| = 2.0$. It also provides signals used to separate electrons from pions.

The calorimeter system covers the range $|\eta| < 4.9$. Within
the region $|\eta|< 3.2$, electromagnetic calorimetry is provided by barrel and
endcap high-granularity lead/liquid-argon (LAr) electromagnetic calorimeters,
with an additional thin LAr presampler covering $|\eta| < 1.8$ to correct for
energy loss in material upstream of the calorimeters. Hadronic calorimetry is
provided by the steel/scintillating-tile calorimeter, segmented into three
barrel structures within $|\eta| < 1.7$, and two copper/LAr hadronic endcap
calorimeters. The solid angle coverage is completed with forward copper/LAr and
tungsten/LAr calorimeter modules optimised for electromagnetic and hadronic
measurements, respectively.

The muon spectrometer (MS) comprises separate trigger and high-precision
tracking chambers measuring the deflection of muons in a magnetic field
generated by the superconducting air-core toroids. The field integral of the toroids
ranges between \num{2.0} and \SI{6.0}{\tesla\metre} across most of the detector.
A set of precision chambers covers the region $|\eta| < 2.7$ with three layers
of drift tubes, complemented by cathode strip chambers in the forward
region, where the background is highest. The muon trigger system covers the
range $|\eta| < 2.4$ with resistive plate chambers in the barrel, and thin gap
chambers in the endcap regions.

A two-level trigger system is used for event 
selection~\cite{PERF-2011-02,TRIG-2016-01}. The first trigger level is implemented in
hardware and uses a subset of detector information to reduce the event rate to a
design value of at most \SI{100}{\kHz}. This is followed by a software-based
trigger that reduces the event rate to about \SI{1}{\kHz}.

%% file: sections/mc_modelling.tex
\FloatBarrier
\section{Monte Carlo simulation}
\label{sec:mc_modelling}
Monte Carlo simulations are used in three ways in this analysis: to estimate the signal
and background composition of the selected data samples, to determine correction
factors for detector and acceptance effects for unfolding, and finally to
estimate systematic uncertainties. In addition, theoretical predictions are
compared with the unfolded data. The computer codes used to generate the samples 
and how they were configured are described in the following. The signal MC samples used in the analysis are listed in Table~\ref{tab:mc_signal}.
\begin{table}
  \caption{Summary of the MC sample set-ups used for modelling the signal processes ($\ttbar
      + \ttbar V + \ttbar H$) for the data analysis and for comparisons with the measured cross-sections and differential distributions.
      All samples used the NNPDF3.0NLO PDF set with the exception of the two
      \SHERPA samples, which used NNPDF3.0NNLO. The different blocks indicate from
      top to bottom the samples used as nominal MC (nom.), systematic variations
      (syst.) and for comparison only (comp.). For details see Section~\ref{sec:mc_modelling}.}
  \footnotesize
  \begin{center}
    \begin{tabular}{lllll}
      \toprule
      Generator sample                      & Process                & Matching                                        & Tune         & Use   \\
      \midrule
      \POWHEGBOX v2 + \PYTHIAV{8.210}       & \ttbar NLO             & \powheg $h_{\mathrm{damp}}=1.5 m_{t}$ & A14          & nom.  \\
      \MGMCatNLO + \PYTHIAV{8.210}          & $\ttbar+V/H$ NLO       & MC@NLO                                          & A14          & nom.  \\
      \cmidrule(lr){1-5}
      \POWHEGBOX v2 + \PYTHIAV{8.210} RadLo & \ttbar NLO             & \powheg $h_{\mathrm{damp}}=1.5 m_{t}$ & A14Var3cDown & syst. \\
      \POWHEGBOX v2 + \PYTHIAV{8.210} RadHi & \ttbar NLO             & \powheg $h_{\mathrm{damp}}=3.0 m_{t}$ & A14Var3cUp   & syst. \\
      \POWHEGBOX v2 + \HERWIGV{7.01}        & \ttbar NLO             & \powheg $h_{\mathrm{damp}}=1.5 m_{t}$ & H7UE         & syst. \\
      \SHERPAV{2.2.1}     \ttbar            & \ttbar+0,1 parton at NLO & \textsc{MePs@Nlo}                               & \SHERPA      & syst. \\
                                            & \phantom{\ttbar}+2,3,4 partons at LO                                                                \\
      \cmidrule(lr){1-5}
      \MGMCatNLO + \PYTHIAV{8.210}            & \ttbar NLO             & MC@NLO                                          & A14          & comp. \\
      \SHERPAV{2.2.1} \ttbb   (4FS)                 & \ttbb NLO              & MC@NLO                                          & \SHERPA      & comp. \\
      \powhel   + \PYTHIAV{8.210} (5FS)             & \ttbb NLO              & \powheg  $h_{\mathrm{damp}}=\HT/2$               & A14          & comp. \\
      \powhel   + \PYTHIAV{8.210} (4FS)             & \ttbb NLO              & \powheg  $h_{\mathrm{damp}}=\HT/2$              & A14          & comp. \\
      \POWHEGBOX v2 + \PYTHIAV{8.210}    \ttbb (4FS) & \ttbb NLO              & \powheg $h_{\mathrm{damp}}=\HT/2$                & A14          & comp. \\
      \bottomrule
    \end{tabular}
  \end{center}
  \label{tab:mc_signal}
\end{table}

The nominal \ttbar sample was generated using the \POWHEGBOX generator (version~2,
r3026)~\cite{Nason:2004rx, Frixione:2007vw, Alioli:2010xd, Frixione:2007nw}
at next-to-leading-order (NLO) in $\alpha_s$ with
the NNPDF3.0NLO set of parton distribution functions (PDF) in the matrix element
calculation. The parton shower, fragmentation, and the underlying event were
simulated using \PYTHIA 8.210~\cite{Sjostrand:Pythia8} with the NNPDF2.3LO PDF
sets~\cite{RichardBalla:2013, RichardBalla:2015} and the corresponding A14 set
of tuned parameters~\cite{ATL-PHYS-PUB-2014-021}. The $h_{\mathrm{damp}}$
parameter, which controls the \pT of the hardest additional parton emission
beyond the Born configuration, was set to $1.5
m_t$~\cite{ATL-PHYS-PUB-2016-020}, where $m_t$ denotes the top-quark mass. 
The \powheg hardness criterion used in the matching
(\texttt{POWHEG:pTdef}) is set to 2 following a study in
Ref.~\cite{ATL-PHYS-PUB-2016-020}. The renormalisation and factorisation scales
were set to $\mu = \sqrt{m_t^2 + p_{\mathrm{T}, t}^2}$, where $p_{\mathrm{T},
  t}$ is the transverse momentum of the top quark. Additional jets, including
$b$-jets, were generated by the hardest additional parton emission and from
parton showering. This sample is called \ppyeight in the following.

Processes involving the production of a $W, Z$ or Higgs boson in addition to a
\ttbar pair were simulated using the \MGMCatNLO generator~\cite{Alwall:2014hca,ATL-PHYS-PUB-2016-005} at \ac{NLO} in $\alpha_s$ in the matrix
element calculation. The parton shower, fragmentation and underlying event were
simulated using \PYTHIAV{8} with the \textsc{A14} parton shower tune. A dynamic
renormalisation and factorisation scale set to
$H_{\mathrm{T}}/2$ was used, where $H_{\mathrm{T}}$ is defined as the scalar sum of the transverse mass, $m_{\mathrm{T}}=\sqrt{m^2 + \pt^2}$, of all partons in the partonic final state. The \textsc{NNPDF3.0NLO} PDF set was used in the matrix element calculation while the \textsc{NNPDF2.3LO} PDF set was used in the parton
shower. In the case of $\ttbar H$, the Higgs boson mass was set to 125~\GeV\ and
all possible Higgs decay modes were allowed, with the branching fractions
calculated with \textsc{HDECAY}~\cite{yellowreport,Djouadi:1997yw}. The $\ttbar
W$ and $\ttbar Z$ samples are normalised to cross-sections calculated to
\ac{NLO} in $\alpha_s$ with \MGMCatNLO. The $\ttbar H$ sample is normalised to a
cross-section calculated to \ac{NLO}
accuracy in QCD, including NLO electroweak corrections~\cite{yellowreport}.

Alternative \ttbar samples were generated to assess the uncertainties due to a
particular choice of QCD MC model for the  production of the additional $b$-jets and to compare with unfolded data, as listed in Table~\ref{tab:mc_signal}. 
In order to investigate the effects of initial- and final-state radiation, two
samples were generated using \ppyeight with the renormalisation and factorisation scales
varied by a factor of 2 (0.5) and using low-radiation (high-radiation) variations of the
A14 tune and an $h_{\mathrm{damp}}$ value of $1.5 m_t$ ($3.0 m_t$),
corresponding to less (more) parton shower
radiation~\cite{ATL-PHYS-PUB-2016-020}. These samples are called \ppyeight
(RadLo) and \ppyeight (RadHi) in the following. To estimate the effect of the choice of parton shower and hadronisation algorithms, a
\ac{MC} sample was generated by interfacing \powheg with
\hwseven~\cite{herwig,herwig7tune} (v7.01) using the H7UE set of tuned
parameters~\cite{herwig7tune}.

In order to estimate the effects of QCD scales, and matching and merging
algorithms used in the NLO \ttbar matrix element calculation and the parton shower to
predict additional $b$-jets, events were generated with the \SHERPAV{2.2.1}
generator~\cite{Gleisberg:2008ta}, which models the zero and one
additional-parton process at NLO accuracy and up to four additional partons at
LO accuracy, using the \textsc{MePs@Nlo} prescription~\cite{Hoeche:2012yf}.
Additional $b$-quarks were treated as massless and the \textsc{NNPDF3.0NNLO} PDF
set was used. The calculation uses its own parton shower tune. This sample is
referred to as \SHERPAtt.

In addition to the \ttbar samples described above, a \ttbar sample was generated
using the \MGMCatNLO~\cite{Alwall:2014hca} (v2.3.3) generator, interfaced to
\PYTHIAV{8.210} and is referred to as \amcnlopyeight hereafter. As with the
nominal \ppyeight \ttbar sample, the \textsc{NNPDF3.0NLO} PDF set was used in
the matrix element calculation and the \textsc{NNPDF2.3LO} PDF set was used in
the parton shower. This sample is used to calculate the fraction of \ttbar
+$V$/$H$ events in \ttbar events and to compare with the data. The A14 set of
tuned parameters was used for \PYTHIA.

The \ttbar samples are normalised to a cross-section of $\sigma_{\ttbar} =
832^{+46}_{-51}$ pb as calculated with the Top++2.0 program to
\ac{NNLO} in perturbative QCD, including soft-gluon
resummation to next-to-next-to-leading-log (NNLL) order (see Ref.~\cite{Czakon:2011xx}
and references therein), and assuming ${m_t=172.5}$~\GeV.
The uncertainty in the theoretical cross-section comes
from independent variations of the factorisation and renormalisation scales
and variations in the PDF and $\alphas$, following the \textsc{PDF4LHC}
prescription with the MSTW 2008 NNLO, CT10 NNLO and NNPDF2.3 5f FFN PDF
sets (see Ref.~\cite{Botje:2011sn} and references therein, and
Refs.~\cite{Martin:2009bu,Gao:2013xoa,Ball:2012cx}).

Four more predictions were calculated only for comparisons with data and are
all based on \ttbb matrix element calculations. These predictions all use the same
renormalisation and factorisation scale definitions as the study presented in
Ref.~\cite{yellowreport}. The renormalisation scale, $\mu_{\mathrm{R}}$, is set
to $\mu_{\mathrm{R}}~=~\prod_{i=t,\bar{t},b,\bar{b}}~E_{\mathrm{T}_i}^{1/4}$,
where $ E_{\mathrm{T}_i}$ refers to the transverse energy of the parton $i$ in
the partonic final state, and the factorisation scale, $\mu_{\mathrm{F}}$, is set to $H_{\mathrm{T}}/2$ which is defined as
\begin{linenomath}
\begin{equation*}
    \mu_{\mathrm{F}} = H_{\mathrm{T}}/2 = \frac{1}{2}  \sum_{i=t,\bar{t},b,\bar{b},j}{E_{\mathrm{T},i}}\,,
\end{equation*} 
\end{linenomath}
 where $j$ refers to the additional QCD-radiated partons at NLO. 
 
  Three of the four predictions are based on the \powheg method, and use the
  \PYTHIAV{8} parton shower with the same parton shower tune and the same
  matching settings as the nominal \ppyeight sample, with the exception of the
  \hdamp parameter, which is set to the same value as the factorisation scale,
  i.e. $H_{\mathrm{T}}/2$. In the \ttbb matrix element calculations with massive $b$-quarks, the
  $b$-quark mass is set to ${m_b=4.75}$~\GeV. The set-up of the four dedicated samples are described below.

  %
  A sample of \ttbb events was generated using
  \SHERPAOL~\cite{Cascioli:2013era}. The \ttbb matrix elements were calculated with
  massive $b$-quarks at \ac{NLO}, using the \textsc{Comix}~\cite{Gleisberg:2008fv} and \textsc{OpenLoops}~\cite{Cascioli:2011va} matrix
  element generators, and merged with the \SHERPA parton shower, tuned by the
  authors~\cite{Schumann:2007mg}. The four-flavour \ac{NNLO} NNPDF3.0 PDF set was
  used. The resummation scale,  $ \mu_{\mathrm{Q}} $, was set to the same value
  as $\mu_{\mathrm{F}}$.
  This sample is referred to as \SHERPAttbb (4FS).
  %
  %
  A sample of \ttbb events was generated using the \powhel
  generator~\cite{Garzelli:2014aba}, where the matrix elements were calculated at
  \ac{NLO} assuming massless $b$-quarks and using the five-flavour
  NLO~NNPDF3.0~PDF set. Events were required to have the invariant mass,
  $m_{bb}$, of the $b\bar{b}$ system to be larger than 9.5 GeV and the $\pt$ of
  the $b$-quark larger than 4.75 GeV as described in Ref.~\cite{yellowreport}. These events were matched to the \PYTHIAV{8} parton shower
  using the \powheg method. 
  This sample is referred to as \powhelpyeight (5FS).

  A sample of \ttbb events using the \powhel generator where the
  matrix elements were calculated at \ac{NLO} with massive $b$-quarks and using
  the four-flavour NLO~NNPDF3.0~PDF set~\cite{Bevilacqua:2017cru}. 
  Events were matched to the \PYTHIAV{8} parton shower using the \powheg method. 
 This sample is referred to as \powhelpyeight (4FS).

  A sample of \ttbb events using the \powheg generator where \ttbb
  matrix elements were calculated at \ac{NLO} with massive $b$-quarks and using
  the four-flavour NLO~NNPDF3.0~PDF set~\cite{Jezo:2018yaf}. Events were matched
  to the \PYTHIAV{8} parton shower using the \powheg method. This sample is
  referred to as \ppyeightttbb (4FS) to distinguish it from the nominal
  \ppyeight sample mentioned above. 

 For all samples involving top quarks, $m_t$ was set to 172.5~\GeV\ and the
 \textsc{EvtGen} v1.2.0 program~\cite{EvtGen} was used for properties of the
 bottom and charm hadron decays except for the \SHERPA samples. To preserve the
 spin correlation information, top quarks were decayed following the method of
 Ref.~\cite{Frixione:2007zp} which is implemented in \POWHEGBOX and by \MADSPIN~\cite{Artoisenet:2012st} in the \amcnlopyeight samples. \SHERPA performs its
 own calculation for spin correlation. Both of the \powhelpyeight samples used
 \PYTHIA to decay the top quarks, with a top-quark decay width of $1.33~\GeV$, and hence
 these predictions do not include \ttbar spin correlations.

The production of single top-quarks in the $tW$- and $s$-channels was simulated
using the \POWHEGBOX (v2, r2819) \ac{NLO} generator with the CT10 PDF set in
the matrix element calculations. Electroweak $t$-channel single-top-quark
events were generated using the \POWHEGBOX (v1, r2556) generator. This
generator uses the four-flavour scheme for the \ac{NLO} matrix elements
calculation together with the fixed four-flavour PDF set CT10f4. For all top
processes, top-quark spin correlations are preserved (in the case of the
$t$-channel, top quarks were decayed using \MADSPIN). The interference between \ttbar
and $tW$ production is accounted for using the
diagram-removal scheme~\cite{Frixione:2008yi}.
The parton shower, fragmentation, and the underlying event were simulated using
\PYTHIAV{6.428}~\cite{Sjostrand:2006za} with the CTEQ6L1 PDF sets and the
Perugia 2012 tune (P2012)~\cite{Pumplin:2002vw,Skands:2010ak}.
The single-top \ac{MC} samples for the $t$- and $s$-channels are normalised to
cross-sections from NLO predictions~\cite{Aliev:2010zk,Kant:2014oha},
while the $tW$-channel \ac{MC} sample is normalised to
approximate NNLO~\cite{Kidonakis:2010ux}.

Events containing $W$ or $Z$ bosons with associated jets were simulated using the
\SHERPAV{2.2.1} generator. Matrix elements were calculated for up to two partons
at \ac{NLO} and up to four partons at \ac{LO} using the \textsc{Comix} and
\textsc{OpenLoops} matrix element generators and merged with the \SHERPA parton
shower using the \textsc{MePs@Nlo} prescription. The NNPDF3.0NNLO PDF set was used in
conjunction with parton shower tuning developed by the \SHERPA authors. The
$W/Z+$jets events are normalised to \ac{NNLO} cross-sections, computed using
\textsc{Fewz}~\cite{Anastasiou:2003ds} with the MSTW 2008 NNLO PDF set.

Diboson processes were simulated using the \SHERPAV{2.1.1} generator.
Matrix elements were calculated using the \textsc{Comix}
and \textsc{OpenLoops} matrix element generators and merged with the \SHERPA
parton shower using the \textsc{MePs@Nlo} prescription.
In the case of both bosons decaying leptonically, matrix
elements contain all diagrams with four electroweak vertices and were calculated
for up to one (four charged leptons or two charged leptons and two neutrinos)
or zero partons (three charged leptons and one neutrino) at \ac{NLO},
and up to three partons at \ac{LO}. In the cases where one of the bosons decays
hadronically and the other leptonically, matrix elements were calculated with up
to one ($ZZ$) or zero ($WW, WZ$) additional partons at \ac{NLO} and up to three
additional partons at \ac{LO}. The CT10 PDF set was used in conjunction with parton shower tuning developed by the \SHERPA authors. 

In all \ac{MC} simulation samples, the effect of multiple $pp$ interactions per bunch
crossing (pile-up) was modelled by adding multiple minimum-bias events simulated
with \PYTHIAV{8.186}~\cite{Sjostrand:Pythia8}, the A2 set of tuned
parameters~\cite{ATL-PHYS-PUB-2012-003} and the MSTW2008LO set of
PDFs~\cite{MartinStirlingThrone:2009}. The MC simulation samples are re-weighted to reproduce the distribution of the mean number of interactions per bunch crossing observed in the data.
%


%% file: sections/object_selection.tex
\section{Object reconstruction and identification}
\label{sec:object_selection}
\subsection{Detector-level object reconstruction}
\label{sec:det_level_object_selection}
A description of the main reconstruction and identification criteria applied for
electrons, muons, jets and $b$-jets is given below.

Electrons are reconstructed~\cite{ATLAS-CONF-2016-024} by matching \ac{ID} tracks to clusters in the
electromagnetic calorimeter. Electrons must satisfy the \textit{tight}
identification criterion, based on a likelihood discriminant combining
observables related to the shower shape in the calorimeter and to the track
matching the electromagnetic cluster, and are required to be isolated in both
the \ac{ID} and the EM calorimeter using the \pt-dependent isolation working
point. Electrons are required to have
$\pt > 25$~\GeV\ and $|\eta_{\mathrm{cluster}}| < 2.47$. Electrons that fall in
the transition region between the barrel and endcap calorimeters ($1.37 <
|\eta_{\mathrm{cluster}}| < 1.52$) are poorly measured and are therefore not
considered in this analysis.

Muon candidates are reconstructed~\cite{PERF-2015-10} by matching \ac{ID} tracks to tracks in
the muon spectrometer. Track reconstruction is performed independently in the
\ac{ID} and MS before a combined track is formed with a global re-fit to hits
in the \ac{ID} and MS.
Muon candidates are required to have $\pT > 25$~\GeV\ and $|\eta| < 2.5$,
must satisfy the \textit{medium} identification criteria and are required to
be isolated using the \pt-dependent isolation working
point.

Electron and muon tracks are required to be associated with the primary vertex. This
association requires the electron (muon) track to have
$|d_0|/\sigma_{d_0} < 5~(3)$ and $|\Delta z_0 \sin\theta | < 0.5 \mathrm{~mm}$,
where $d_0$ and $z_0$ are the transverse and longitudinal impact parameters of
the electron (muon) track, respectively, $\sigma_{d_0}$
is the uncertainty in the measurement of $d_0$, and $\theta$ is the angle of
the track relative to the axis parallel to the beamline.

Reconstruction, identification and isolation efficiencies of electrons (muons)
are corrected in simulation to match those observed in data using
$Z\to e^+ e^- (\mu^+ \mu^-)$ events, and the position and width of the observed
$Z$ boson peak
is used to calibrate the electron (muon) energy (momentum) scale and
resolution.

  The \antikt algorithm~\cite{Cacciari:2008gp} with a radius parameter
  of $R=0.4$ is used to
  reconstruct jets with a four-momentum recombination scheme,
  using energy deposits in topological clusters in the calorimeter
  as inputs~\cite{PERF-2014-07}. Jets are calibrated using a series of
  simulation-based corrections and \textit{in situ} techniques~\cite{PERF-2016-04}. Calibrated jets are required to have $\pT > 25$~\GeV\ and
  $|\eta| < 2.5$ so that data from the \ac{ID} is available for determining whether they
  contain $b$-hadrons. Jets with $\pt < 60$~\GeV\ and $|\eta| < 2.4$ are required
  to be identified as originating from the primary vertex using a \ac{JVT}
  algorithm~\cite{ATLAS-CONF-2014-018}.

 Jets containing $b$-hadrons are identified exploiting the lifetimes of $b$-hadrons and their masses. 
 A multivariate algorithm, MV2c10, that combines track and secondary-vertex information is used to distinguish
 $b$-jets from other jets~\cite{ATL-PHYS-PUB-2016-012}. 
 Four working points are defined by
 different $b$-tagging discriminant output thresholds corresponding to efficiencies
 of 85\%, 77\%, 70\% and 60\% in simulated \ttbar events for $b$-jets with $\pt
 > 20$~\GeV\ and rejection factors ranging from 3--35 for $c$-jets and 30--1500
 for light-flavour jets~\cite{ATL-PHYS-PUB-2016-012, PERF-2016-05}.

  After selecting electrons, muons and jets as defined above, several criteria
  are applied to ensure that objects do not overlap. If a selected electron and muon
  share a track then the electron is rejected. If an electron is within $\Delta
  R = 0.2$ of one or more jets then the closest jet to the electron is removed.
  If there are remaining jets within $\Delta R = 0.4$ of an electron then the
  electron is removed. When a jet is within $\Delta R = 0.4$ of a muon, it is
  removed if it has fewer than three tracks, otherwise the muon is removed.

\subsection{Particle-level object definitions}
\label{sec:plevel_object_selection}
Particle-level objects are selected in simulated events using definitions that
closely match the detector-level objects defined in
Section~\ref{sec:det_level_object_selection}.
 Particle-level objects are defined
using stable particles having a proper lifetime greater than 30~ps.

This analysis considers electrons and muons that do not come from hadron decays
for the fiducial definition.\footnote{Electrons and muons from $\tau$ decays are
  thus included.} In order to take into account final-state photon radiation, the
four-momentum of each lepton is modified by adding to it the four-momenta of all photons,
not originating from a hadron, that are located within a $\Delta R = 0.1$ cone around the lepton. 
Electrons and muons are required to have $\pT > 25$~\GeV\ and $|\eta| < 2.5$.

Jets are clustered using the \antikt algorithm with a radius parameter of 0.4.
All stable particles are included except those identified as
electrons and muons, and the photons added to them, using the
definition above and neutrinos not from hadron decays.
 These jets do not
include particles from pile-up events but do include those from the underlying
event. The decay products of hadronically decaying $\tau$-leptons are
therefore included. Jets are required to have $\pT > 25$~\GeV\ and $|\eta| <
2.5$.

Jets are identified as $b$-jets by requiring that at least one $b$-hadron with
$\pT > 5$~\GeV\ is matched to the jet by ghost association~\cite{Cacciari:2008gn}. 
Here, the ghost-association procedure includes $b$-hadrons in the jet clustering after
scaling their \pt to a negligible value. A similar procedure is followed
to define $c$-jets, with the $b$-jet definition taking precedence, i.e.\ a jet
containing one $b$-hadron and one $c$-hadron is defined as a $b$-jet. Jets
that do not contain either a $b$-hadron or a $c$-hadron are considered to be light-flavour jets.

Electrons and muons that meet the selection criteria defined above are
required to be separated from selected jets by 
${\Delta R(\mathrm{lepton}, \mathrm{jet}) > 0.4}$.
This ensures compatibility with the detector-level selection defined in
Section~\ref{sec:det_level_object_selection}.

%% file: sections/event_selection.tex
\FloatBarrier
\section{Event selection and definition of the fiducial phase space}
\label{sec:event_selection_fiducial_def}
\subsection{Data event selection}
\label{sec:event_selection}

The data analysed were collected by the ATLAS detector in 2015 and 2016 during
stable $pp$ collisions at $\sqrt{s}=13$~\TeV\ while all components of the ATLAS
detector were fully operational.
The total integrated luminosity recorded in this period is \lumi.

In order to ensure events originate from $pp$ collisions, events are required to
have at least one primary vertex with at least two tracks. The primary vertex is
defined as the vertex with the highest $\sum \pt^2$ of tracks assigned to it.

Single-electron or single-muon triggers are used to select the events. 
They require a \pt of at least 20 (26)~\GeV\ for muons and 24 (26)~\GeV\ for 
electrons for the 2015 (2016) data set and also include requirements on
the lepton quality and isolation. These triggers are complemented by others
with higher \pt requirements but loosened isolation requirements to ensure
maximum efficiencies at higher lepton \pt.

In the $e\mu$ channel, events are required to have exactly one electron and one
muon of $\pt > 27$~\GeV\ and with opposite electric charge. At least one of the two leptons must be matched in flavour and angle to a trigger object.
In the \ljets channel, exactly one selected lepton of $\pt > 27$~\GeV\ is required and must be
matched to the trigger object that triggered the event.

In the $e\mu$ channel, at least two jets are required and at least two of these
must be $b$-tagged at the 77\% efficiency $b$-tagging working point for
the baseline selection. The measurement of the fiducial cross-section with
one (two) additional $b$-jets requires at least three (at least four) jets to be $b$-tagged.
For the measurement of the $b$-jet multiplicity distribution, at least two jets
are required and at least two of them must be $b$-tagged. All other
differential cross-section measurements in the $e\mu$ channel require at least three
jets  and at least three of these must be $b$-tagged.

In the \ljets channel, at least five jets are required and at least two of these
must be $b$-tagged for the baseline selection. For the measurement of the
fiducial cross-section with one (two) additional $b$-jets, five (six) jets are
required, of which at least three (at least four) must be $b$-tagged. For the
measurement of the differential cross-sections, at least six jets, at least four
of which are $b$-tagged, are required. In this channel, $b$-jets are identified
using the tighter 60\% efficiency $b$-tagging working point to better suppress
$c$-jets from $W^- \to \bar{c}s$ or $W^+ \to c\bar{s}$ decays.
 
\subsection{Fiducial phase-space definition}
\label{sec:fiducialphasespace}
The phase space in which the fiducial cross-section is measured is defined using particle-level objects with
kinematic requirements similar to those placed on reconstructed objects in the
event selection. The definitions of the fiducial phase spaces used for the cross-sections measurements are given below. The data are corrected to particle level using slightly
different definitions of the fiducial phase space depending on the top-pair
decay channel and on the observable.

In the $e\mu$ channel, fiducial cross-sections are determined by requiring exactly
one electron and one muon with opposite-sign charge at particle level and at least
three (at least four) $b$-jet(s) for the fiducial cross-section with one (two)
additional $b$-jets. The normalised differential cross-sections are measured in
the fiducial volume containing the leptons and at least two $b$-jets for the
distribution differential in number of $b$-jets and at least three $b$-jets for
all other differential measurements.

In the \ljets channel, the fiducial phase space for the measurement of the
integrated cross-section with one (two) additional $b$-jet(s) is defined as containing
exactly one particle-level electron or muon and five (six) jets, at least three
(four) of which are $b$-jets. Differential cross-sections are measured
in a fiducial volume containing at least six jets and where at least four of them are
required to be $b$-jets.

%% file: sections/background_estimation.tex
\FloatBarrier
\section{Background estimation}
\label{sec:background_estimation}
The baseline selection with at least two $b$-tagged jets results in a sample with
only small backgrounds from processes other than \ttbar production.
As mentioned before, events with additional \bjets produced in \ttV or \ttH production
are treated as signal.
The estimation of \ttbar production in association with additional light-flavour jets or
$c$-jets is described in Section~\ref{sec:heavy_flavour_scale_factors} and is
performed simultaneously with the extraction of fiducial cross-sections.

The remaining background events are classified into two types: those with prompt leptons from single top, $W$ or $Z$ decays (including those produced via
leptonic $\tau$ decays), which are discussed in Section~\ref{sec:promptbackground},
and those where at least one of the reconstructed lepton candidates is
non-prompt or ``fake'' (NP \& fake lep.), i.e.\ a non-prompt lepton from the
decay of a $b$- or $c$-hadron, an electron from a photon conversion, hadronic
jet activity misidentified as an electron, or a muon produced from an in-flight
decay of a pion or kaon. This is estimated using a combined data-driven and
simulation-based approach in the $e\mu$ channel, and a data-driven approach in
the \ljets channel, both of which are described in Section~\ref{sec:fakes}.
\subsection{Background from single-top, $Z/\gamma^* + $ jets and \Wjets events}
\label{sec:promptbackground}
The background from single top-quark production is estimated from the MC simulation 
predictions in both the $e\mu$ and \ljets channels. 
This background contributes 3\% of the event yields in both channels, with 
slightly smaller contributions in the four \bjets selections.

In the $e\mu$ channel, a very small number of events from Drell--Yan production and
$Z/\gamma^* (\rightarrow\tau\tau)$+jets
fulfil the selection criteria. This background is estimated from MC simulation
scaled to the data with separate scale factors for the two-$b$-tagged jets and three-$b$-tagged jets cases.
The scale factors are derived from data events that have a reconstructed mass of the
dilepton system corresponding to the $Z$ boson mass and that fulfil the
standard selection except that the lepton flavour is $ee$ or $\mu\mu$. The
fraction of background events from $Z/\gamma^* (\rightarrow\tau\tau)$+jets
is below two per mill for all $b$-tagged jet multiplicities. A small number of $Z/\gamma^*$+jets events, where the $Z/\gamma^*$ is decaying into any lepton flavour pair, can enter in the \ljets channel and is estimated from MC simulation. 


In the \ljets channel, a small background from \Wjets remains
after the event selection; however, this contribution is below 2\% in events that have
at least three $b$-tagged jets. This background is estimated directly from MC simulation.

\subsection{Background from non-prompt and fake leptons}
\label{sec:fakes}
In the $e\mu$ channel, the normalisation of this background is estimated from
data using events in which the electron and muon have the same-sign electric charge. The method is described in Ref.~\cite{TOPQ-2015-09}. Known
sources of same-sign prompt leptons are subtracted from the data and the
non-prompt and fake background is extracted by scaling the remaining data events
by a transfer factor determined from MC simulation. This transfer factor is
defined as the ratio of predicted opposite-sign to predicted same-sign
non-prompt and fake leptons.

In the \ljets channel, the background from non-prompt and fake leptons
is estimated using the \textit{matrix method}~\cite{ATLAS-CONF-2014-058}. A sample enriched in non-prompt and fake leptons is obtained by
removing the isolation and impact parameter requirements on the lepton
selections defined in Section~\ref{sec:object_selection}. The efficiency for
these leptons, hereafter referred to as \textit{loose} leptons,
to meet the identification criteria defined in Section~\ref{sec:det_level_object_selection} is then measured
separately for prompt and fake leptons. \footnote{Here fake leptons also include non-prompt leptons.} For both electrons and muons the
efficiency for a prompt loose lepton to pass the identification criteria defined in Section~\ref{sec:det_level_object_selection} is measured
using a sample of $Z$ boson decays. The efficiency for fake loose leptons
to pass the identification criteria is measured using events that have low missing transverse
momentum for electrons and high lepton impact-parameter significance for muons.
These efficiencies allow the number of fake leptons selected in the
signal region to be estimated.
\begin{figure}
  \centering
  \subcaptionbox{\label{fig:nbjets_emu_prefit}}{
    \includegraphics[width=0.45\textwidth]{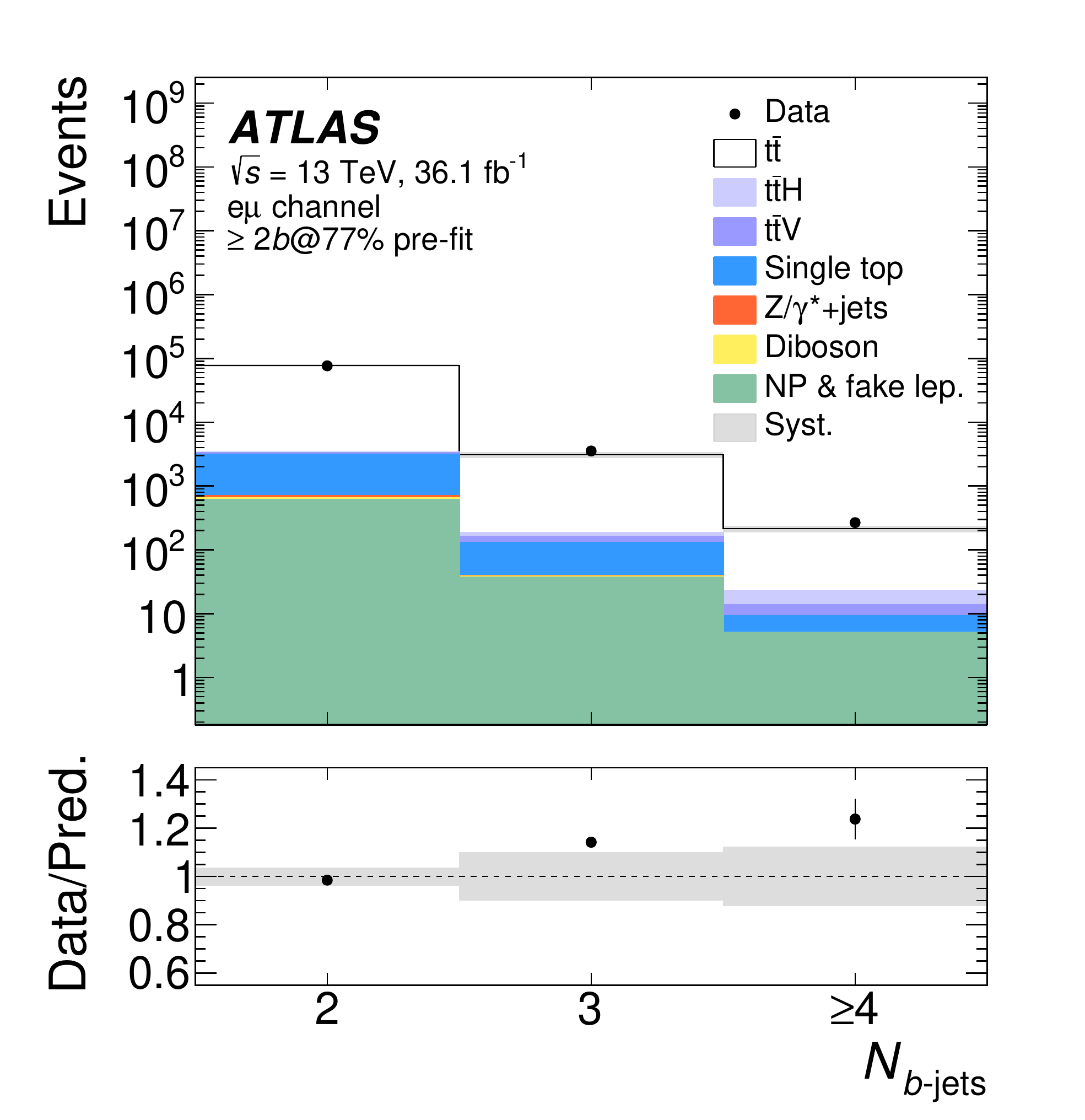}
  }
  \subcaptionbox{\label{fig:nbjets_ljets_prefit}}{
    \includegraphics[width=0.45\textwidth]{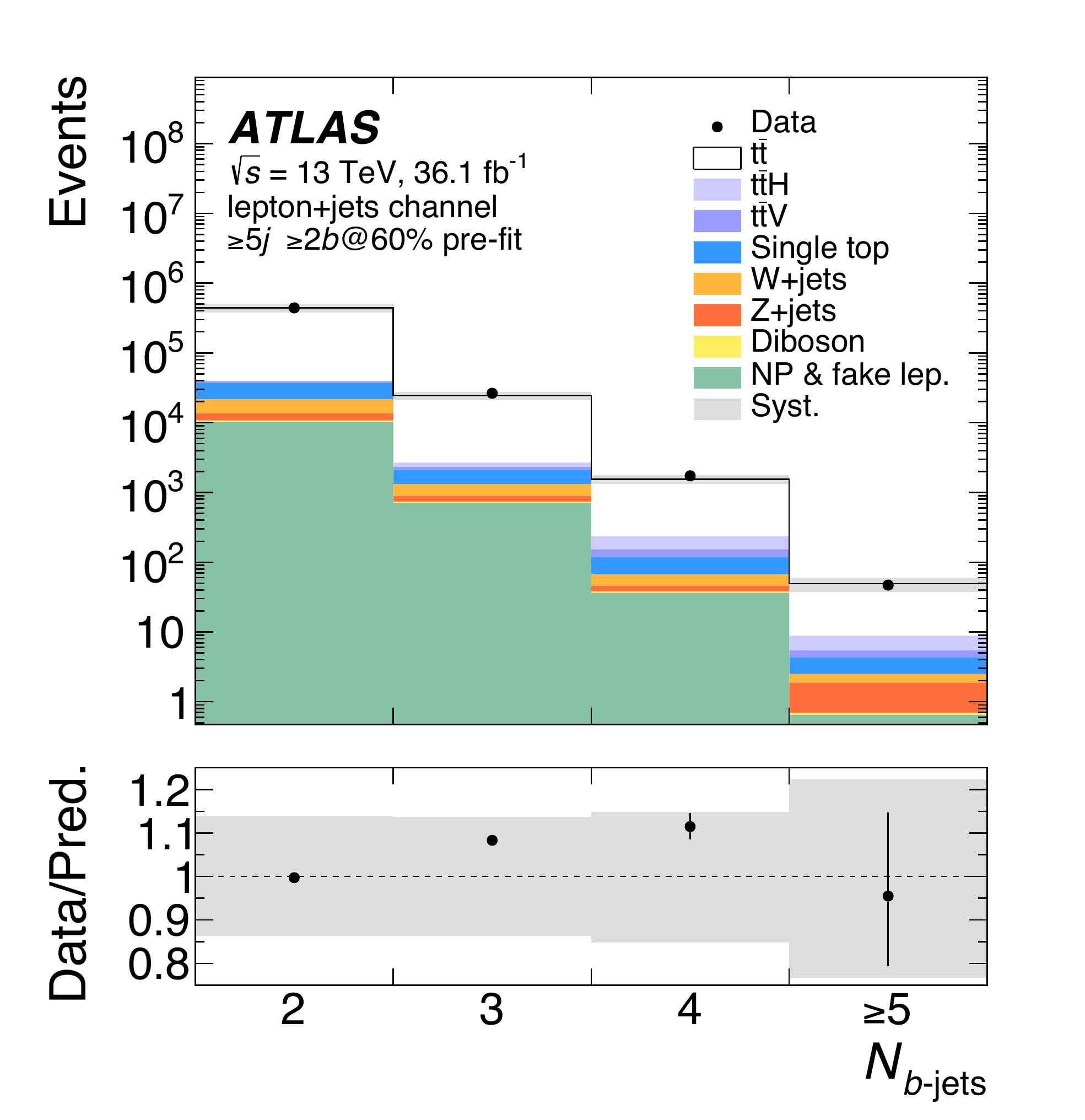}
  }
  \caption{Comparison of the data distributions with predictions for the
    number of $b$-tagged jets, in events with at least 2 $b$-tagged jets, in the
    \subref{fig:nbjets_emu_prefit} $e\mu$ and \subref{fig:nbjets_ljets_prefit}
    \ljets channels. The systematic uncertainty band, shown in grey, includes
    all uncertainties from experimental sources.}
  \label{fig:reco_prefit_nbjets}
  \vspace*{\floatsep}
  \centering
  \subcaptionbox{\label{fig:leadbjetpt_emu_prefit}}{
    \includegraphics[width=0.45\textwidth]{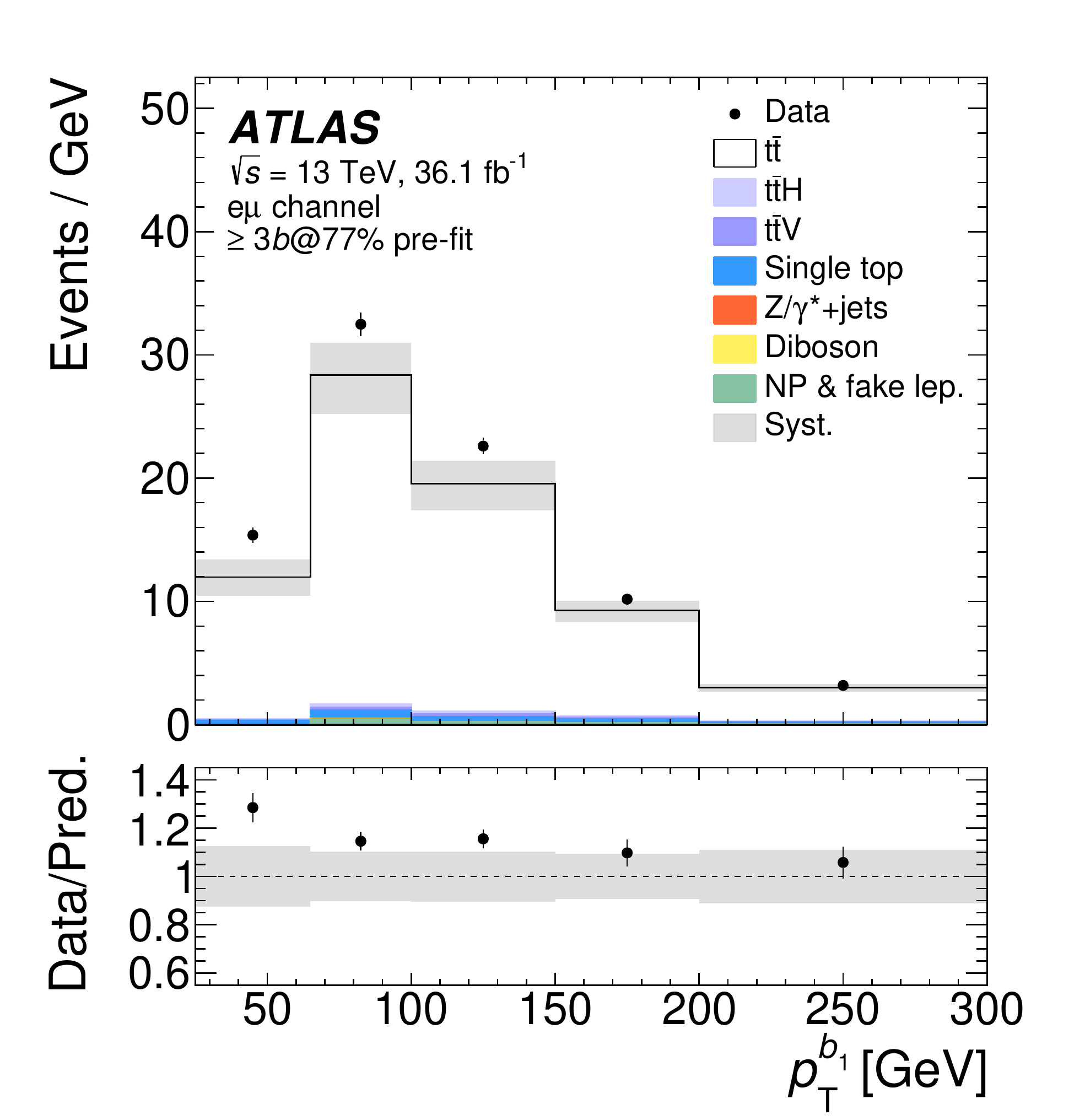}
  }
  \subcaptionbox{\label{fig:leadbjetpt_ljets_prefit}}{
    \includegraphics[width=0.45\textwidth]{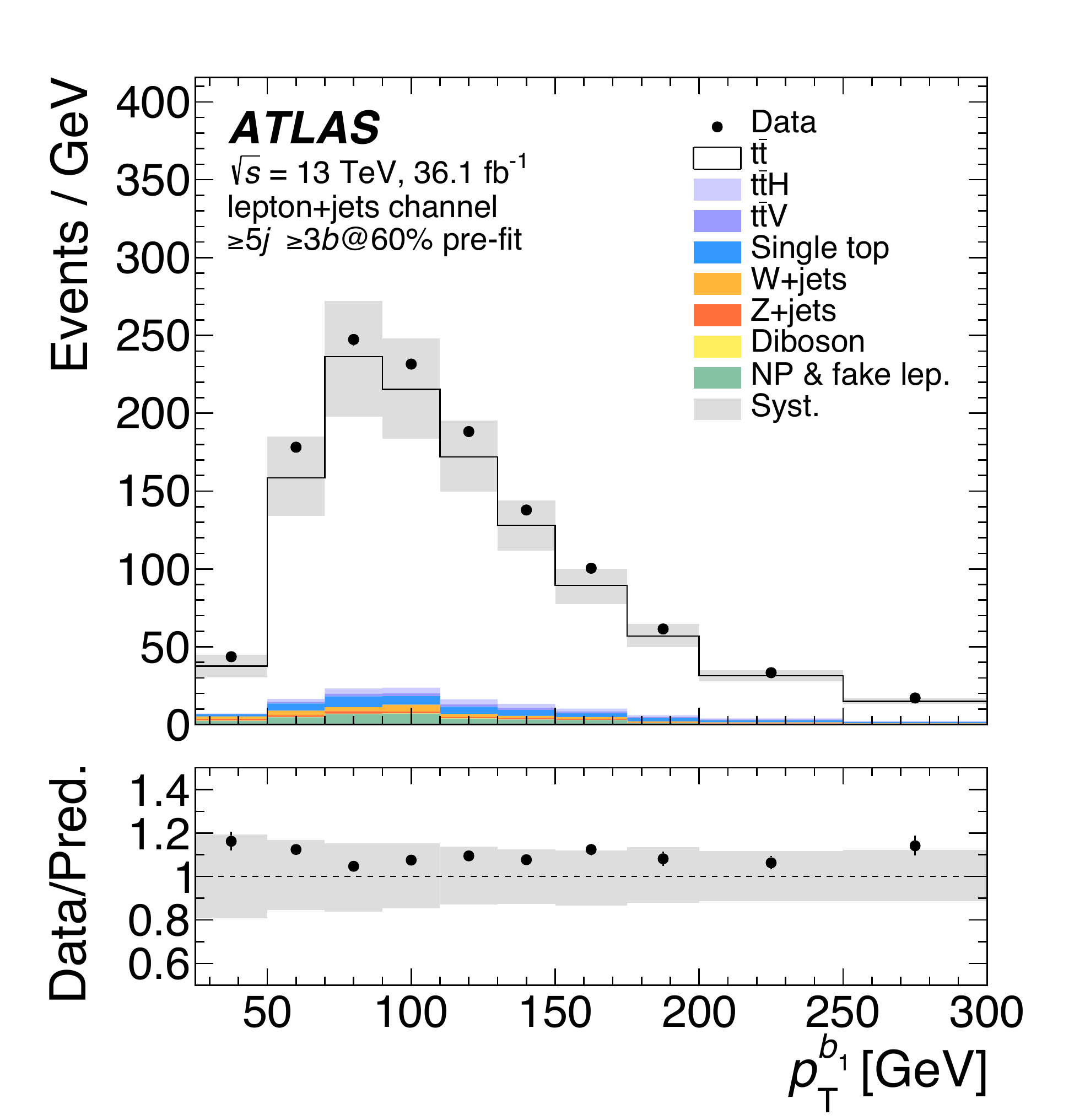}
  }
  \caption{
    Comparison of the data distributions with predictions for the
    leading $b$-tagged jet \pt, in events with at least 3 $b$-tagged jets, in the
    \subref{fig:leadbjetpt_emu_prefit} $e\mu$ and \subref{fig:leadbjetpt_ljets_prefit}
    \ljets channels. The systematic uncertainty band, shown in grey, includes
    all uncertainties from experimental sources. Events that fall outside of the
    range of the $x$-axis are not included in the plot.}
  \label{fig:reco_prefit_bjet_pt}
\end{figure}
%
%
\input{tables/eventYieldTabledilepton}
\input{tables/eventYieldTablelepjets}
\FloatBarrier
\subsection{Data and prediction comparison of baseline selection}
The overall number of events fulfilling the baseline selection is well
described by the prediction in both channels, as seen in
Tables~\ref{tab:emu_yields} and~\ref{tab:ljets_yields} and
Figure~\ref{fig:reco_prefit_nbjets}, where $b$ and $j$ denote a $b$-jet and a jet of any flavour, respectively. However, the number of events with more
than two $b$-tagged jets is slightly underestimated, as shown in
Figures~\ref{fig:reco_prefit_nbjets} and~\ref{fig:reco_prefit_bjet_pt}.
Therefore, data-driven scale factors are derived to correct the predictions of
additional $c$-jets or light jets in the \ttbar MC simulation, as
described in the next section.

%% file: tables/eventYieldTabledilepton.tex
\begin{table}
  \centering
  \caption{Predicted and observed $e\mu$ channel event yields in 2$b$,
    $\geq$ 3$b$ and $\geq$ 4$b$ selections. The quoted errors are symmetrised and indicate total
    statistical and systematic uncertainties in predictions due to experimental sources.
  }
  \label{tab:emu_yields}
  \sisetup{round-mode=figures, round-precision=2,group-digits=integer, group-minimum-digits=4}
  \renewcommand{\arraystretch}{1.15}
  \begin{tabular}{l
    S[table-format=5.1, table-number-alignment=right]@{$\,\pm\,$}S[table-format=4.1]
    S[table-format=4.1, table-number-alignment=right]@{$\,\pm\,$}S[table-format=3.1]
    S[table-format=3.2, table-number-alignment=right]@{$\,\pm\,$}S[table-format=2.2]}
    \toprule
    Process                                 & \multicolumn{2}{c}{$2b$}                        & \multicolumn{2}{c}{$\geq3b$}                   & \multicolumn{2}{c}{$\geq4b$}                                                 \\
    \midrule
    Signal ($\ttbar + \ttbar H + \ttbar V$) & \sisetup{round-precision=3} 74449.21               & 2885.66                                          & 3177.31                             & 313.13 & 206.50                            & 29.36 \\
    \hspace{3mm} $\ttbar$                                & \sisetup{round-precision=3} 74213.52               & 2885.61                                           & 3107.32                             & 312.98 & 192.67                            & 29.09 \\
    \hspace{3mm} $\ttbar H$                              & \sisetup{round-precision=3}  45.34                 & 6.59                                              & \sisetup{round-precision=3} 36.48                               & 7.04   & 9.41                              & 3.29  \\
    \hspace{3mm} $\ttbar V$                              & \sisetup{round-precision=3} 190.35                 & 15.64                                             & \sisetup{round-precision=3} 33.51                               & 6.68   & 4.41                              & 2.18  \\
    \midrule 
    Background                              & \sisetup{round-precision=3} 3149.32                & 809.96                                            & 141.07                              & 53.43  & 9.24                             & 5.64  \\
    \hspace{3mm} Single top                              & \sisetup{round-precision=3} 2455.22                & \sisetup{round-precision=2} 539.49                & 95.75                               & 31.70  & 4.10                              & 2.49  \\
    \hspace{3mm} NP and fake lep.                        & 603.67                                             & 603.67                                            & 43.00                               & 43.00  & 5.06                              & 5.06  \\
    \hspace{3mm} $Z/\gamma^*$+jets                                & \sisetup{round-precision=2} 53                                                 & 13                                                & 1.3                                & \sisetup{round-precision=1} 0.3   & \sisetup{round-precision=1} 0.07                              & \sisetup{round-precision=1} 0.02  \\
    \hspace{3mm} Diboson                                 & \sisetup{round-precision=3} 38                                                 & 20                                                & 1.0                                & 1.1   & \multicolumn{2}{c}{$<0.01$}  \\
    \midrule
    Expected                                & \sisetup{round-precision=3} 77598.53               & 2997.18                                           & \sisetup{round-precision=3} 3318.37 & 317.66 & \sisetup{round-precision=3}215.74 & 29.90 \\
    Observed                                & \multicolumn{2}{l}{\num[round-mode=off]{76425}} & \multicolumn{2}{l}{\num[round-mode=off]{3809}} & \multicolumn{2}{l}{\num[round-mode=off]{267}}                                \\
    \bottomrule
  \end{tabular}
\end{table}

%% file: tables/eventYieldTablelepjets.tex
\begin{table}
  \small
  \centering
  \caption{Predicted and observed \ljets event yields in the $\geq 5j \geq2b$,
    $\geq5j \geq3b$, $\geq5j =3b$, and $\geq6j \geq4b$ selections. The quoted
    uncertainties are symmetrised and indicate total statistical and systematic uncertainties in
    predictions due to experimental sources.}
  \sisetup{round-mode=figures, round-precision=3, group-digits=integer, group-minimum-digits=4}
  \renewcommand{\arraystretch}{1.15}
  \setlength{\tabcolsep}{9pt}
  \begin{tabular}{
    l
    S[table-format=6.0, table-number-alignment=right]@{$\,\pm\,$}S[table-format=5.0]
    S[table-format=5.1, table-number-alignment=right]@{$\,\pm\,$}S[table-format=4.1]
    S[table-format=5.1, table-number-alignment=right]@{$\,\pm\,$}S[table-format=4.1]
    S[table-format=4.2, table-number-alignment=right]@{$\,\pm\,$}S[table-format=3.2, round-precision=2]
    }
    \toprule
        Process
    & \multicolumn{2}{c}{$\geq5j$, $\geq2b$}
    & \multicolumn{2}{c}{$\geq5j$, $\geq3b$}
    & \multicolumn{2}{c}{$\geq5j$, $=3b$}
    & \multicolumn{2}{c}{$\geq6j$, $\geq4b$}\\
    \midrule
    Signal ($\ttbar + \ttbar H + \ttbar V$)
        & 429000 & 42000
        & 23700  & 2200
        & 22300  & 2100
        & 1130   & 110  \\
    \hspace{3mm} \ttbar
        & 426000 & 42000
        & 23000  & 2200
        & 21700  & 2100
        & 1030   & 110  \\
    \hspace{3mm} $\ttbar H$
        & 1250   & 58
        & 437    & 23
        & 351    & 18
        & 68.3   & 5.8  \\
    \hspace{3mm} $\ttbar V$
        & 2020   & 110
        & 250    & 16
        & 215    & 14
        & 28.3   & 2.8  \\
    \midrule
    Background
        & 39500  & 7900
        & 2230   & 470
        & 2110   & 450
        & 87     & 23   \\
    \hspace{3mm} Single top
        & 16400  & 2000
        & 856    & 99
        & 803    & 94
        & 35.7   & 6.5  \\
    \hspace{3mm} NP and fake lep.
        & 11000  & 5500
        & 740    & 380
        & 710    & 360
        & 32     & 21   \\
    \hspace{3mm} $W$+jets
        & 8600   & 5300
        & 440    & 270
        & 410    & 260
        & 11.0   & 6.9  \\
    \hspace{3mm} $Z/\gamma^*$+jets
        & 2960   & 480
        & 164    & 26
        & 155    & 26
        & \sisetup{round-precision=2}5.9    & 1.5  \\
    \hspace{3mm} Diboson
        & 529    & 80
        & 34.0   & \sisetup{round-precision=2} 5.6
        & 32.0   & \sisetup{round-precision=2} 5.5
        & 1.79   & 0.58 \\
    \midrule
    Expected
        & 469000 & 42000
        & 26000  & 2300
        & 24400  & 2200
        & 1220   & 110  \\
    Observed
        & \multicolumn{2}{l}{\num[round-mode=off]{469793}}
        & \multicolumn{2}{l}{\num[round-mode=off]{28167}}
        & \multicolumn{2}{l}{\num[round-mode=off]{26389}}
        & \multicolumn{2}{l}{\num[round-mode=off]{1316}}    \\
    \bottomrule
  \end{tabular}
  \label{tab:ljets_yields}
\end{table}

%% file: sections/ttbar_bkg_fit.tex
\section{Extraction of the fiducial cross-sections}
\label{sec:fiducial_xsec}
Fiducial cross-sections in the phase spaces defined in Section~\ref{sec:fiducialphasespace} for the
different observables are extracted from detector-level distributions obtained
after the event selections described in Section~\ref{sec:event_selection} and
subtracting the number of background events produced by the non-\ttbar\ processes
described in Section~\ref{sec:background_estimation}.
After the subtraction of non-\ttbar\ background, the data suffer from backgrounds from \ttbar
events with additional light-flavour jets (\ttlight) or $c$-jets (\ttc) that are
misidentified as \bjets by the $b$-tagging algorithm. The correction
factors for these backgrounds are measured in data, as presented in Section~\ref{sec:heavy_flavour_scale_factors}.
The data are then unfolded using the
corrected MC simulation as described in Section~\ref{sec:unfolding}.
\subsection{Data-driven correction factors for flavour composition of additional jets in \ttbar events}
\label{sec:heavy_flavour_scale_factors}
The measurement of $\ttbar + b$-jets production is dependent on the
determination of the background from other \ttbar processes. For example,
according to simulation studies in the $e\mu$ channel, only about 50\% of the
events selected at detector level with at least three $b$-tagged jets at the
$77\%$ efficiency working point
and within the fiducial phase space of the analysis, also have
at least three \bjets at particle level. The other events contain at
least one \cjet or light-flavour jet which is misidentified as a \bjet.
The cross-section of \ttbar with additional jet production has been
measured with $10\%~(16\%)$ uncertainty for events with two (three)
additional jets~\cite{TOPQ-2015-17}. However, these measurements did
not determine the flavours of the additional jets.
Due to the lack of precise
measurements of these processes, template fits to data are performed
to extract the \ttb signal yields and estimate the \ttc and \ttlight
backgrounds as described in the following.  The templates are
constructed from \ttbar, \ttH and \ttV MC simulated samples, as the
signal includes the contributions from \ttV and \ttH.

The events in the $e\mu$ channel are selected within an analysis region consisting of
at least three $b$-tagged jets at the 77\%
$b$-tagging working point as specified in
Section~\ref{sec:event_selection}.  This avoids
extrapolation of the background shapes determined outside the selected region
into the analysis region. The fit in the \ljets channel is performed on a sample
with at least five jets, at least two of which are $b$-tagged with a $b$-tagging
efficiency of $60\%$. While this means that the \ac{MC} simulation is needed
to extrapolate the results of the fit into the signal regions, it allows the
\ttlight background to be extracted in what is effectively a control region. The
\ljets channel suffers from an additional background due to $W^+\rightarrow c \bar{s}$
 or corresponding $W^-$ decays in the inclusive \ttbar process,
 where the $c$-jet is misidentified as a $b$-jet.
In order to separate this background from
$t\bar{t}$+$c$-jets events, events containing only one particle-level $c$-jet
are attributed to this background and
grouped into a \ttlight class, while those with two particle-level $c$-jets are
placed into a \ttc class, as summarised in Table~\ref{tab:fitclass}.
In this sample, $85\%$ of the events with exactly one particle-level $c$-jet are
found to contain $W \to c\bar{s} (\bar{c}s)$ decays,
according to \ttbar \ac{MC} simulation.
\begin{table}
  \caption{Event categorisation (for the definition of the MC templates) based on the particle-level selections of $b$-jets,
    $c$-jets and light-flavour jets.}
  \small
  \begin{center}
    \begin{tabular}{ccc}
      \toprule
      Category & $e\mu$                                 & \ljets                       \\
      \midrule
      \ttb     & $\ge$3 $b$-jets                        & $\ge$3 $b$-jets                        \\
      \ttc     & $<$ 3 $b$-jets and $\ge$ 1 $c$-jet     & $<$ 3 $b$-jets and $\ge$ 2 $c$-jets    \\
      \ttlight & events that do not meet above criteria & events that do not meet above criteria \\
      \bottomrule
    \end{tabular}
  \end{center}
  \label{tab:fitclass}
\end{table}
Templates are created for events in the different categories described in
Table~\ref{tab:fitclass} using the $b$-tagging discriminant value of the jet with the third-highest $b$-tagging discriminant  in the  $e\mu$ channel, and the two jets with the third- and fourth-highest $b$-tagging discriminant values in the \ljets channel. 
%
%
The discriminant values are divided into five $b$-tagging discriminant bins
such that each bin corresponds to a certain range of $b$-tagging efficiencies
defined by the working points. The bins range from 1 to 5, corresponding to
efficiencies of 100\%--85\%, 85\%--77\%, 77\%--70\%, 70\%--60\%,
and $<60\%$ respectively. In the $e\mu$ channel, one-dimensional templates with three bins are formed
corresponding to $b$-tagging efficiencies between 77\% and 0\% for the jet with the
third highest $b$-tagging discriminant value. In the
\ljets channel, two-dimensional templates are created using the $b$-tagging
discriminant values of the two jets with the third- and fourth-highest
$b$-tagging discriminant values, corresponding to $b$-tagging efficiencies
between 100\% and 0\% for the two jets.

%
%

In both channels, one template is created from the sum of all backgrounds
described in Section~\ref{sec:background_estimation} and three templates are
created from \ttbar, \ttV and \ttH MC simulations, to account for \ttb, \ttc and \ttlight
events, as detailed in Table~\ref{tab:fitclass}. These templates are then fitted
to the data using a binned maximum-likelihood fit, with a Poisson
likelihood
\begin{equation*}
  \label{eq:poisson_likelihood}
  \mathcal{L}(\vec{\alpha} | x_1, \ldots, x_n) =
  \prod_{k}^{n}\frac{{\mathrm{e}}^{-\nu_k(\vec{\alpha})} \nu_k(\vec{\alpha})^{x_k}}{x_k!}~,
\end{equation*}
where $x_k$ is the number of events in bin $k$ of the data template and
$\nu_k(\vec{\alpha})$ is the expected number of events, and depends
upon a number of free parameters, $\vec{\alpha}$.

\begin{figure}
  \subcaptionbox{\label{fig:fit_dilepton}}{
    \includegraphics[width=0.48\textwidth]{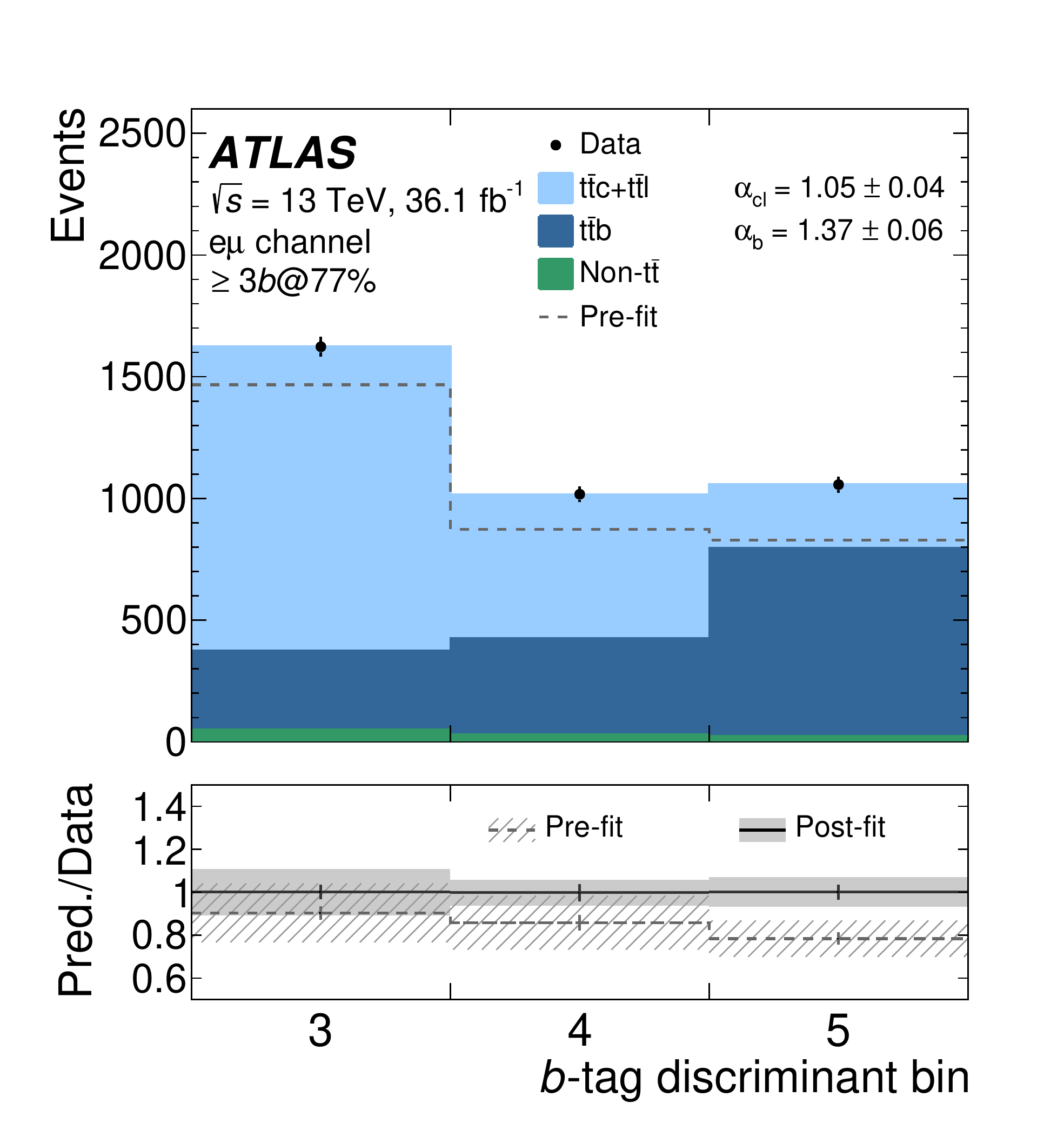}
  }
  \subcaptionbox{\label{fig:fit_lepjets}}{
    \includegraphics[width=0.48\textwidth]{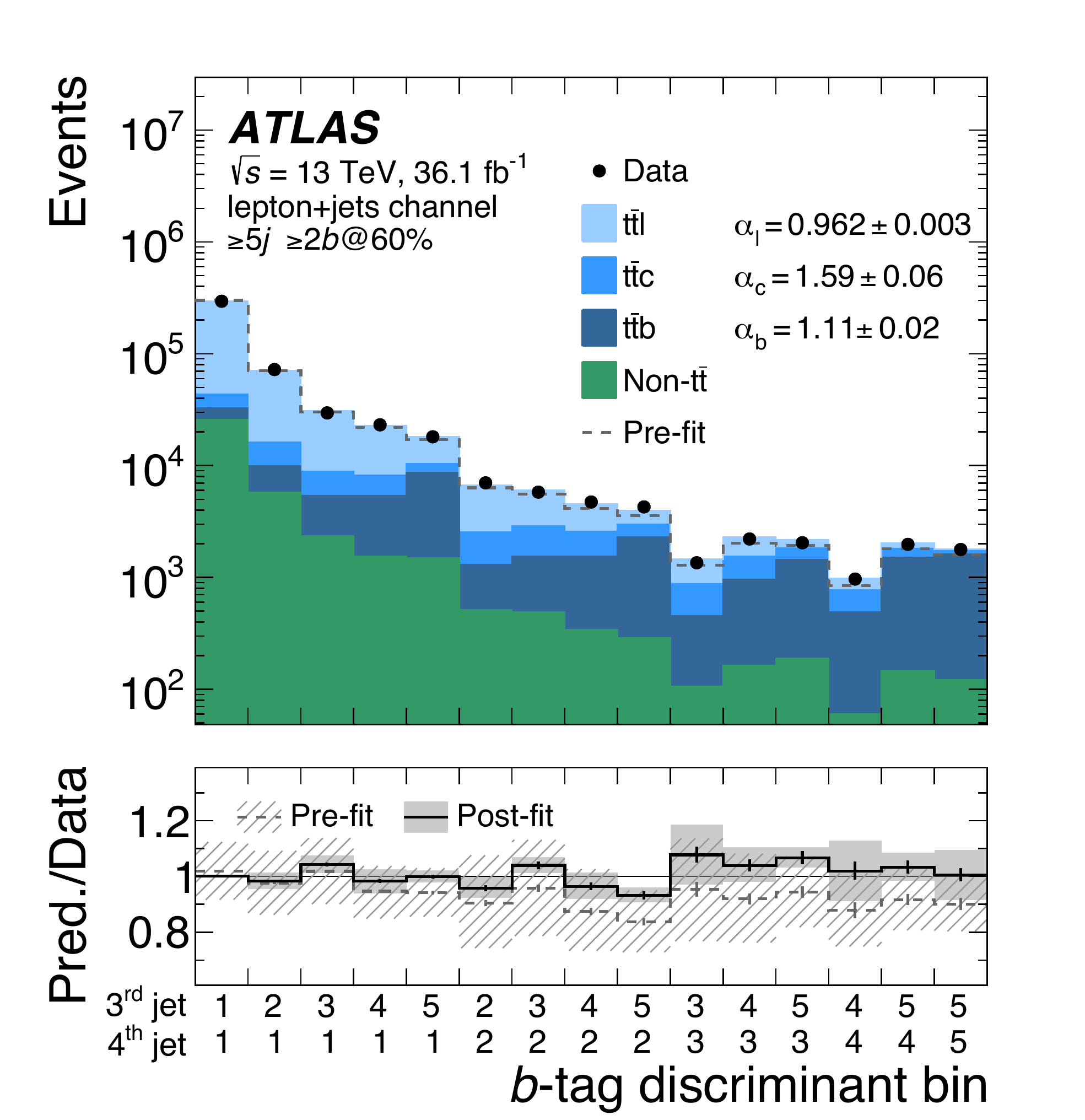}
  }
  \caption{The $b$-tagging distribution of the third-highest $b$-tagging
    discriminant-ranked jet for the \protect\subref{fig:fit_dilepton} $e\mu$
    channel, and of the third and fourth $b$-tagging discriminant-ranked jet for the
    \protect\subref{fig:fit_lepjets} \ljets channel. For clarity, the
    two-dimensional \ljets templates have been flattened into one dimension. The
    ratios of total predictions before and after the fit to the data are shown in the
    lower panel. The vertical bar in each ratio represents only the statistical
    uncertainty, and the grey bands represent the total error including
    systematic uncertainties from experimental sources. The extracted scale
    factors $\alpha_b,\alpha_c,\alpha_l,\alpha_{cl} $ are given
    considering only statistical uncertainties.}
    \label{fig:fit}
\end{figure}

In the $e\mu$ channel, two free parameters are used, such that the expected
number of events in bin $k$ is
\begin{equation*}
  \nu_k(\alpha_b, \alpha_{cl}) =
  \alpha_b N^k_{\ttb} + \alpha_{cl} \left(N^k_{\ttc} +  N^k_{\ttlight}\right)
  + N^k_{\textrm{non-}\ttbar}~,
\end{equation*}
where $N^k_{\ttb}$, $N^k_{\ttc}$, $N^k_{\ttlight}$ and
$N^k_{\textrm{non-}\ttbar}$ are the numbers of events in bin $k$ of the \ttb,
\ttc, \ttlight and non-\ttbar background templates, respectively. The scale
factors obtained from the fit are $\alpha_{b}=1.37 \pm 0.06$ and
$\alpha_{cl}=1.05 \pm 0.04$, where the quoted uncertainties are statistical
only. Figure~\ref{fig:fit_dilepton} shows the distributions of the templates
before and after scaling the templates by these scale factors.

In the \ljets channel, three free parameters, $\alpha_b$, $\alpha_c$ and
$\alpha_l$, are used in the maximum-likelihood fit, such that the expected
number of events in bin $k$ is
\begin{equation}
  \label{eq:full_likelihood}
  \nu_k(\alpha_b, \alpha_c, \alpha_l) =
  \alpha_b N^k_{\ttb} + \alpha_{c} N^k_{\ttc} +  \alpha_l N^k_{\ttlight}
  + N^k_{\textrm{non-}\ttbar}~.
\end{equation}

The best-fit values of the free parameters are $\alpha_{b}=1.11 \pm 0.02$,
$\alpha_{c} = 1.59 \pm 0.06$ and $\alpha_{l} = 0.962 \pm 0.003$ where the quoted
uncertainties are statistical only. Including systematic uncertainties, the values of $\alpha_b$ extracted in the $e\mu$ and \ljets channels are found to be compatible at a level better than $1.5$ standard deviations.  Some of the dominant common systematic uncertainties have small correlations between the two channels, while the uncertainty in $\alpha_b$ due to the modelling of the \ttc template in the $e\mu$ channel, as discussed in Section~\ref{sec:ttcfit-syst} is uncorrelated between the two channels.  Taking only this uncertainty as uncorrelated, the values of $\alpha_b$ extracted from the two channels are found be compatible at a level better than $1.7$ standard deviations.  Figure~\ref{fig:fit_lepjets} shows the distribution of the $b$-tagging discriminant before and after the fit. For clarity, the
two-dimensional \ljets templates are flattened into a single dimension.
%
\begin{figure}
  \centering
  \subcaptionbox{\label{fig:nbjets_emu_postfit}}{
    \includegraphics[width=0.45\textwidth]{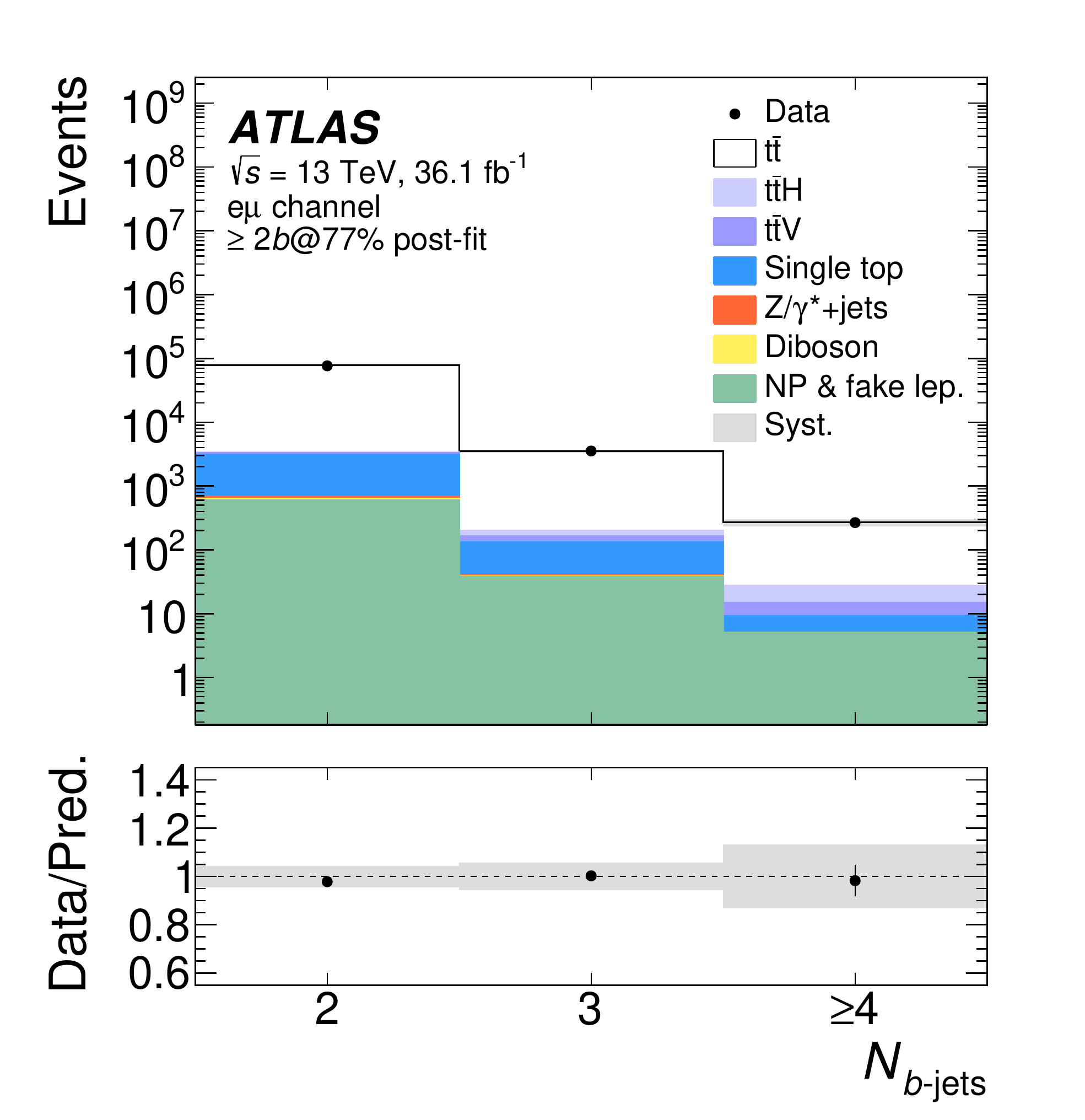}
  }
  \subcaptionbox{\label{fig:nbjets_ljets_postfit}}{
    \includegraphics[width=0.45\textwidth]{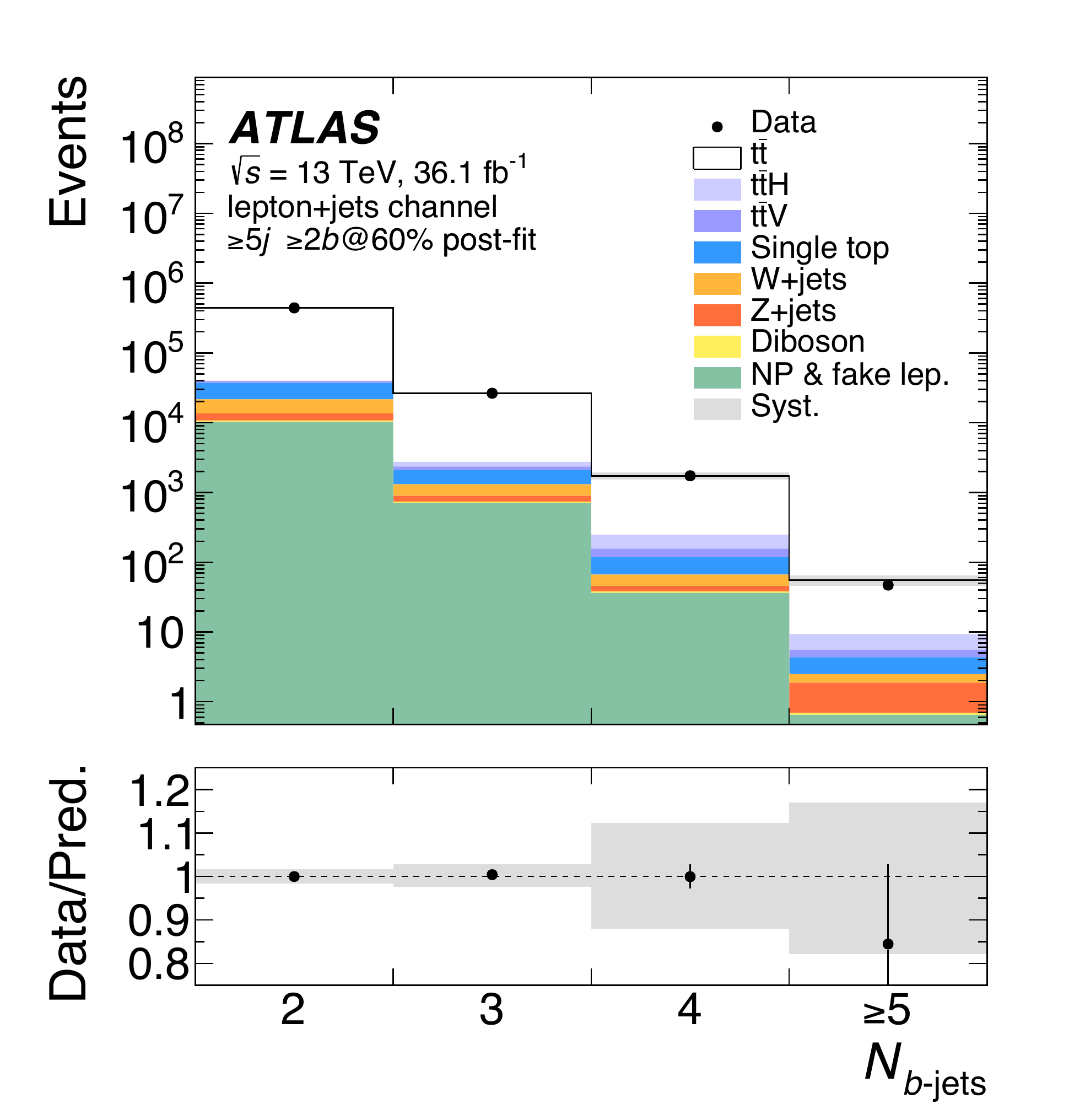}
  }
  \caption{Comparison of the data distributions with predictions, after
    applying scale factors, for the number of $b$-tagged jets, in events with at
    least 2 $b$-tagged jets, in the \subref{fig:nbjets_emu_postfit} $e\mu$ and
    \subref{fig:nbjets_ljets_postfit} \ljets channels. The systematic
    uncertainty band, shown in grey, includes all uncertainties from
    experimental sources.}
  \label{fig:reco_postfit_nbjets}
  \vspace*{\floatsep}
  \subcaptionbox{\label{fig:leadbjetpt_emu_postfit}}{
    \includegraphics[width=0.45\textwidth]{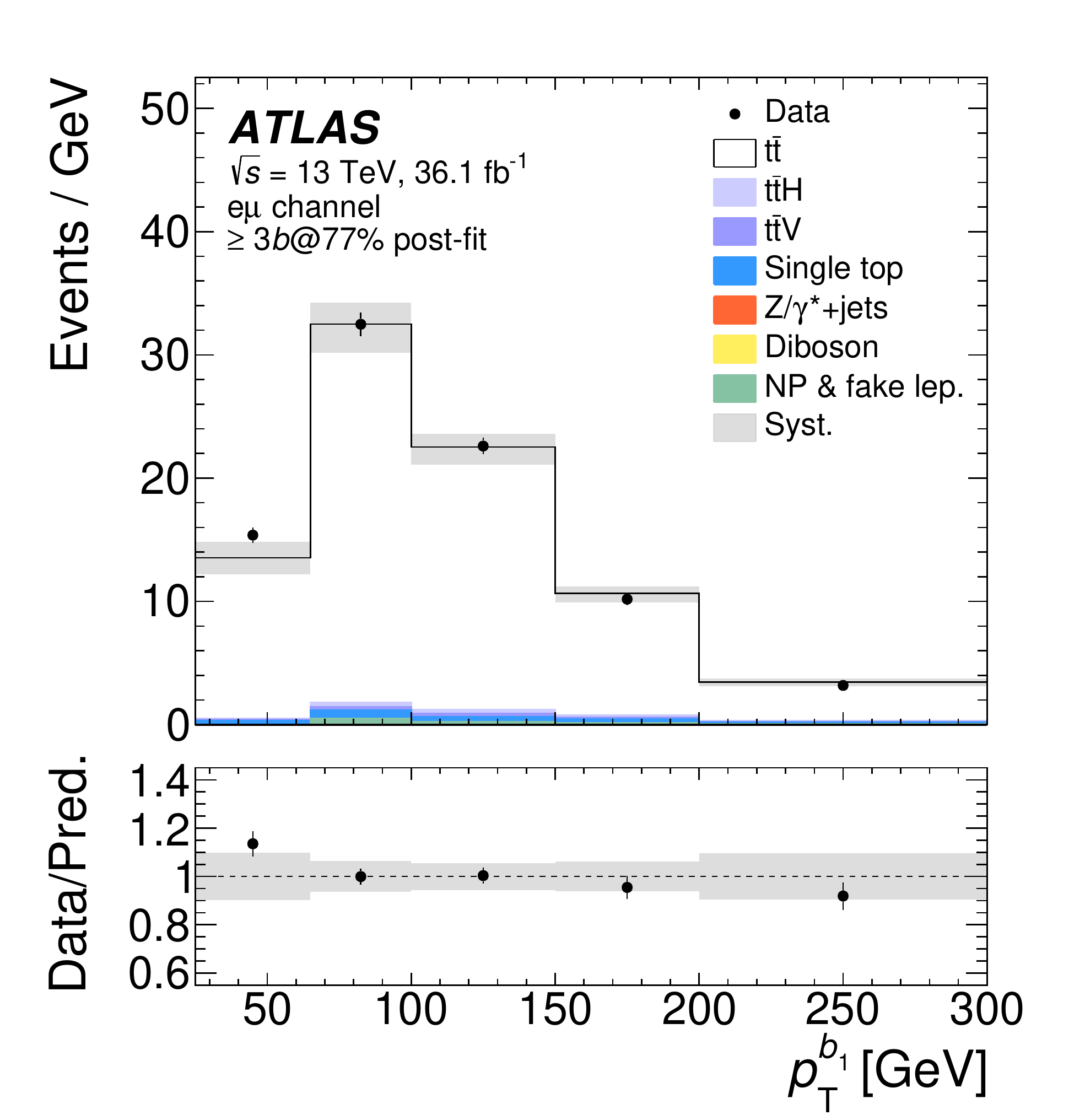}
  }
  \subcaptionbox{\label{fig:leadbjetpt_ljets_postfit}}{
    \includegraphics[width=0.45\textwidth]{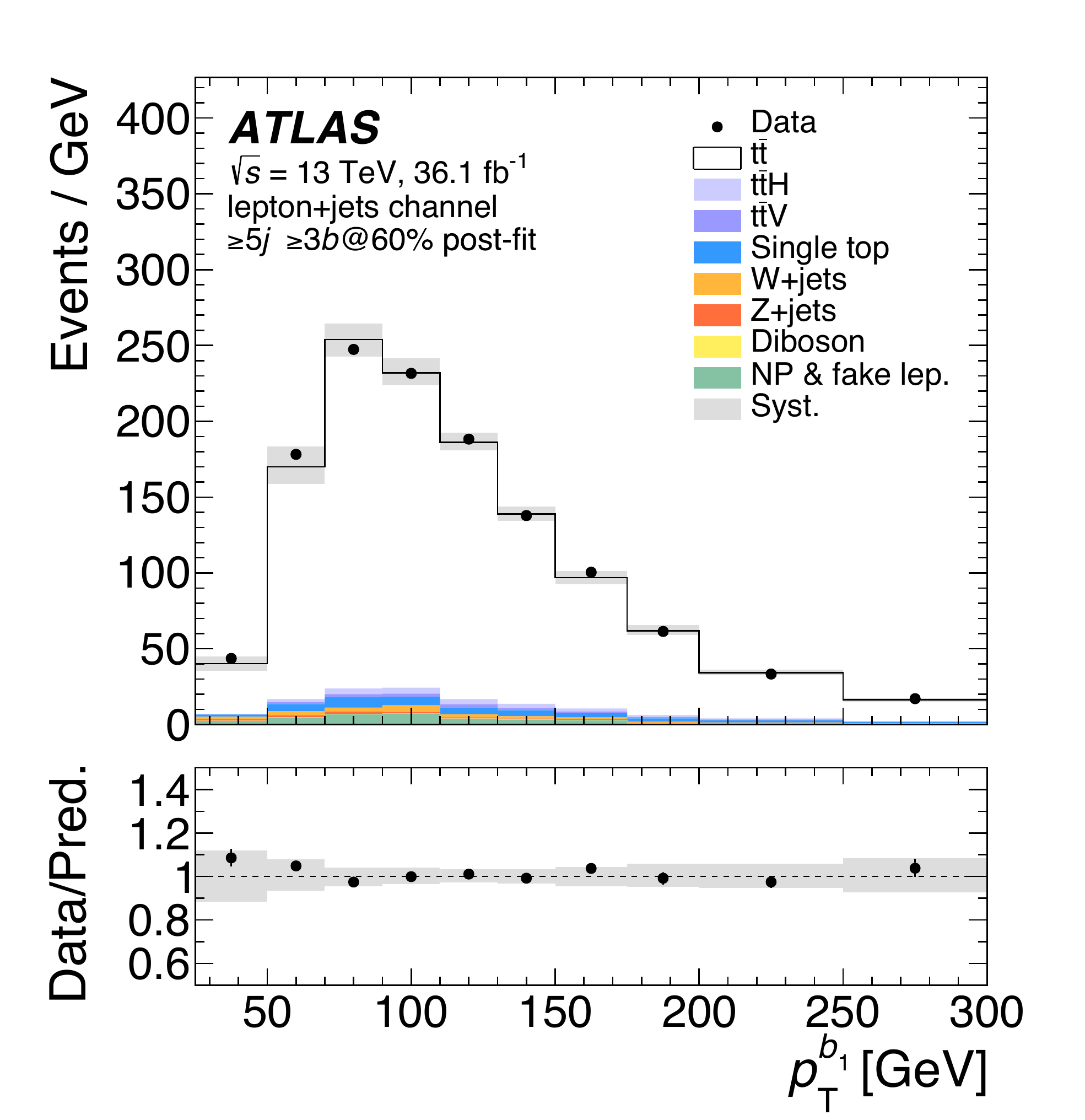}
  }
  \caption{Comparison of the data distributions with predictions for the leading
    $b$-tagged jet \pt, after applying scale factors, in events with at least 3
    $b$-tagged jets, in the \subref{fig:leadbjetpt_emu_postfit} $e\mu$ and
    \subref{fig:leadbjetpt_ljets_postfit} \ljets channels. The systematic
    uncertainty band, shown in grey, includes all uncertainties from
    experimental sources. Events that fall outside of the range of the $x$-axis
    are not included in the plot.}
  \label{fig:reco_postfit_bjet_pt}
\end{figure}
Figures~\ref{fig:reco_postfit_nbjets} and~\ref{fig:reco_postfit_bjet_pt} show
the comparison of data and predictions for the $b$-tagged jet multiplicity and
the leading $b$-tagged jet $p_\mathrm{T}$ in the $e\mu$ and \ljets channels
after the \ttb signal, and the $t\bar{t}c$ and $t\bar{t}l$ backgrounds, are
scaled by the extracted scale factors. The data are described much better
by the prediction after the scaling is applied.

%% file: sections/unfolding.tex
\subsection{Unfolding}
\label{sec:unfolding}

The measured distributions at detector level are unfolded to the particle level.
The unfolding procedure corrects for resolution effects and for detector
efficiencies and
acceptances.

First, the number of non-\ttbar background events in bin $j$
($N^j_{\text{non-}t\bar{t}\text{-bkg}}$), described in
Section~\ref{sec:background_estimation}, is subtracted from the data
distribution at the detector level in bin $j$ ($N^j_{\mathrm{data}}$). This
retains a mixture of signal and \ttbar-related backgrounds, the latter coming
from mis-tagged events as described in
Section~\ref{sec:heavy_flavour_scale_factors}.
A series of corrections are then applied, with all corrections derived from
simulated \ttbar, $\ttbar H$ and $\ttbar V$ events. Following the subtraction of 
non-\ttbar background, the data are first corrected for mis-tagged events by applying a
correction
\begin{equation*}
  f^j_{\ttb} = \frac{\alpha_b N^j_{\ttb, \mathrm{reco}}}{\alpha_b N^j_{\ttb, \mathrm{reco}} + \mathcal{B}^j}~,
\end{equation*}
where $\alpha_b$ is defined in the previous section, $N^j_{\ttb, \mathrm{reco}}$
is the number of detector-level \ttb events predicted by MC simulation, and
$\mathcal{B}^j$ is the number of detector-level \ttc and \ttlight events in bin
$j$, after being scaled by the fit parameters, $\alpha_{cl}$ or $\alpha_c$ and
$\alpha_l$, defined in the previous section. In the $e\mu$ channel,
\begin{equation*}
  \mathcal{B}^j = \alpha_{cl} \left( N_{t\bar{t}c, \mathrm{reco}}^j + N_{t\bar{t}l, \mathrm{reco}}^j \right) \ ,
\end{equation*}
and in the \ljets channel,
\begin{equation*}
  \mathcal{B}^j = \alpha_{c} N_{\ttc, \mathrm{reco}}^j + \alpha_{l} N^j_{\ttlight, \mathrm{reco}}~ ,
\end{equation*}
where $N^j_{\ttc, \mathrm{reco}}$ and $N^j_{\ttlight, \mathrm{reco}}$ are the
numbers of reconstructed \ttc and \ttlight events in bin $j$, as predicted by MC
simulation, respectively.
Next, an acceptance correction, $f^j_{\mathrm{accept}}$, is applied, which
corrects for the fiducial acceptance and is defined as the probability of a \ttb
event passing the detector-level selection in a given bin $j$
($N^j_{\ttb, \mathrm{reco}}$) to also fall within the fiducial particle-level phase
space ($N^j_{\ttb, \mathrm{reco \land part}}$). It is estimated as
\begin{equation*}
  f^j_{\mathrm{accept}} = \frac{N^j_{\ttb, \mathrm{reco \land part}}}{N^j_{\ttb, \mathrm{reco}}}~.
\end{equation*}
The detector-level objects are required to be matched within $\Delta R = 0.4$ to the
corresponding particle-level objects. 
This requirement leads to a better correspondence between the particle and
detector levels and improves the unfolding performance. 
The matching factor $f_{\mathrm{matching}}^j$ is defined as
\begin{equation*}
  f_{\mathrm{matching}}^j = \frac{N^j_{\ttb, \mathrm{reco \land part \land matched}}}{N^j_{\ttb, \mathrm{reco \land part}}}~,
\end{equation*}
where $N^j_{\ttb, \mathrm{reco \land part \land matched}}$ is the subset of
reconstructed events falling in the particle-level fiducial volume which are
matched to the corresponding particle-level objects.

The remaining part of the unfolding procedure consists of effectively inverting
the migration matrix $\mathcal{M}$ to correct for the resolution effects and
subsequently correcting for detector inefficiencies. An iterative Bayesian
unfolding technique~\cite{DAgostini:1994fjx}, as implemented in the
\textsc{RooUnfold} software package~\cite{Adye:2011gm}, is used. The matrix,
$\mathcal{M}$, represents the probability for a particle-level event in bin $i$
to be reconstructed in bin $j$. The chosen binning is optimised for each
distribution to have a migration matrix with a large fraction of events on the
diagonal and a sufficient number of events in each bin. The Bayesian unfolding
technique performs the effective matrix inversion, $\mathcal{M}^{-1}_{ij}$,
iteratively. Four iterations are used for all measured distributions.

Finally, the factor $f^i_{\mathrm{eff}}$ corrects for the reconstruction
efficiency and is defined as
\begin{equation*}
  f^i_{\mathrm{eff}} =   \frac{N^i_{\ttb, \mathrm{part \land reco \land matched}}}{N^i_{\ttb, \mathrm{part}}}~,
\end{equation*}
where $N_{\ttb, \mathrm{part}}^i$ is the number of \ttb events passing the
particle-level selection in bin $i$ and $N^i_{\ttb, \mathrm{part \land reco \land matched}}$
is the number of \ttb events at particle level in bin $i$ that also pass the
detector-level selection, containing matched objects.

The unfolding procedure for an observable $X$ at particle level can be
summarised by the following expression
\begin{equation*}
  \label{eq:unfolded_events}
  \frac{\mathrm{d} \sigmafid}{\mathrm{d}X^i} =
  \frac{N^i_{\mathrm{unfold}}}{\mathcal{L} ~ \Delta X^i } =
  \frac{1}{\mathcal{L} ~ \Delta X^i ~ f^i_{\mathrm{eff}}} ~ \sum_j \mathcal{M}^{-1}_{ij}
  ~ f^j_{\mathrm{matching}} ~ f^j_{\mathrm{accept}} ~ f^j_{\ttb}
  ~ (N_{\mathrm{data}}^j - N_{\textrm{non-}\ttbar\textrm{-bkg}}^j)~,
\end{equation*}
where $\Delta X^i$ is the bin width, $N^i_{\mathrm{unfold}}$ is the number of
events in bin $i$ of the unfolded distribution and $\mathcal{L}$ is the
integrated luminosity. In this paper, the integrated fiducial cross-section
\sigmafid\ is obtained from
\begin{equation*}
 \label{eq:fidcrosssec}
 \sigmafid = \int \frac{ \mathrm{d}\sigmafid} {\mathrm{d}X} \mathrm{d}X = \frac{\sum N^i_{\mathrm{unfold}}}{\mathcal{L} }
\end{equation*}
and is used as a normalisation factor such that results are presented in terms
of a relative differential cross-section as $1 / \sigmafid ~\cdot \mathrm{d}
\sigmafid / \mathrm{d}X^i$.

%% file: sections/systematics.tex
\section{Systematic uncertainties}
\label{sec:systematics}

In this section, the statistical and systematic uncertainties considered in this
analysis are described. Experimental sources of uncertainty are described in
Section~\ref{sec:exp_uncerts}, sources of uncertainty due to \ttbar modelling
are described in Section~\ref{sec:model-syst} and uncertainties due to the
treatment of the \ttbar (\ttc and \ttlight) and non-\ttbar background processes
are described in Sections~\ref{sec:ttcfit-syst} and~\ref{sec:background-syst},
respectively. The method used to propagate the effects of systematics
uncertainties to the final results are described in
Section~\ref{sec:uncert_prop}. The impact of these uncertainties on the fiducial
and differential cross-section measurements are discussed in
Section~\ref{sec:results}.


\subsection{Experimental uncertainties}
\label{sec:exp_uncerts}
The uncertainty in the combined 2015+2016 integrated luminosity is 2.1\%. It is
derived, following a methodology similar to that detailed in
Ref.~\cite{DAPR-2013-01}, and using the LUCID-2 detector for the baseline
luminosity measurements \cite{LUCID2}, from a calibration of the luminosity
scale using $x$--$y$ beam-separation scans.

The uncertainty in the pile-up reweighting of the reconstructed events in the MC
simulation is estimated by comparing the distribution of the number of primary
vertices in the MC simulation with the one in data as a function of the instantaneous
luminosity. Differences between these distributions are adjusted by scaling the
mean number of $pp$ interactions per bunch crossing in the MC simulation and the $\pm
1\sigma$ uncertainties are assigned to these scaling factors. The pile-up
weights are recalculated after varying the scale factors within their
uncertainties.

As discussed in Section~\ref{sec:object_selection}, scale factors to correct
differences seen in the lepton reconstruction, identification and trigger
efficiency between the data and MC simulation are derived using a tag-and-probe
technique in $Z\to e^+e^-$ and $Z \to \mu^+ \mu^-$ events~\cite{PERF-2015-10,ATLAS-CONF-2016-024,ATL-PHYS-PUB-2016-015}. 
The electron (muon) momentum scale and resolution are determined
using the measurement of the position and width of the $Z$ boson peak in
$Z\to e^+e^- (\mu^+ \mu^-)$  events~\cite{PERF-2015-10,ATLAS-CONF-2016-024,ATL-PHYS-PUB-2016-015}. The
lepton uncertainties considered in this analysis  are considerably
smaller than the jet and flavour-tagging uncertainties.

The \ac{JVT} is calibrated using $Z~(\to\mu\mu)$ + jet events where the jet
balances the \pt of the $Z$ boson. Scale factors binned in jet \pt are applied to each
event in order to correct for small differences in the \ac{JVT} efficiency between the data
and MC simulation. The scale factors are $0.963 \pm 0.006$ for jets with $20 <
\pt < 30$~\GeV, getting closer to one with smaller uncertainties as the jet \pt
increases.
The uncertainty in the efficiency to pass the \ac{JVT} requirement is evaluated
by varying the scale factors within their uncertainties~\cite{PERF-2014-03}.

Jets are calibrated using a series of simulation-based corrections and
\textit{in situ} techniques~\cite{PERF-2016-04}. The uncertainties due to the
\ac{JES} are estimated using a combination of simulations, test-beam data and
\textit{in situ} measurements. Contributions from the jet-flavour composition,
$\eta$-intercalibration, leakage of the hadron showers beyond the extent of the
hadronic calorimeters (punch-through), single-particle response, calorimeter
response to different jet flavours, and pile-up are taken into account, resulting
in 21 orthogonal uncertainty components. The total uncertainty due to the
\ac{JES} is one of the dominant uncertainties in this analysis.

The \ac{JER} is measured using both data and simulation. First, the ``true''
resolution is measured by comparing the particle and reconstructed jet \pt in
\ac{MC} simulation as a function of the jet \pt and~$\eta$. Second, an
\textit{in situ} measurement of the \ac{JER} is made using the \textit{bisector}
method in dijet events~\cite{PERF-2011-04}. The resolution in data and \ac{MC}
simulation are compared and the energies of jets in the \ac{MC} simulation are
smeared to match the resolution observed in data. The uncertainties in the
\ac{JER} stem
from uncertainties in both the modelling and the data-driven method.

Differences in the $b$-tagging and $c$-jet mis-tag efficiencies between the data
and \ac{MC} simulation are corrected using scale factors derived from dilepton
\ttbar events and \ljets \ttbar events, respectively. A negative-tag method is
used to calibrate mis-tagged light-flavour ($u$, $d$, $s$)
jets~\cite{ATLAS-CONF-2018-006-FIXED}. The scale factors are measured for
different $b$-tagging working points and as a function of jet kinematics, namely
the jet \pt for the $b$-tagging efficiency and $c$-jet mis-tag scale factors,
and the jet \pt and $\eta$ for the light-flavour jet mis-tag scale factors. The
$c$-jet and light-jet mis-tag scale factors are known to a precision of
6--22\%~\cite{ATLAS-CONF-2018-001} and
15--75\%~\cite{ATLAS-CONF-2018-006-FIXED}, respectively.
The associated flavour-tagging uncertainties, split into eigenvector components,
are computed by varying the scale factors within their uncertainties. In total,
there are 30 components related to the $b$-tagging efficiencies and 15 (80)
components related to the mis-tag rates of $c$-jets (light-flavour jets). Due to the
large number of $b$-tagged jets in each event used in this analysis, the total
uncertainty due to $b$-tagging is one of the dominant uncertainties in this
analysis.

\subsection{Modelling systematic uncertainties}
\label{sec:model-syst}
Uncertainties due to the choice of \ttbar \ac{MC} generator are evaluated by
unfolding alternative \ttbar samples, described in
Section~\ref{sec:mc_modelling} and presented in Table~\ref{tab:mc_signal}, with the nominal unfolding set-up. Uncertainties
related to the choice of matrix element generator (labelled ``generator''
uncertainty) are evaluated using the \SHERPAtt sample. 
This generator comes with its own parton shower and hadronisation model; hence these are included in the variation.
Uncertainties due to the choice of parton shower and hadronisation model are
evaluated using the \phwseven sample, in which only the parton shower and hadronisation
model is varied relative to the nominal \ppyeight sample. Additionally, two
\ac{MC} samples are used to evaluate an uncertainty in the modelling of initial-
and final-state radiation, namely the RadHi and RadLo samples described in
Section~\ref{sec:mc_modelling}.

The uncertainty due to the choice of PDF is evaluated following the
\textsc{PDF4LHC} prescription~\cite{Butterworth:2015oua} using
event weights  that are available in the nominal \ppyeight sample. 
The uncertainty in the $\ttbar H$ cross-section is evaluated
by scaling the $\ttbar H$ component of the prediction by factors of zero and two,
with the nominal values being taken from theoretical predictions.
A factor of two is chosen as this is the current 95\% confidence-level upper limit on the $\ttbar
H \to b\bar{b}$ signal strength as measured by ATLAS~\cite{HIGG-2017-03-FIXED}.

The uncertainty in the $\ttbar V$ cross-section is evaluated by varying the
$\ttbar V$ component of the prediction up and down by 30\% to cover the measured
uncertainty in this process~\cite{TOPQ-2015-22}.

\subsection{Uncertainty in \ttc and $\ttlight$ background}
\label{sec:ttcfit-syst}
Since the $\ttc$ and $\ttlight$ backgrounds in the $e\mu$ channel are determined
within a single fit, the uncertainty in this result is determined by changing
the sample composition.
This is achieved by loosening the $b$-tagging requirement on the jet with the
third-highest $b$-tagging discriminant value, such that it is tagged at the 85\%
$b$-tagging efficiency working point or not required to be $b$-tagged at all.
This results in the templates having more bins and allows the likelihood to be
modified such that three free parameters are used in the fit. The number of
expected events is then given by Eq.~\eqref{eq:full_likelihood}.
With these looser selections the values of $\alpha_c$ vary by about $40\%$ and
this is used as a systematic uncertainty in the \ttc template.
The validity of this uncertainty is checked by investigating the variations in
the values of the \ttc scale factors after fitting to pseudo-data from
alternative MC samples and it is found to cover the
uncertainties in the \ttc template modelling. The values of $\alpha_l$ remain
consistent within the statistical uncertainty in fits with looser selections.
After propagating the uncertainty in the \ttc template through the nominal fit
set-up, by varying the input \ttc template by $\pm 40 \%$ before performing the
fit, the value of $\alpha_b$ is found to change by $\pm 11 \%$, while the value
of $\alpha_{cl}$ changes by $\pm 7 \%$.
When evaluating systematic uncertainties related to the choice of
\ttbar model in the $e\mu$ channel, double counting of these
uncertainties with uncertainties associated with the difference of
\ttb, \ttc and \ttlight fractions in the alternative \ac{MC} samples
is avoided by repeating the flavour-composition fits for each
systematic model.

In the \ljets channel uncertainties in the flavour
composition are taken directly from the samples used to evaluate
systematic uncertainties in the modelling, as described in
Section~\ref{sec:model-syst}.

\subsection{Uncertainty in non-\ttbar background estimation}
\label{sec:background-syst}
The uncertainty in the single-top background is evaluated by comparing the
nominal single-top $tW$ sample (with overlap with \ttbar removed via the
diagram-removal scheme) with an alternative sample generated using the
diagram-subtraction scheme~\cite{Frixione:2008yi}. Potential effects of QCD
radiation on the single-top background are estimated using MC simulation predictions where
the renormalisation and factorisation scales were varied by factors of 0.5 and 2.
The uncertainty in the inclusive single-top cross-section~\cite{Kidonakis:2010ux} is taken to be
$^{+5\%} _{-4\%}$.

The uncertainty attributed to the \Wjets background normalisation is evaluated
by varying the renormalisation and factorisation scales in the MC simulation prediction by a
factor of two up and down. Furthermore, the uncertainty due to PDFs is estimated
by using a set of 100 different PDF eigenvectors recommended
in Ref.~\cite{Butterworth:2015oua}. An additional uncertainty of 30\% is assumed for
the normalisation of the $W+$heavy-flavour jets cross-section, based on MC simulation 
comparisons performed in the context of Ref.~\cite{HIGG-2017-03-FIXED}.

The uncertainty in the non-prompt or fake lepton background is obtained by
varying the estimate of this background by a factor of $\pm 50\%$ ($\pm 100\%$)
in the \ljets ($e\mu$) channel. No shape uncertainty is applied, as this background is small in both channels. 

The uncertainty in the Drell--Yan background normalisation is evaluated by
varying the estimate of this background by $\pm 25\%$. It accounts for the
impact of the reconstructed-mass resolution of the $Z$ boson in the $Z \rightarrow ee$ and $Z \rightarrow \mu\mu$ events, for the background
contribution of the \ttbar events in the \Zjets selection, and for differences in
the scale factors obtained from each of the individual $Z \rightarrow ee$ and $Z
\rightarrow \mu\mu$ decay channels relative to the nominal scale factor obtained from the combined $Z \rightarrow ee$ and $Z
\rightarrow \mu\mu$ sample.

\subsection{Propagation of uncertainties}
\label{sec:uncert_prop}
Pseudo-experiments based on 10\,000 histogram replicas are performed to evaluate statistical
uncertainties for each distribution considered. Each entry for every event is
given a random weight drawn from a Poisson distribution with a mean of one.
Each of these histograms is then unfolded using the unfolding procedure
described in Section~\ref{sec:unfolding}. The standard deviation of each bin
across all unfolded histogram replicas is then taken as the statistical
uncertainty in that bin. This procedure is similar to simply obtaining
pseudo-experiments by directly Poisson-fluctuating the measured data
distributions, but has the added advantage that correlations between bins of
different distributions are conserved.

This procedure is extended to include all experimental systematic uncertainties.
For each systematic uncertainty effect considered, the relative variation due to
that uncertainty is obtained at the detector level, using the nominal MC sample.
Rather than unfolding each shifted histogram individually, each Poisson-fluctuated 
data distribution is smeared by all experimental systematic
uncertainties simultaneously. For each pseudo-experiment, and for each
uncertainty considered, the size of the shift applied is obtained randomly from
a Gaussian distribution with a mean of zero and width equal to the relative
shift at detector level in each bin due to that uncertainty, producing a new
detector-level distribution. The same procedure that is followed for the
statistical uncertainty alone is then followed to get the sum of the statistical and
experimental systematic uncertainty.
When evaluating the systematic uncertainties in this way, the
data-driven correction factors are not extracted for each individual
pseudo-experiment and instead the values obtained in
Section~\ref{sec:heavy_flavour_scale_factors} are used.

In the case of \ttbar modelling systematic uncertainties, detector-level
distributions from alternative \ac{MC} samples are unfolded using the unfolding
procedure described in Section~\ref{sec:unfolding}, with the unfolding
corrections derived from the nominal \ppyeight sample. The unfolded
distributions are compared with the particle-level distribution from the
alternative sample and the relative difference in each bin is taken as the
systematic uncertainty.

%% file: sections/results.tex
\section{Inclusive and differential fiducial cross-section results}
\label{sec:results}
\begin{figure}
  \centering
  \includegraphics[width=\textwidth]{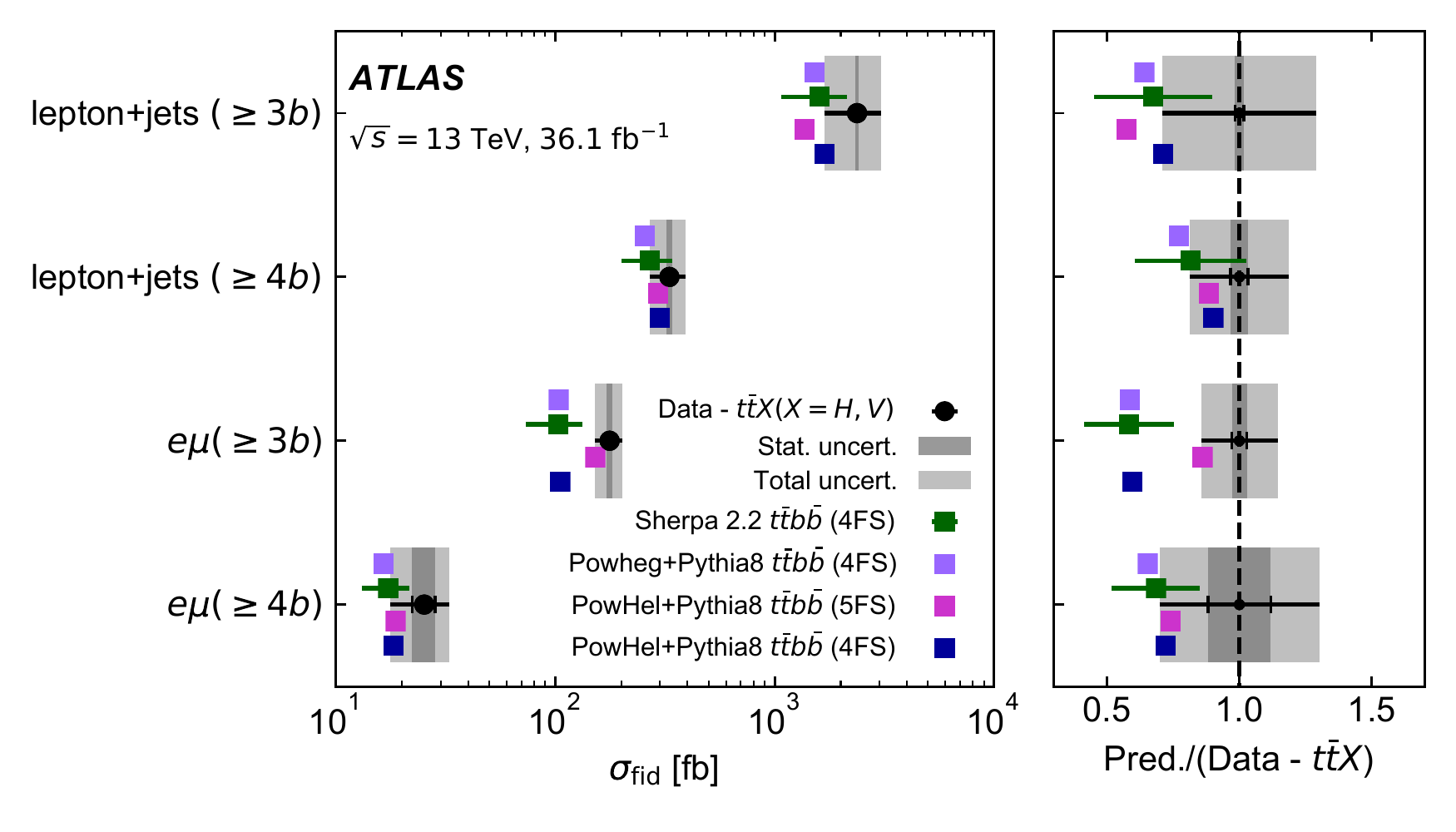}
  \caption{The measured fiducial cross-sections, with \ttH and \ttV
    contributions subtracted from data, compared with $t\bar{t}b\bar{b}$
    predictions obtained using \SHERPAttbb with uncertainties obtained by varying
    the renormalisation and factorisation scales by factors of 0.5 and 2.0 and
    including PDF uncertainties. Comparisons with the central values of the
    predictions of \ppyeight and \powhelpyeight are also made. No uncertainties are
    included in the subtraction of the \ttH or \ttV predictions.}
  \label{fig:fid_xsec_summary}
\end{figure}

The unfolded results are presented in this section as inclusive fiducial cross-sections
and as normalised differential fiducial cross-sections as a function of the \bjet
multiplicity, global event properties and kinematic variables. Table~\ref{tab:res_xsec}
lists the measured fiducial cross-sections for \ttbar production in association with
additional at least one and at least two \bjets and Table~\ref{tab:xsec_unc} lists the contributions
to the uncertainty in these cross-sections.
The most precise cross-section measurements are for the $\geq 3b$ phase space in the
$e\mu$ channel, which has an uncertainty of 13\%, and the $\geq 6j$, $\geq 4b$ phase
space in the \ljets channel, which has an uncertainty of 17\%.
The uncertainties are dominated by systematic uncertainties, which are mainly caused by the uncertainties due to
\ttbar modelling and the uncertainties related to $b$-tagging and the 
jet energy scale. In the $e\mu$ channel, the uncertainty due to the $t\bar{t}c$ fit variations is also significant. 
This measurement is more precise than the uncertainties in the
theoretical predictions of the inclusive cross-section for this process, which are
20\%--30\%~\cite{yellowreport}. The results are summarised in
Figure~\ref{fig:fid_xsec_summary} after subtracting the \amcnlopyeight predicted values of \ttH and \ttV cross-sections from the measured fiducial \ttbb cross-section, and compared with \ttbb predictions from \SHERPAttbb, \ppyeight and \powhelpyeight. This procedure of \ttH and \ttV subtraction is also employed for all following figures showing the normalised differential distributions.\\ 
%
\begin{table}[htb]
  \centering
  \caption{Measured and predicted fiducial cross-section results for additional
    \bjet production in the $e\mu$ and the \ljets decay channels.}
  \input{tables/fid_xsec_results}
  \label{tab:res_xsec}
\end{table}
\input{tables/uncert_fidxsec_table}

\begin{figure}
  \centering
  \includegraphics[width=0.45\textwidth]{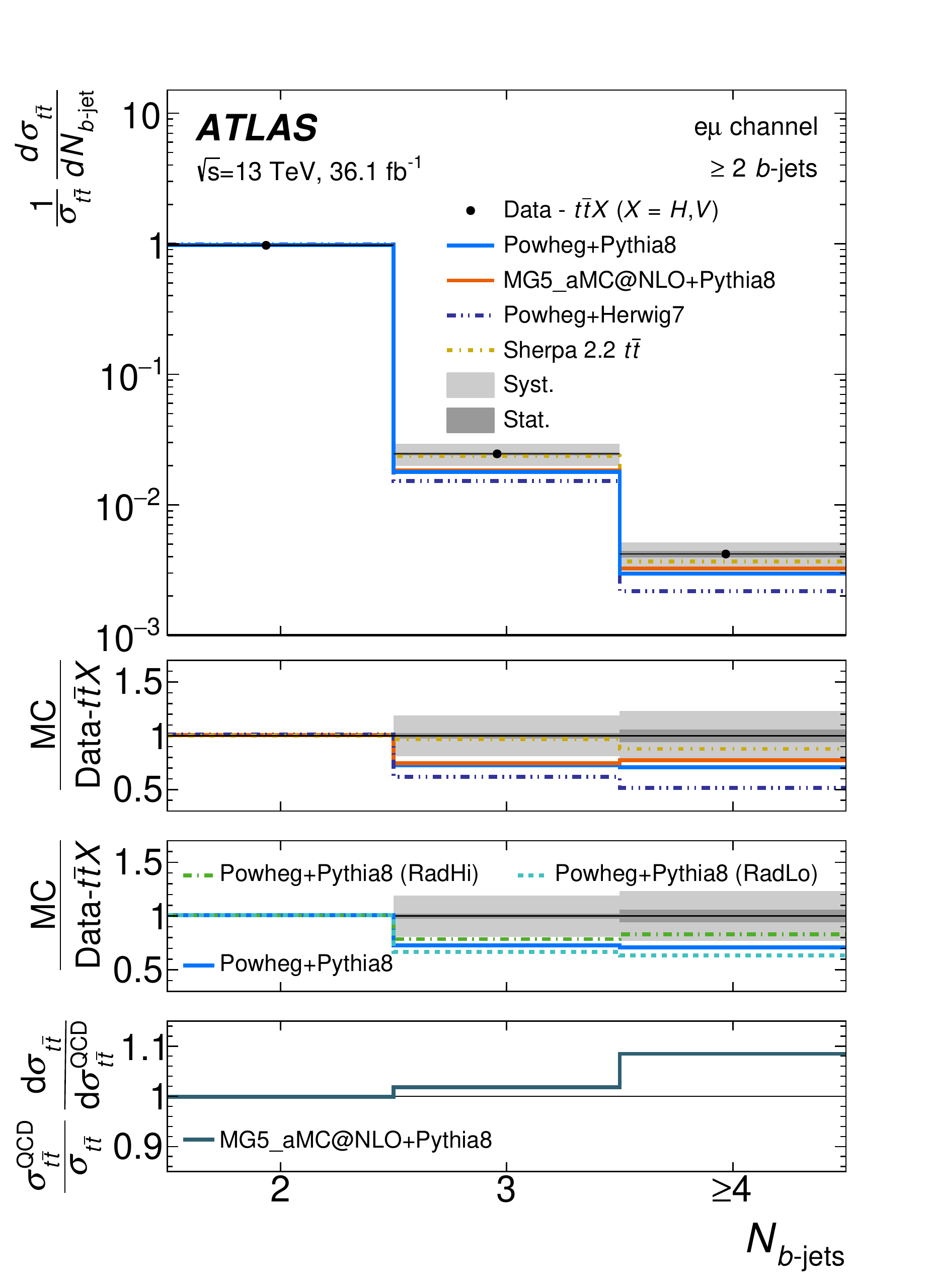}
  \caption{The relative differential cross-section as a function of the $b$-jet
    multiplicity in events with at least two $b$-jets in the $e\mu$ channel
    compared with various MC generators. The \ttH and \ttV contributions are subtracted from data. 
    Three ratio panels are shown, the first
    two of which show the ratios of various predictions to data. The third panel
    shows the ratio of predictions of normalised differential cross-sections
    from \amcnlopyeight including (numerator) and not including (denominator)
    the contributions from \ttV and \ttH production. Uncertainty bands represent the
    statistical and total systematic uncertainties as described in
    Section~\ref{sec:systematics}.}
  \label{fig:res_nbjets}
\end{figure}

Figure~\ref{fig:res_nbjets} shows the normalised fiducial cross-section as a
function of the $b$-jet multiplicity compared with predictions from various
\ac{MC} generator set-ups. A quantitative assessment of the level of agreement between
data and the various predictions is performed by calculating a $\chi^2$ for each
prediction. The $\chi^2$ is defined as
\begin{equation*}
  \chi^2 = S_{b-1}^T~V^{-1}~S_{b-1} ~,
\end{equation*}
where $V^{-1}$ is the inverse of the covariance matrix $V$, calculated for each
variable including all statistical and systematic uncertainties and $S_{b-1}$ is
a vector of the differences between the measured and predicted cross-sections
being tested. The resulting value of the $\chi^2$ calculation is converted into a $p$-value using the number of degrees of freedom for each variable,
 which is the number of bins minus one in the
case of the normalised differential cross-sections to reflect the
 normalisation constraint.

 As normalised distributions are used, one element of $S_{b-1}$ is discarded in
 the calculation along with the corresponding row and column of the covariance
 matrix. The resulting $\chi^2$ does not depend on the element of $S_{b-1}$ or
 the row and column of the covariance matrix that is discarded. The resulting
 $\chi^2$ values are shown in Table~\ref{tab:localchi2_nbjets}, where the second
 column is for the normalised \bjets multiplicity distribution with
 $N_{b\mathrm{\textrm{-}jets}} \geq 2$ and the last column is for the normalised \bjets
 multiplicity distribution with $N_{b\mathrm{\textrm{-}jets}} \geq 3$. All
 \ac{MC} predictions that calculate the top-quark pair production matrix element
 at NLO, but rely on the parton shower for high jet multiplicities, predict too
 few events with three or four \bjets. This suggests that the \bjet production
 by the parton shower is not optimal in these set-ups. The situation does not
 improve significantly when the renormalisation and factorisation scales in the
 matrix element calculation and in the parton shower are changed by factors of
 0.5 and 2, as shown in the middle ratio panel of Figure~\ref{fig:res_nbjets}.
 \SHERPAtt, which models one additional-parton process at NLO accuracy and up to four additional partons at LO accuracy, is the only one of the presented generators that describes the \bjet production well over the full phase space.

 Predictions that include additional massive $b$-quarks in the matrix element
 calculation (\SHERPAttbb~(4FS), \powhelpyeight~(4FS), \ppyeightttbb~(4FS)) 
 do not provide top-pair production without
 additional \bjets and cannot be compared with the region with less than
 three \bjets. Table~\ref{tab:localchi2_nbjets} therefore also includes   $\chi^2$ values where the  
 total additional $b$-jet production has been adjusted through the normalisation
 to $N_{b\mathrm{\textrm{-}jets}} \geq 3$.
 The relative rate of one, two and more than two additional $b$-jets is described well by all  predictions.
   It is also interesting to note that parton shower generators
 predict the relative rate of one and two additional $b$-jets well once the
 total additional $b$-jet production has also been adjusted through the normalisation
 to $N_{b\mathrm{\textrm{-}jets}} \geq 3$. 

The comparison of the predictions from various MC generators with the data are made after subtracting the simulation-estimated contributions of \ttV and \ttH production from the data.
The third ratio panel of Figure~\ref{fig:res_nbjets} shows the ratio of predictions of normalised
differential cross-sections from \amcnlopyeight including (numerator) and not
including (denominator) the contributions from the \ttV and \ttH processes. The
impact of including these processes in the prediction increases with \bjet
multiplicity, resulting in a change of about 10\% relative to the QCD
\ttbar prediction alone in the inclusive four-\bjet bin.
\begin{table}
  \caption{Values of $\chi^{2}$ per degree of freedom and $p$-values between the
    unfolded normalised cross-section and the predictions for $b$-jet
    multiplicity measurements in the $e\mu$ channel. The number of degrees of
    freedom is equal to the number of bins minus one.  Calculations are performed after subtracting estimated contributions from \ttH and
\ttV from the data. In the two right columns, data and predictions are normalised to cross-section for $N_{b\mathrm{\textrm{-}jets}} \geq3$ before calculating 
    $\chi^{2}$ per degree of freedom and $p$-values.} 
  \small
  \centering
  \input{tables/combined_tables_ttXsubtracted/nbjets}
\label{tab:localchi2_nbjets}
\end{table}

Observables sensitive to the details of the QCD modelling of additional \bjet
production are studied in events with at least three \bjets in the $e\mu$
channel and in events with at least four \bjets in the \ljets channel. While the
sample with at least four \bjets has high signal purity, leading to smaller
dependence on the MC models, the $e\mu$ channel benefits from an order of
magnitude larger size of the sample containing at least three \bjets.

\begin{figure}
  \centering
  \subcaptionbox{\label{fig:HT_emu}}{
    \includegraphics[width=0.45\textwidth]{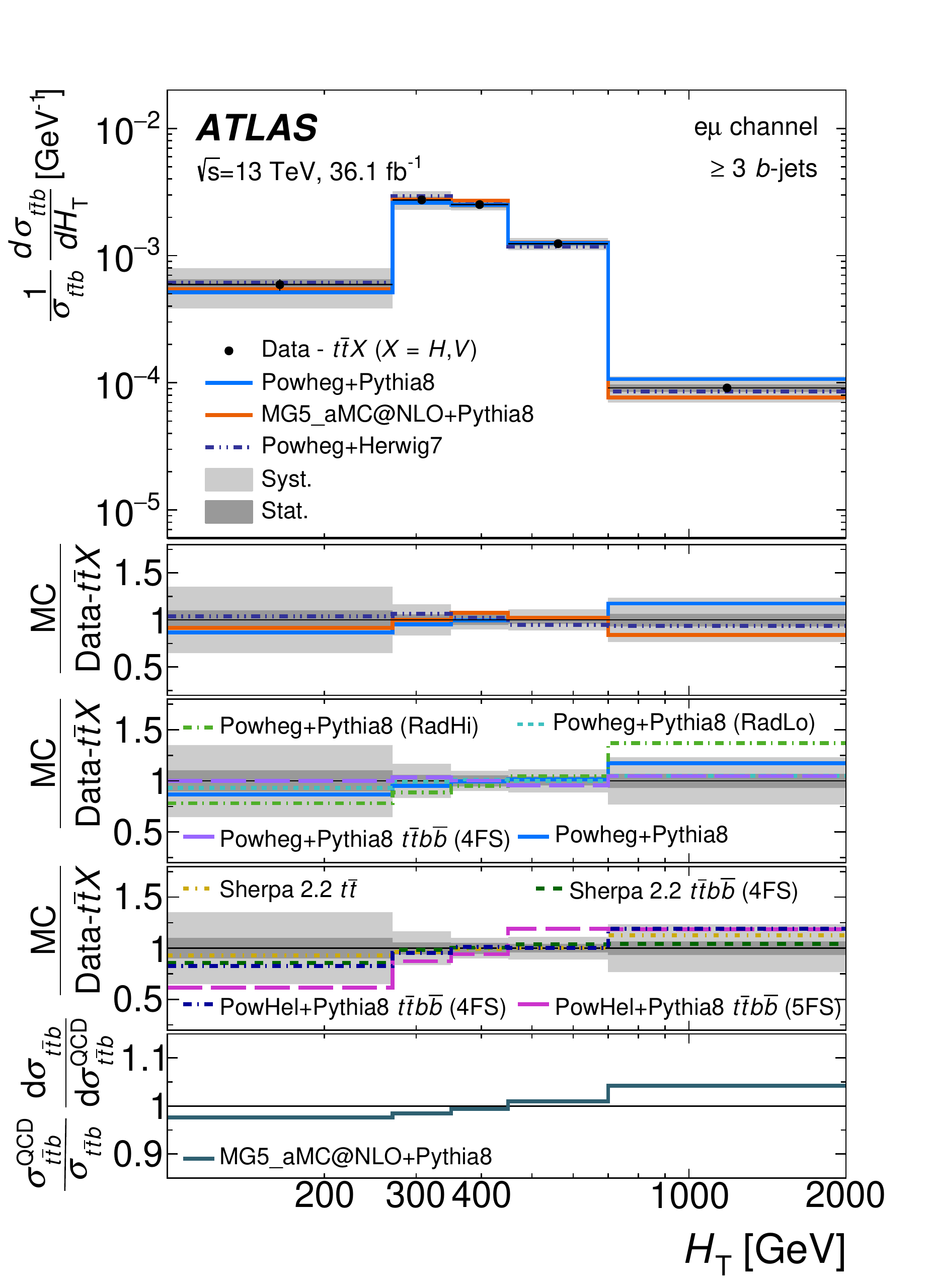}
  }
  \subcaptionbox{\label{fig:HThad_emu}}{
    \includegraphics[width=0.45\textwidth]{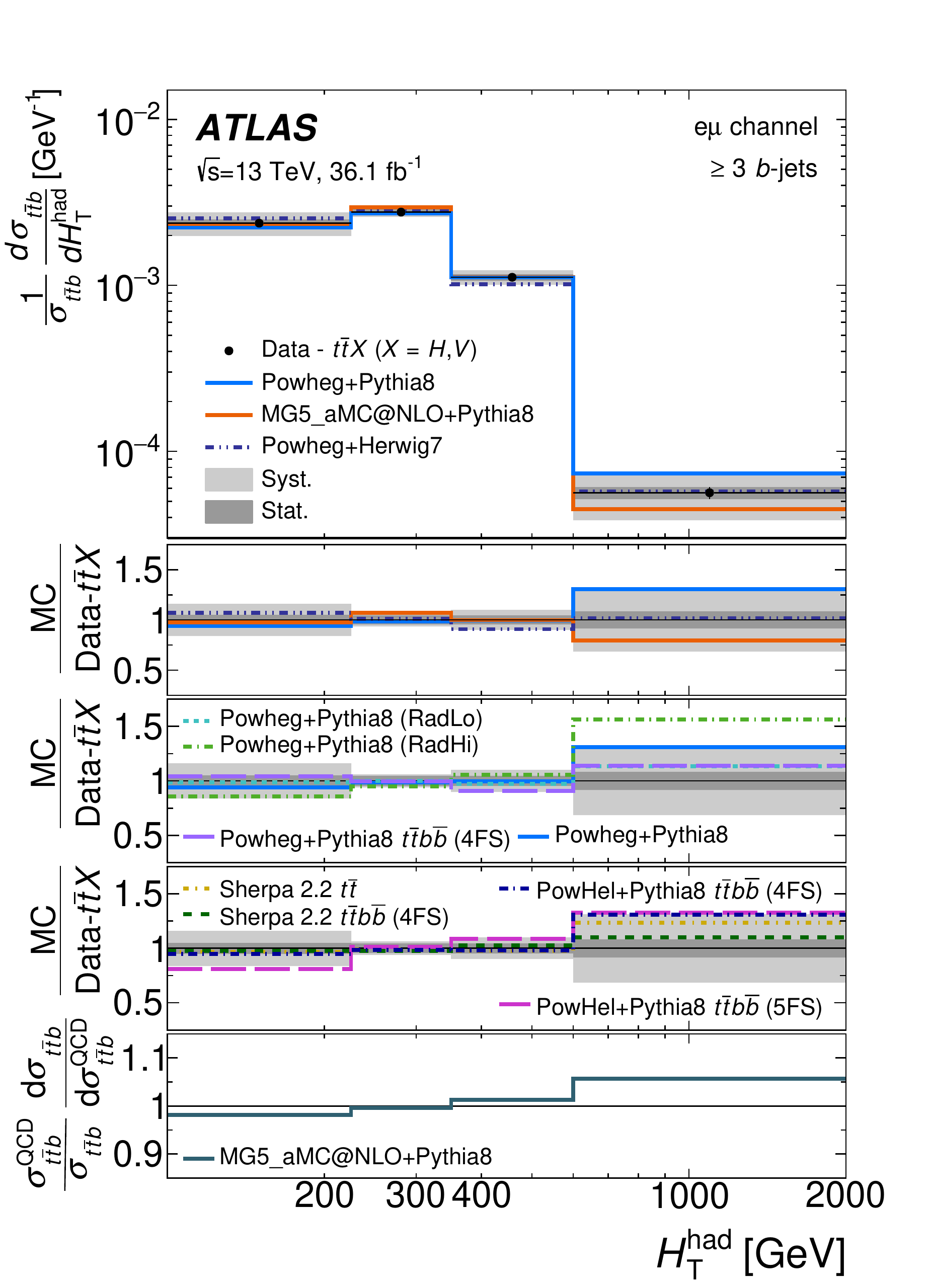}
  }
  \caption{Relative differential cross-sections as a function of
    \protect\subref{fig:HT_emu} $H_{\mathrm{T}}$, \protect\subref{fig:HThad_emu}
    $H_{\mathrm{T}}^{\mathrm{had}}$ in events with at least three $b$-jets in
    the $e\mu$ channel compared with various MC generators. The \ttH and \ttV contributions are subtracted from data. 
    Four ratio panels are shown, the first three of which show the ratios of
    various predictions to data. The last panel shows the ratio of predictions of
    normalised differential cross-sections from \amcnlopyeight including
    (numerator) and not including (denominator) the contributions from \ttV and
    \ttH production. Uncertainty bands represent the statistical and total systematic uncertainties as described in Section~\ref{sec:systematics}.  Events with $H_{\mathrm{T}}$ ($H_{\mathrm{T}}^{\mathrm{had}}$) values outside the axis range are not included in the plot.}
  \label{fig:res_ht}
\end{figure}

\begin{figure}
  \centering
  \subcaptionbox{\label{fig:HT_ljets}}{
    \includegraphics[width=0.45\textwidth]{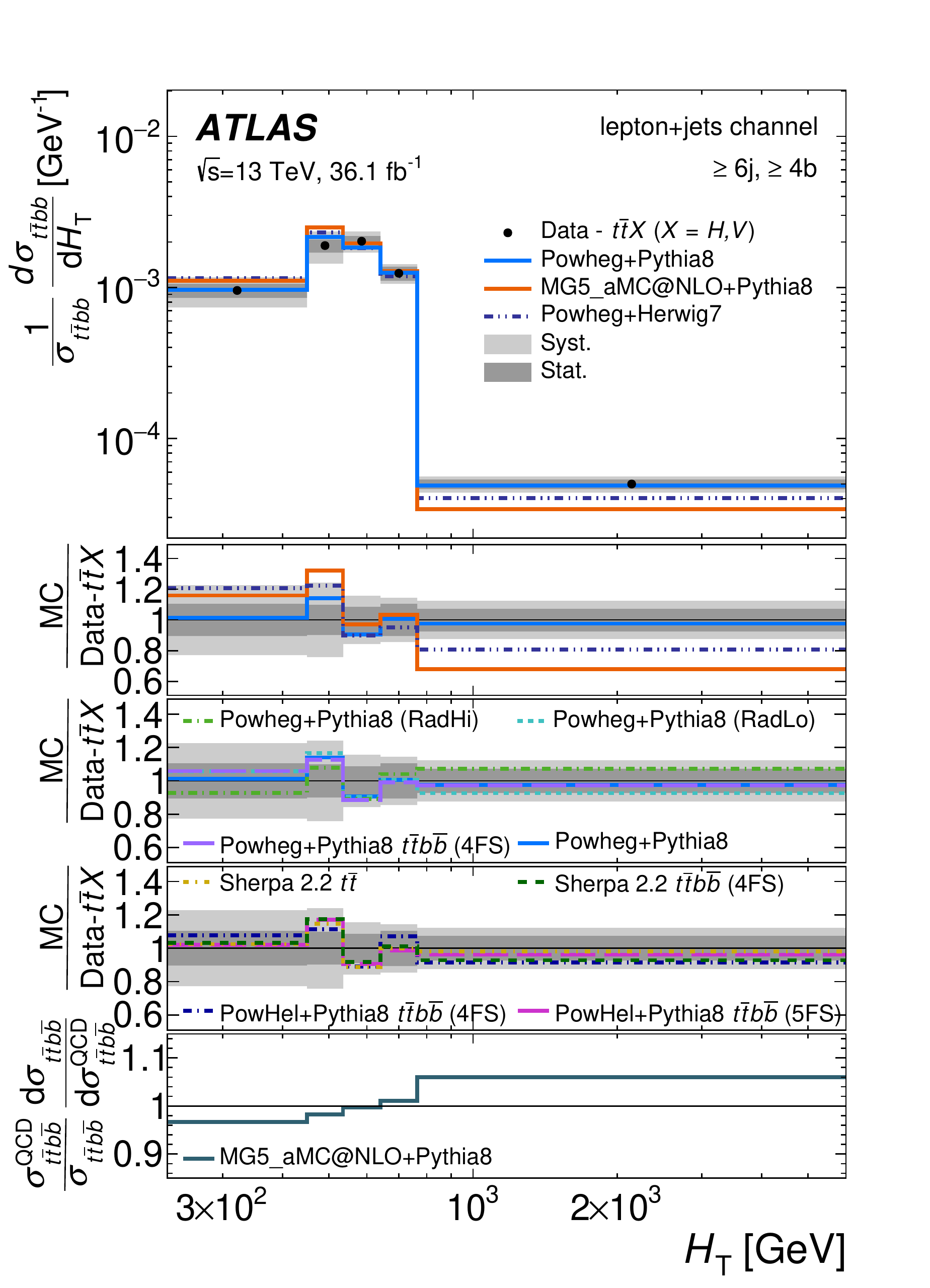}
  }
  \subcaptionbox{\label{fig:HThad_ljets}}{
    \includegraphics[width=0.45\textwidth]{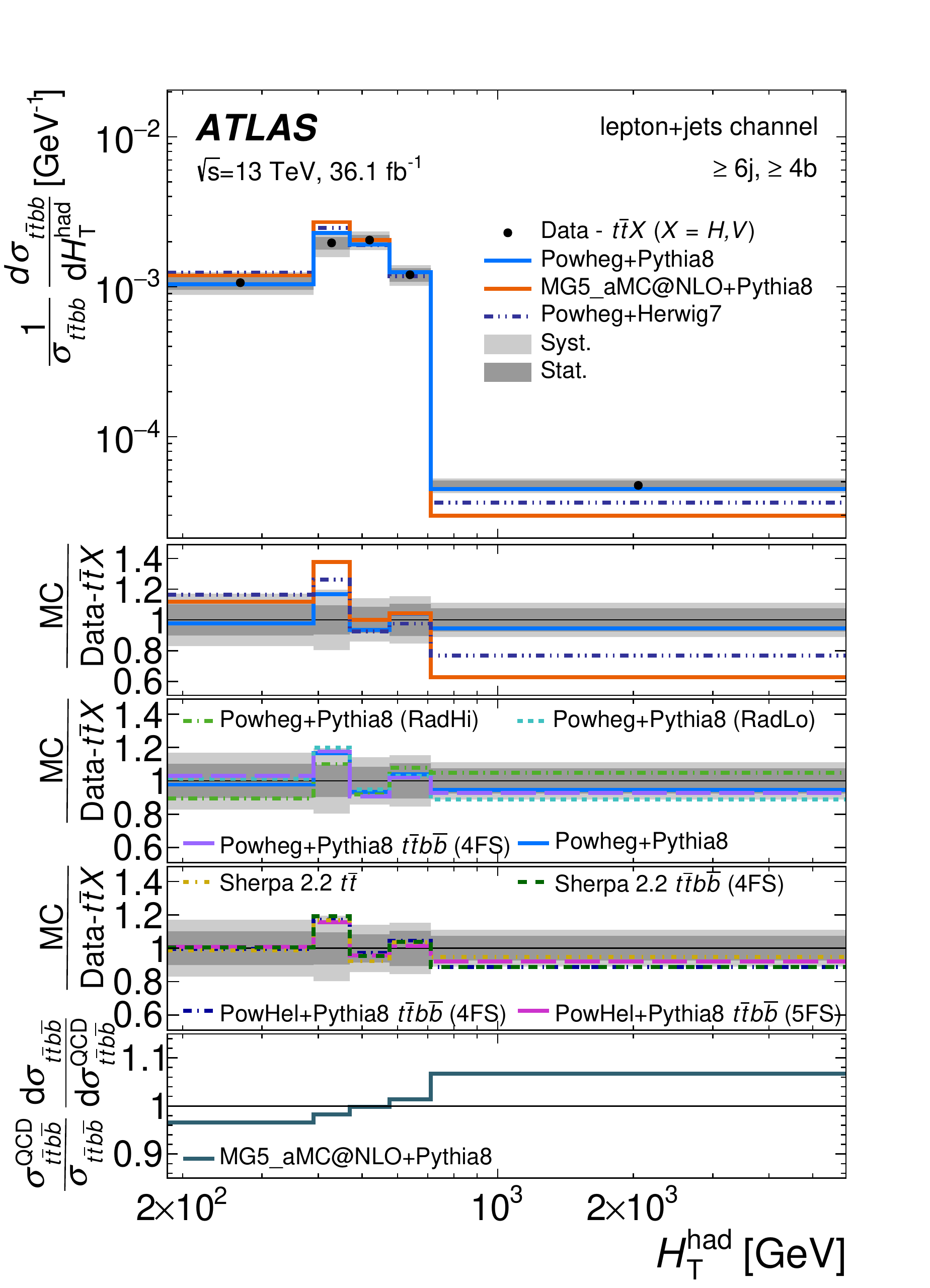}
  }
  \caption{Relative differential cross-sections as a function of
    \protect\subref{fig:HT_ljets} $H_{\mathrm{T}}$,
    \protect\subref{fig:HThad_ljets} $H_{\mathrm{T}}^{\mathrm{had}}$ in events
    with at least four $b$-jets in the \ljets channel compared with various MC
    generators. The \ttH and \ttV contributions are subtracted from data. Four ratio panels are shown, the first three of which show the ratios of
    various predictions to data. The last panel shows the ratio of predictions of
    normalised differential cross-sections from \amcnlopyeight including
    (numerator) and not including (denominator) the contributions from \ttV and
    \ttH production. Uncertainty bands represent the statistical and total systematic uncertainties as described in Section~\ref{sec:systematics}. Events with $H_{\mathrm{T}}$ ($H_{\mathrm{T}}^{\mathrm{had}}$) values outside the axis range are not included in the plot.}
  \label{fig:res_htljets}
\end{figure}

\begin{figure}
  \centering
  \subcaptionbox{\label{fig:syst_hthad:emu}}{
    \includegraphics[width=0.45\textwidth]{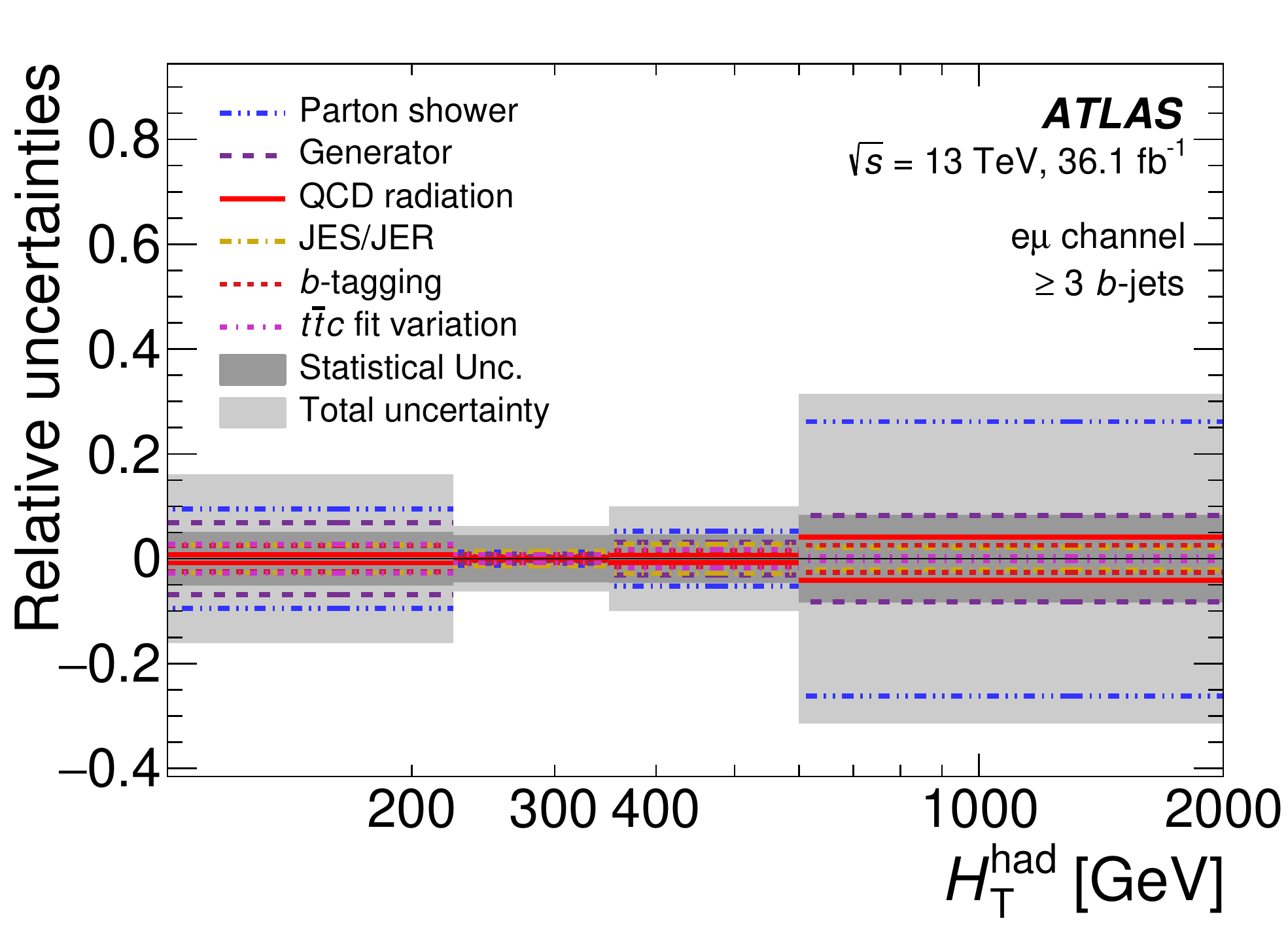}
  }
  \subcaptionbox{\label{fig:syst_hthad:ljets}}{
    \includegraphics[width=0.45\textwidth]{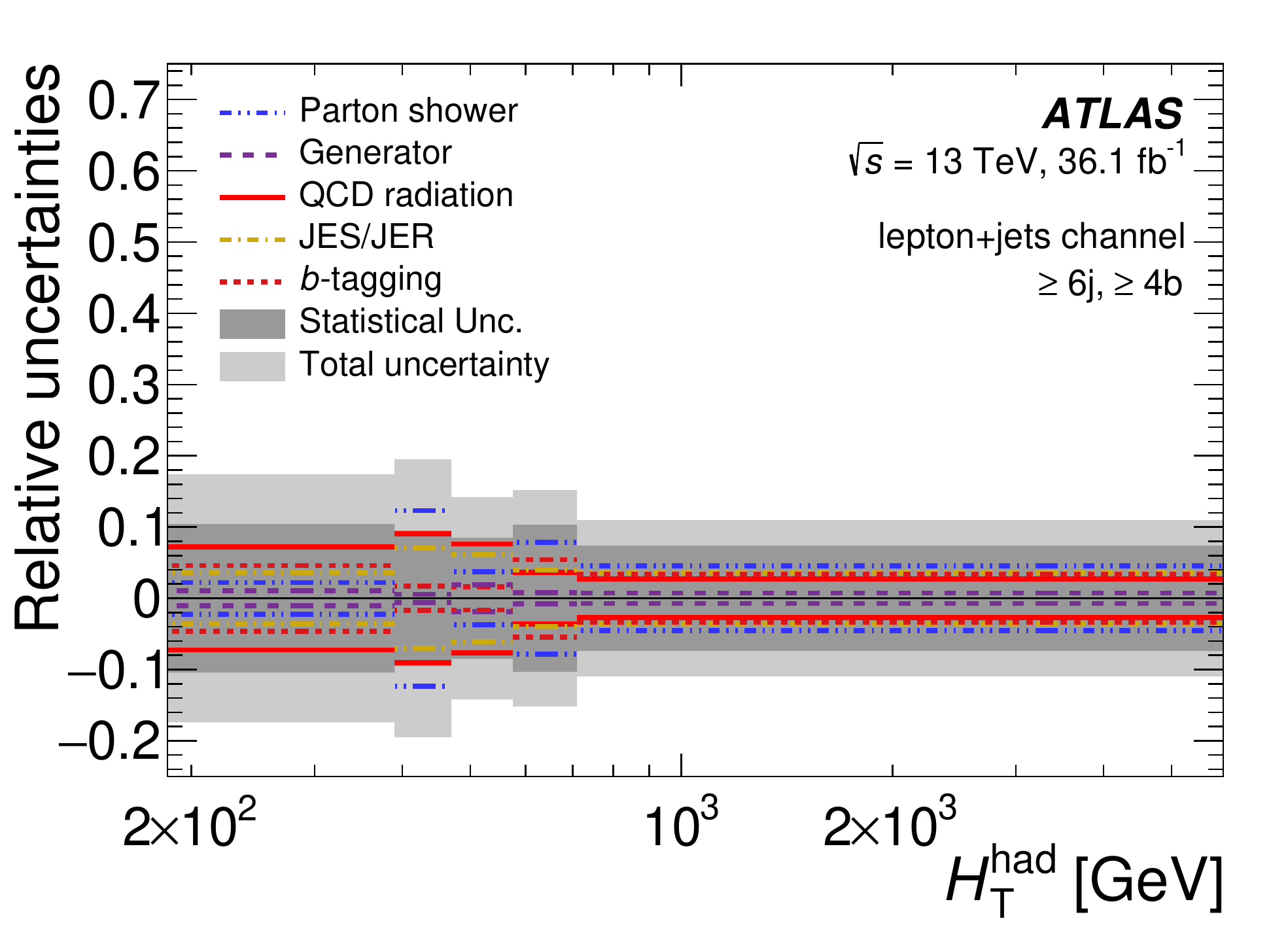}
  }
  \caption{Relative systematic uncertainties from various theoretical and
    experimental sources for $H_{\mathrm{T}}^{\mathrm{had}}$ variable measured
    in the \protect\subref{fig:syst_hthad:emu} $e\mu$ and
    \protect\subref{fig:syst_hthad:ljets} \ljets channels.}
  \label{fig:syst_hthad}
\end{figure}

Distributions for $H_{\mathrm{T}}$ and \hthad are shown
in Figures~\ref{fig:res_ht} and~\ref{fig:res_htljets}. Assessments of the
level of agreement between data and the various \ac{MC} predictions are
presented in Table~\ref{tab:comb_ht_chi2}. The data are well described by all
MC models in both channels within uncertainties of $10$\%--$30$\%, except for  \amcnlopyeight, which shows poor agreement  in the \ljets channel.  Major
contributions of systematics uncertainties in the measurement from various
sources are illustrated in Figure~\ref{fig:syst_hthad}. Parton shower
modelling is the dominant uncertainty in most regions of \hthad. Similar
uncertainties are found in the measurement of $H_{\mathrm{T}}$, where the low
$H_{\mathrm{T}}$ region has relatively larger uncertainties due to QCD radiation
scale variations because of softer jets contributing to this region.
\begin{table}
  \caption{Values of $\chi^2$ per degree of freedom and $p$-values between the
    unfolded normalised cross-sections and the various predictions for the
    $H_{\mathrm{T}}$ and $H_{\mathrm{T}}^{\mathrm{had}}$ measurements in the
    $e\mu$ and \ljets channels.  The number of degrees of freedom is
    equal to the number of bins in the measured distribution minus one.}
  \small
  \centering
  \input{tables/combined_tables_ttXsubtracted/ht}
  \label{tab:comb_ht_chi2}
\end{table}

\begin{figure}
  \centering
  \subcaptionbox{\label{fig:leadpt_emu}}{
    \includegraphics[width=0.42\textwidth]{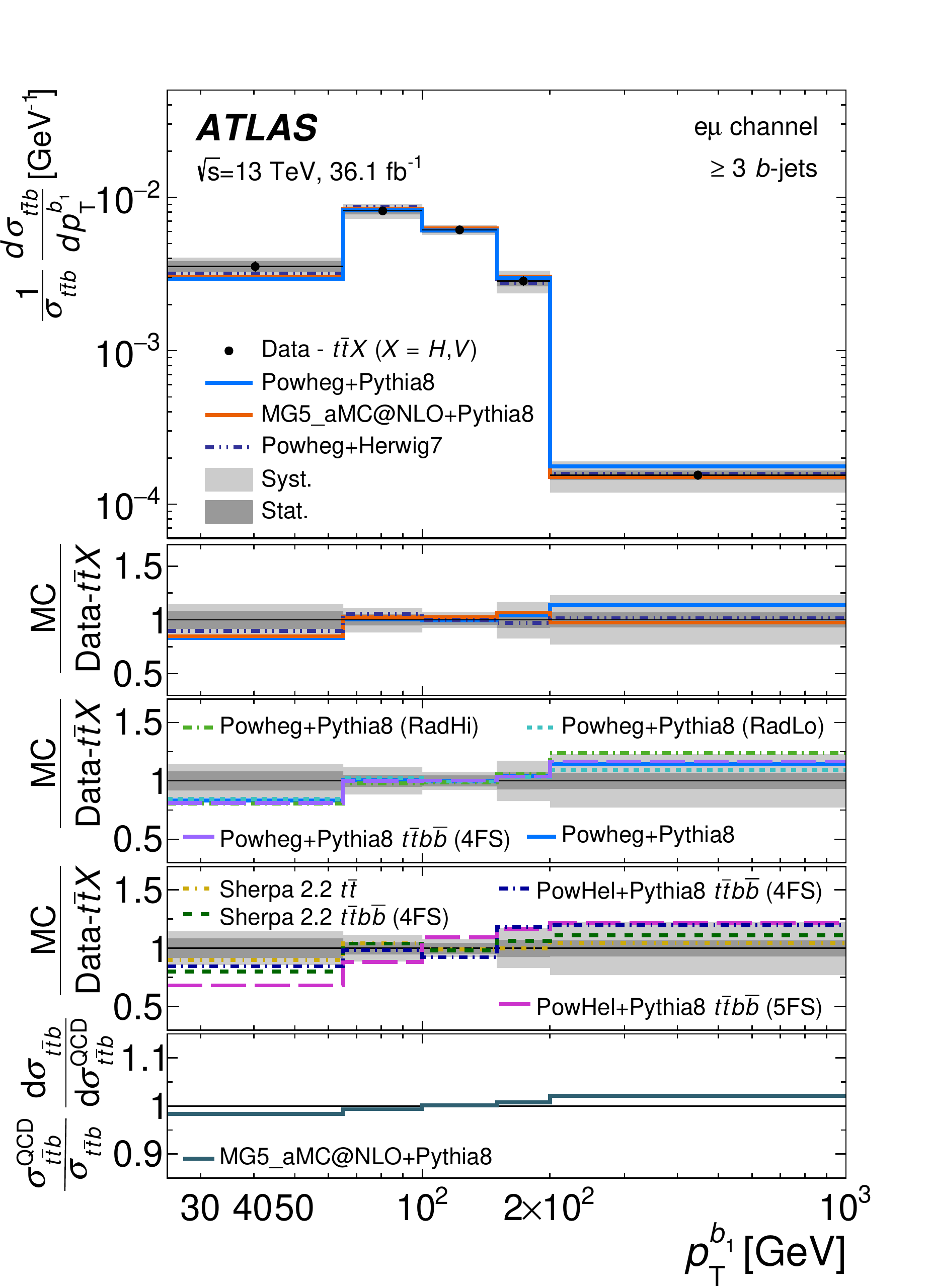}
  }
  \subcaptionbox{\label{fig:subleadpt_emu}}{
    \includegraphics[width=0.42\textwidth]{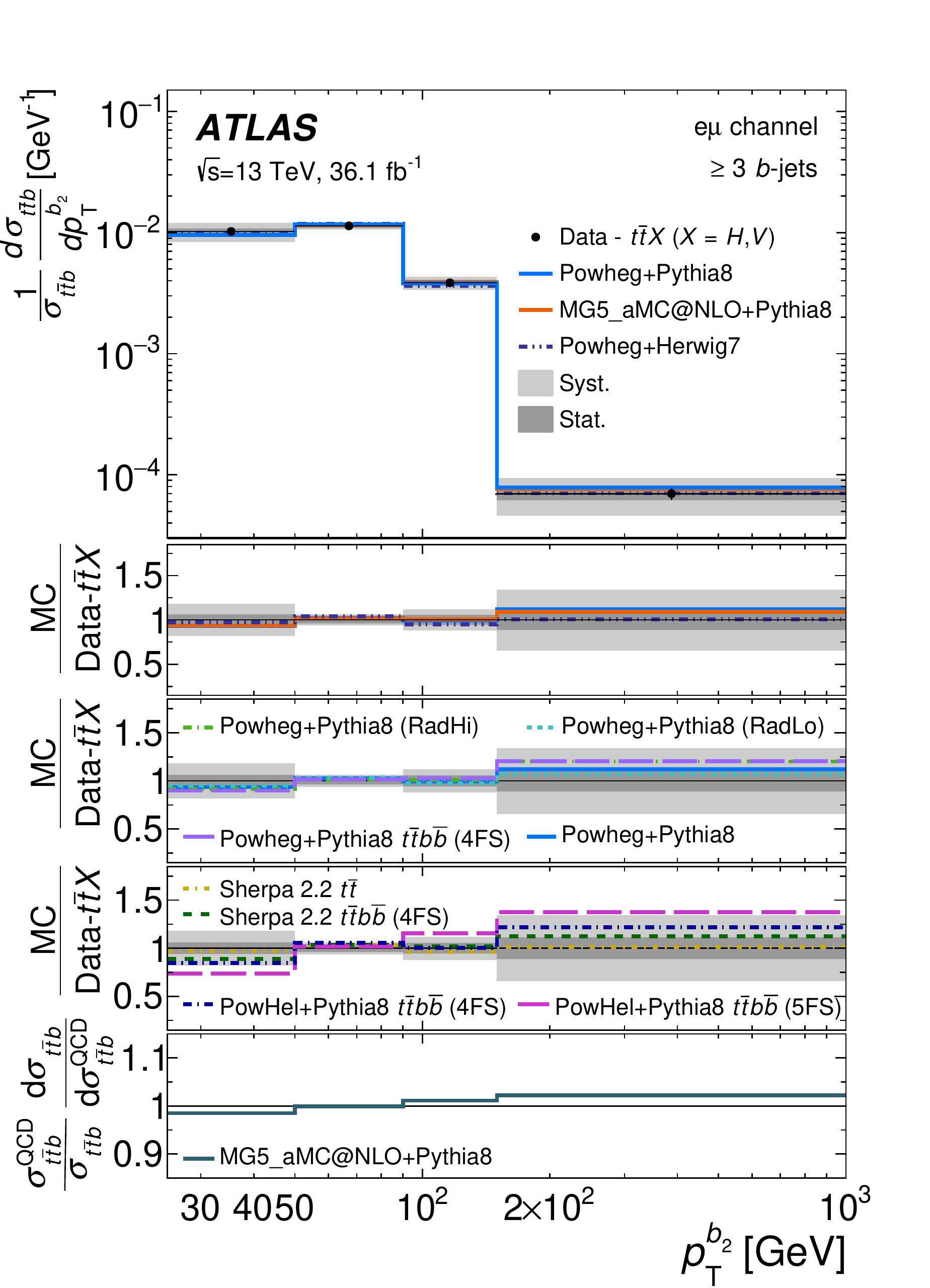}
  } \\
  \subcaptionbox{\label{fig:thirdleadpt_emu}}{
    \includegraphics[width=0.42\textwidth]{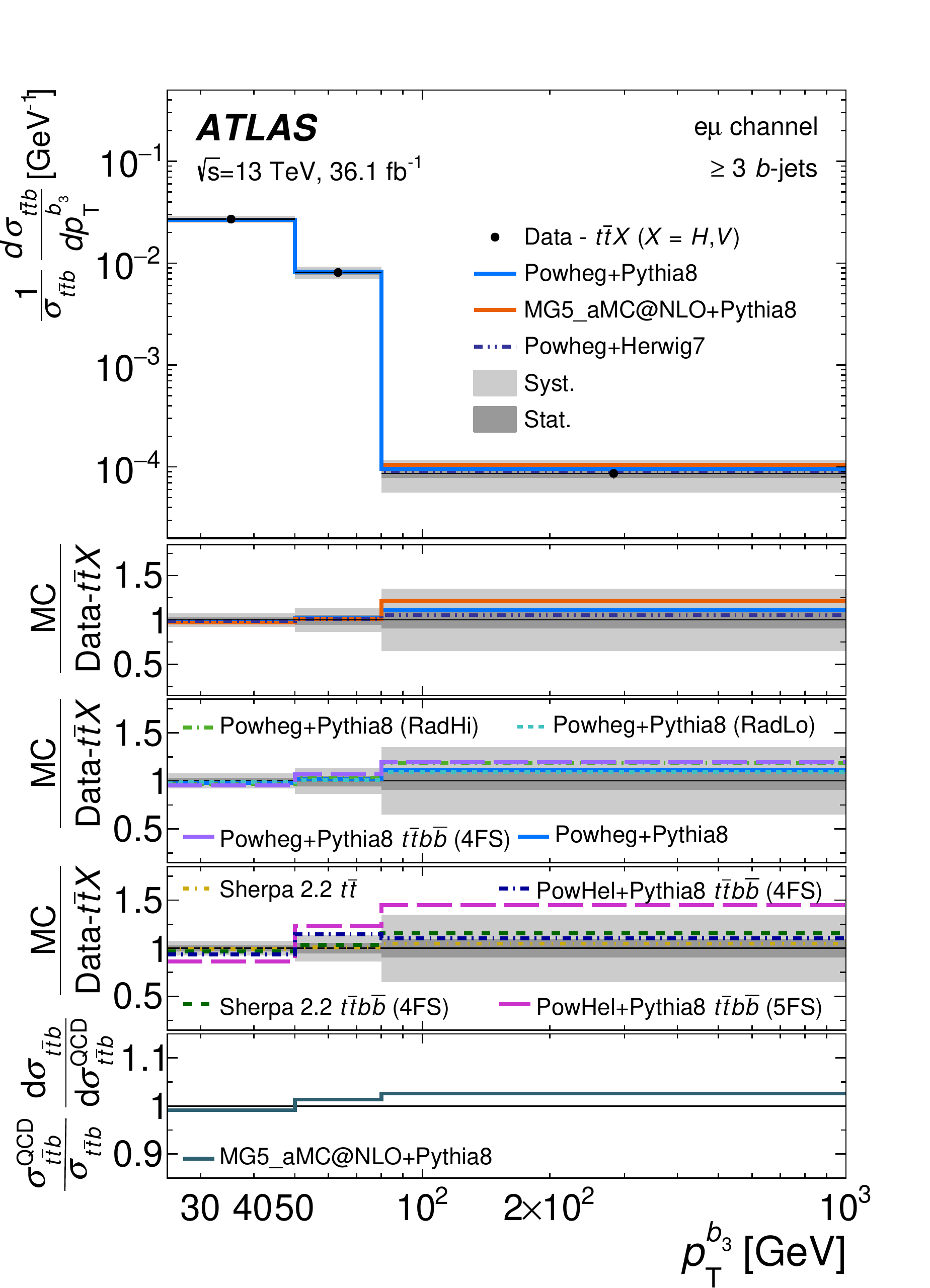}
  }
  \caption{Relative differential cross-sections as a function of $b$-jets
    $p_{\mathrm{T}}$ for \pt-ordered $b$-jets in events with at least three
    $b$-jets in the $e\mu$ channel compared with various MC generators. The \ttH and \ttV contributions are subtracted from data. 
    \protect\subref{fig:leadpt_emu} leading $b$-jet \pt,
    \protect\subref{fig:subleadpt_emu} sub-leading $b$-jet \pt,
    \protect\subref{fig:thirdleadpt_emu} third-leading $b$-jet \pt.
    Four ratio panels are shown, the first three of which show the ratios of
    various predictions to data. The last panel shows the ratio of predictions of
    normalised differential cross-sections from \amcnlopyeight including
    (numerator) and not including (denominator) the contributions from \ttV and
    \ttH production. Uncertainty bands represent the statistical and total systematic uncertainties as described in Section~\ref{sec:systematics}.  Events with $b$-jets $p_{\mathrm{T}}$ values outside the axis range are not included in the plot.}
    \label{fig:res_pt}
\end{figure}

\begin{figure}
  \centering
  \subcaptionbox{\label{fig:leadpt_ljets}}{
    \includegraphics[width=0.417\textwidth]{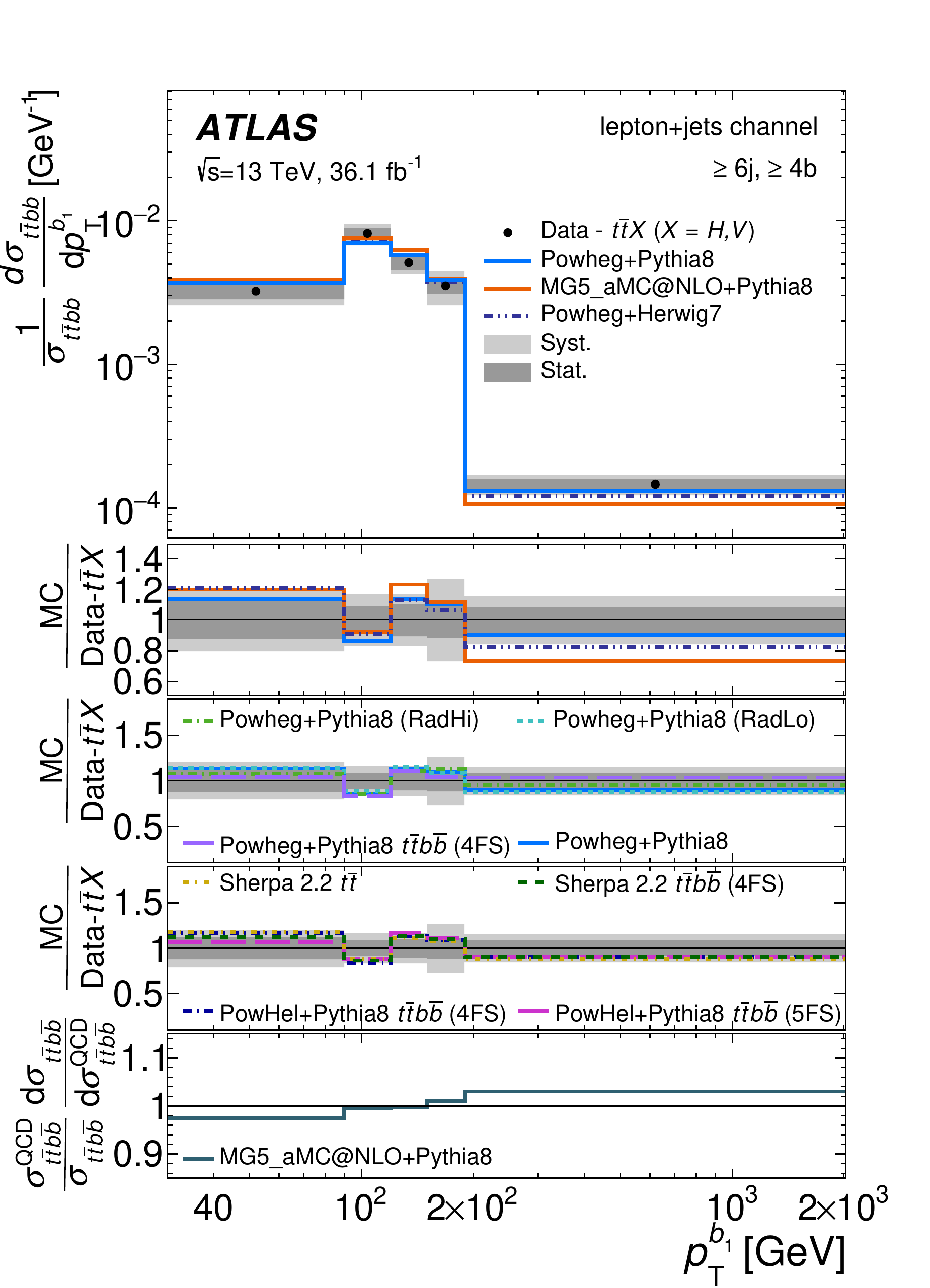}
  }
  \subcaptionbox{\label{fig:subleadpt_ljets}}{
    \includegraphics[width=0.417\textwidth]{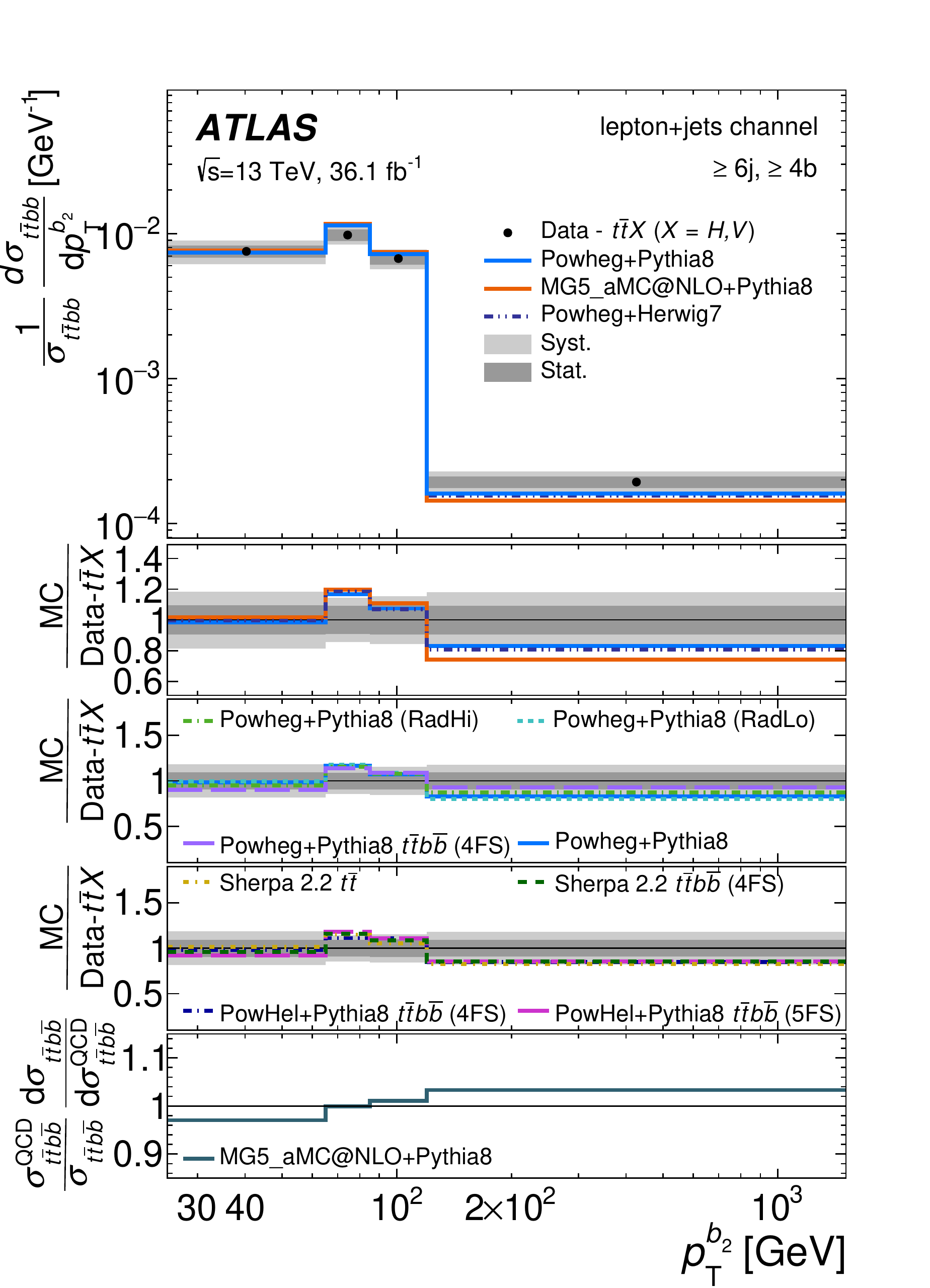}
  }
  \subcaptionbox{\label{fig:thirdleadpt_ljets}}{
    \includegraphics[width=0.417\textwidth]{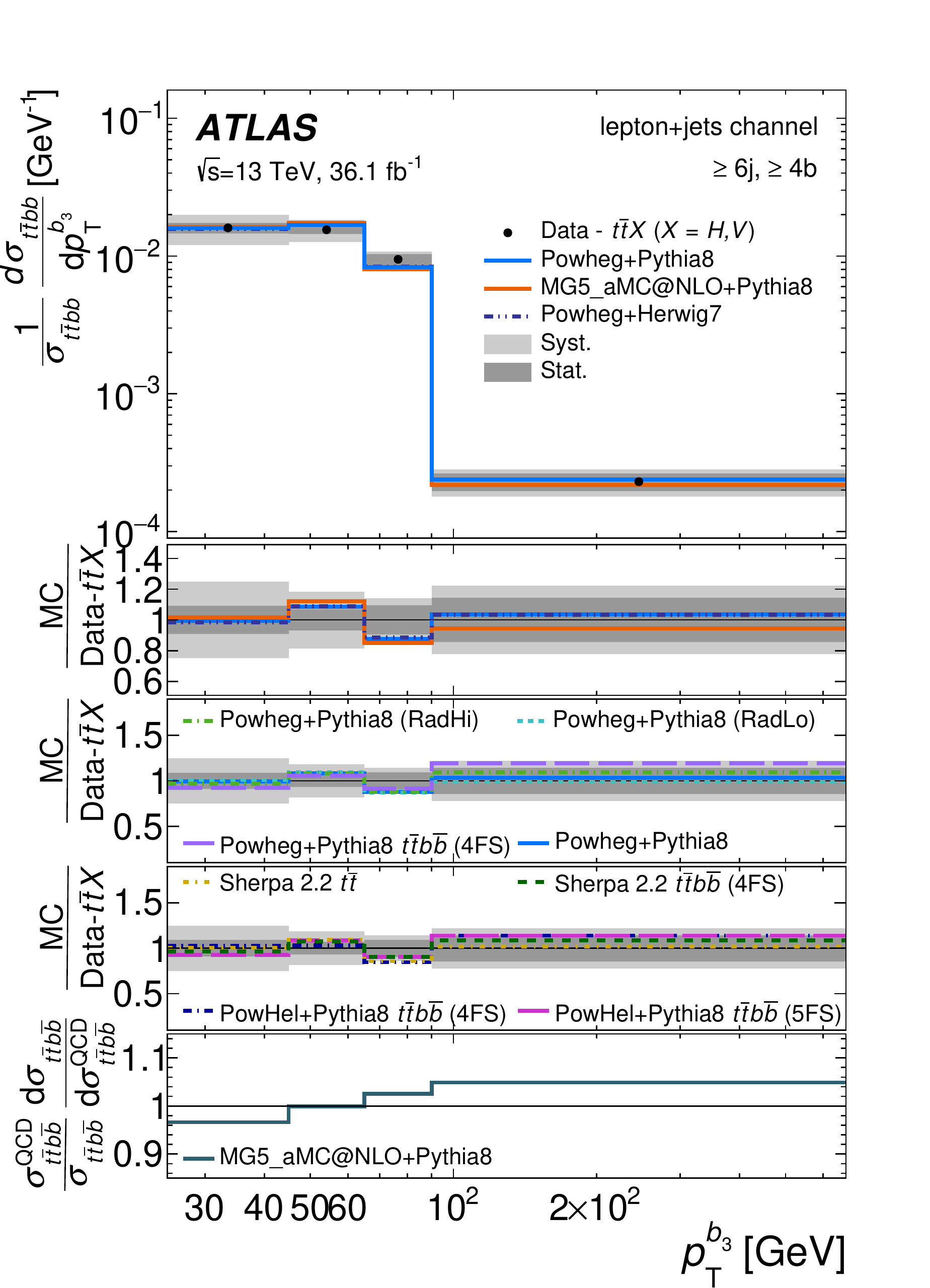}
  }
  \subcaptionbox{\label{fig:fourthleadpt_ljets}}{
    \includegraphics[width=0.417\textwidth]{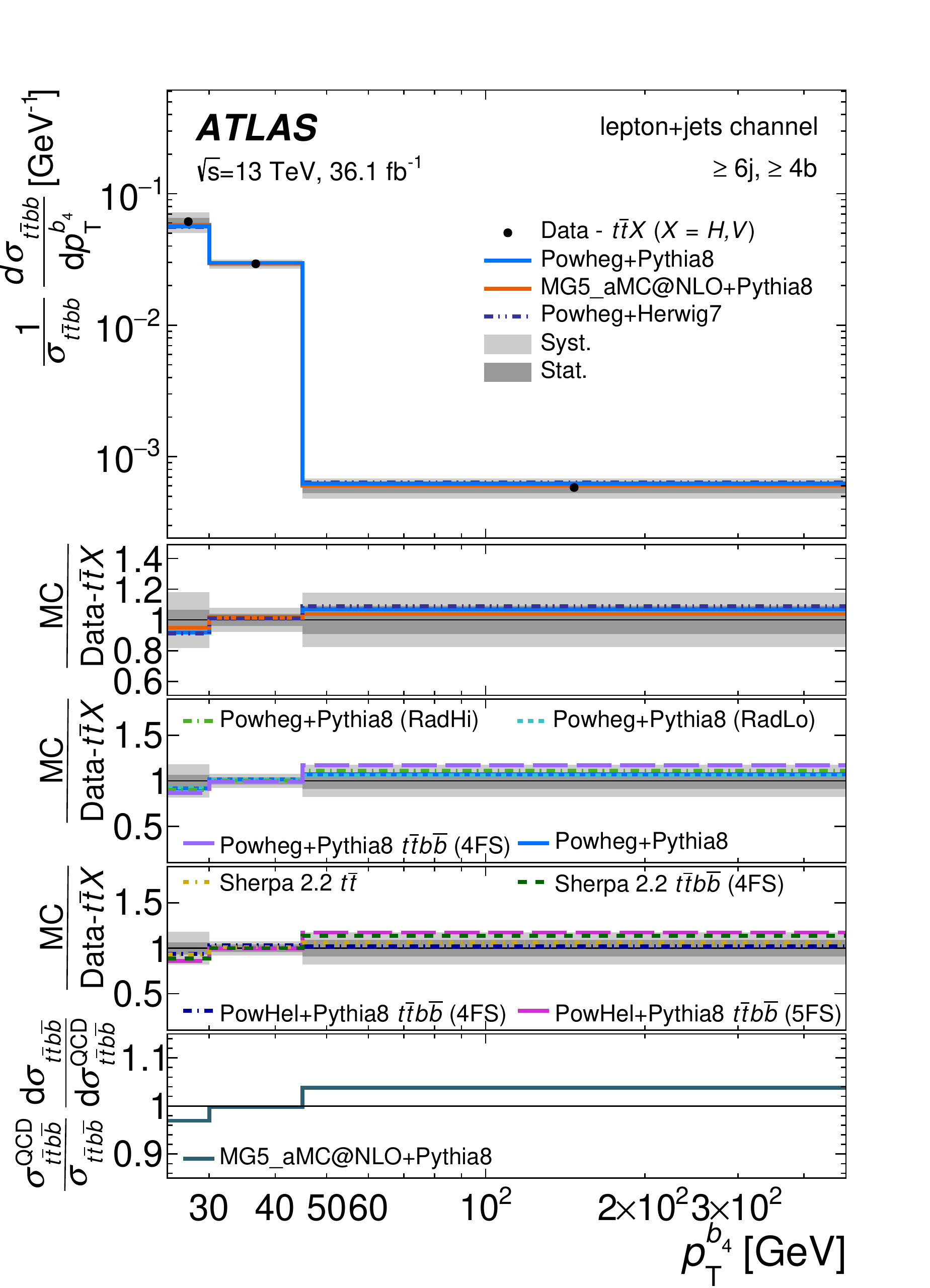}
  }
  \caption{Relative differential cross-sections as a function of $b$-jets
    $p_{\mathrm{T}}$ for \pt-ordered $b$-jets in events with at least four
    $b$-jets in the \ljets channel compared with various \ac{MC} generators. The \ttH and \ttV contributions are subtracted from data. 
    \protect\subref{fig:leadpt_ljets} leading $b$-jet \pt,
    \protect\subref{fig:subleadpt_ljets} sub-leading $b$-jet \pt,
    \protect\subref{fig:thirdleadpt_ljets} third-leading $b$-jet \pt,
    \protect\subref{fig:fourthleadpt_ljets} fourth-leading $b$-jet \pt.
    Four ratio panels are shown, the first three of which show the ratios of
    various predictions to data. The last panel shows the ratio of predictions of
    normalised differential cross-sections from \amcnlopyeight including
    (numerator) and not including (denominator) the contributions from \ttV and
    \ttH production. Uncertainty bands represent the statistical and total systematic uncertainties as described in Section~\ref{sec:systematics}. Events with $b$-jets $p_{\mathrm{T}}$ values outside the axis range are not included in the plot.}
  \label{fig:res_ptljets}
\end{figure}

The \pt distributions of the \pt-ordered \bjets are shown in
Figure~\ref{fig:res_pt} and Figure~\ref{fig:res_ptljets} for events with $\geq 3$
\bjets in the $e\mu$ channel and
$\geq 4$ \bjets in the \ljets channel, respectively, with quantitative assessments of the
level of data--MC agreement shown in Table~\ref{tab:comb_bjets_chi2}. Most MC
predictions describe the data well, except \powhelpyeight (5FS) for the leading
and third-highest \pt $b$-jets  in events with $\geq 3$ \bjets in the $e\mu$ channel.  As the
\bjets from the top-quark decays have a tendency to be harder than the \bjets from
additional $b$-quark production via gluon splitting, the leading and sub-leading
\bjet distributions have relatively higher probability to contain the \bjets
from the top-quark
decays, while the third and the fourth \bjet distributions contain mainly jets
from gluon splitting. The measurement uncertainties are between 10\% and 25\%
depending on the \pt of the jet and the top-quark decay channel. Statistical
uncertainties are dominant in only the highest \pt bins. The uncertainties are
dominated by systematic uncertainties in the jet-energy scale and the $b$-tagging
algorithm.

\begin{sidewaystable}
  \caption{Values of $\chi^2$ per degree of freedom and $p$-values between the
    unfolded normalised cross-sections and the various predictions for the three
    (four) leading $b$-jet \pt measurements in the $e\mu$ (\ljets) channel. 
    The number of degrees of freedom is equal to the number of bins in the
    measured distribution minus one.}
  \small
  \centering
  \input{tables/combined_tables_ttXsubtracted/bjet_pt}
  \label{tab:comb_bjets_chi2}
\end{sidewaystable}

\begin{figure}
  \centering
  \subcaptionbox{\label{fig:mb1b2_emu}}{
    \includegraphics[width=0.42\textwidth]{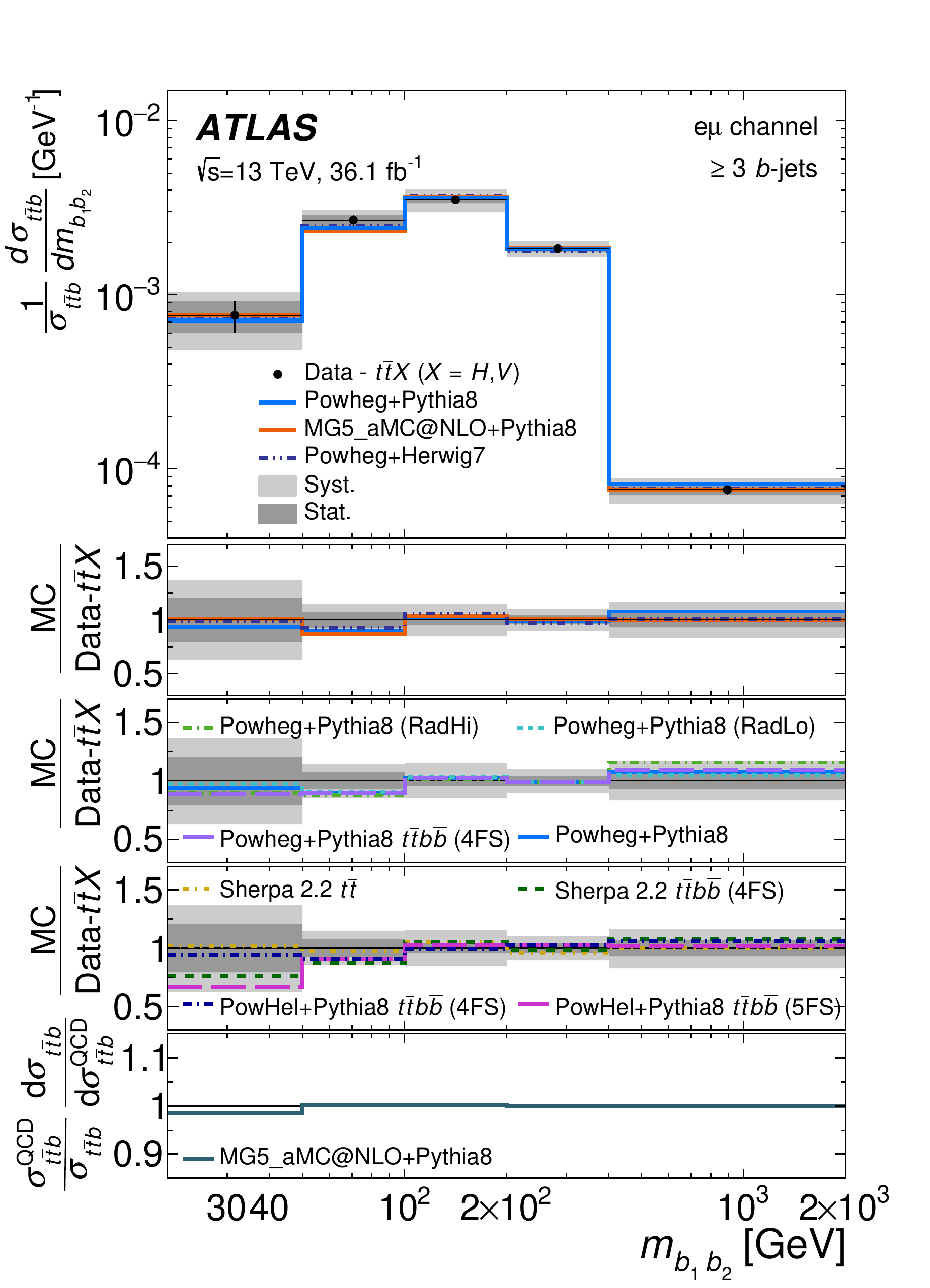}
  }
  \subcaptionbox{\label{fig:ptb1b2_emu}}{
    \includegraphics[width=0.42\textwidth]{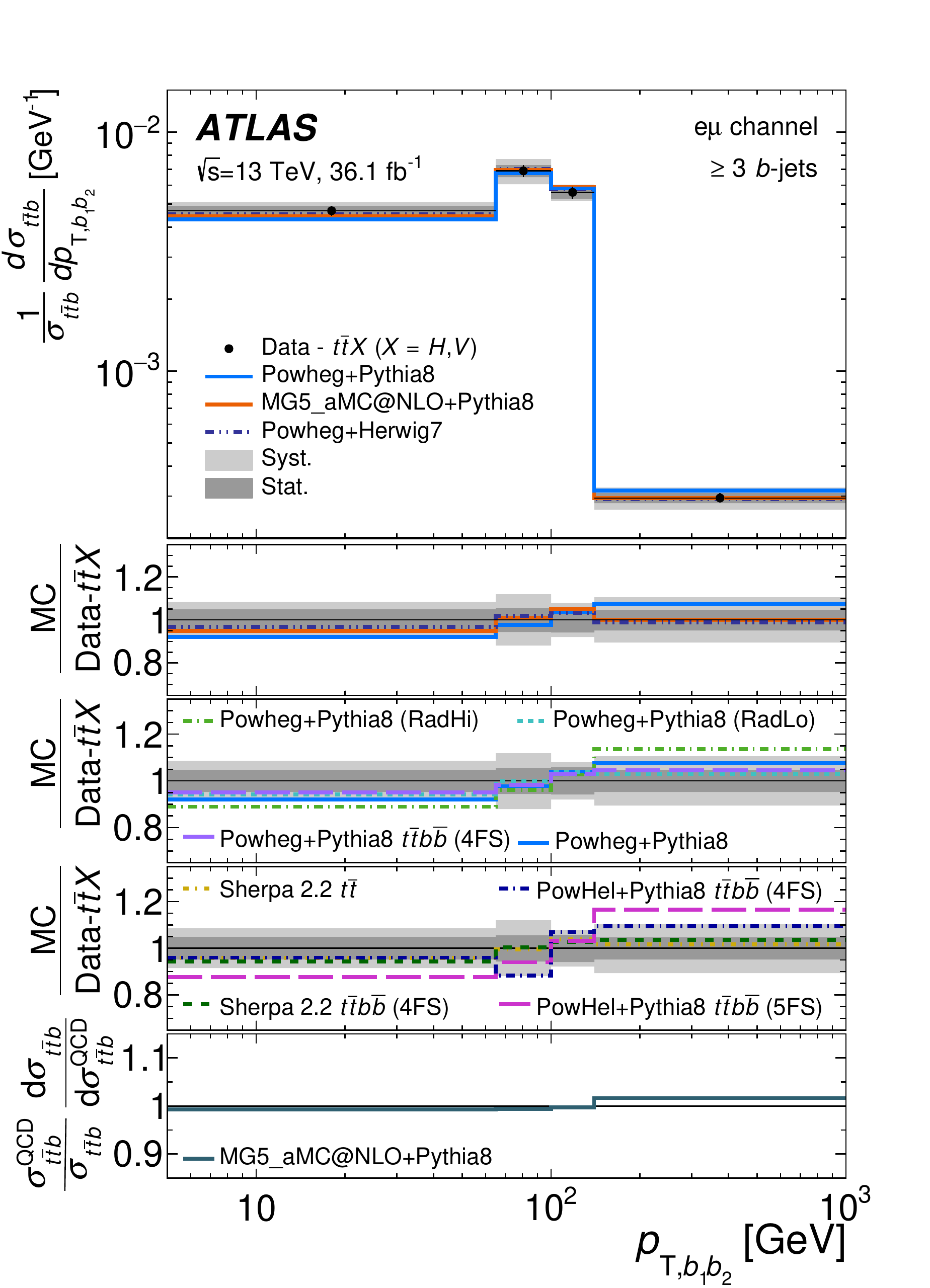}
  }
  \subcaptionbox{\label{fig:drb1b2_emu}}{
    \includegraphics[width=0.42\textwidth]{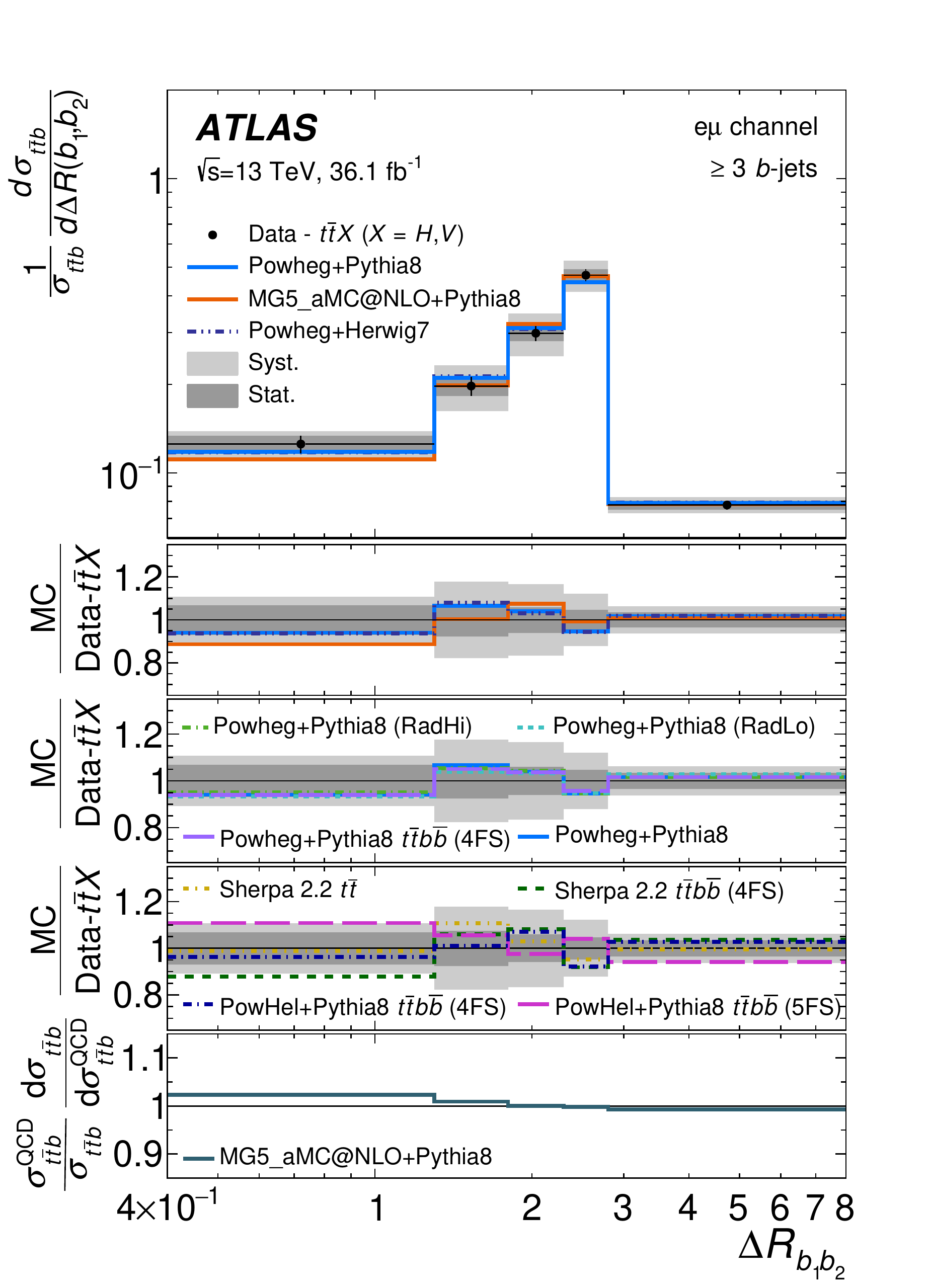}
  }
  \caption{Relative differential cross-sections as a function of
    \protect\subref{fig:mb1b2_emu} $m_{b_1b_2}$, \protect\subref{fig:ptb1b2_emu}
    $p_{\mathrm{T},b_1b_2}$, and \protect\subref{fig:drb1b2_emu} $\Delta
    R_{b_1,b_2}$ of two highest-\pt $b$-jets in events with at least three
    $b$-jets in the $e\mu$ channel compared with various MC generators. The \ttH and \ttV contributions are subtracted from data. 
    Four ratio panels are shown, the first three of which show the ratios of
    various predictions to data. The last panel shows the ratio of predictions of
    normalised differential cross-sections from \amcnlopyeight including
    (numerator) and not including (denominator) the contributions from \ttV and
    \ttH production. Uncertainty bands represent the statistical and total systematic uncertainties as described in Section~\ref{sec:systematics}.  Events with observable values outside the axis range are not included in the plot.}
  \label{fig:res_dRpt}
\end{figure}

\begin{figure}
  \centering
  \subcaptionbox{\label{fig:mb1b2_ljets}}{
    \includegraphics[width=0.42\textwidth]{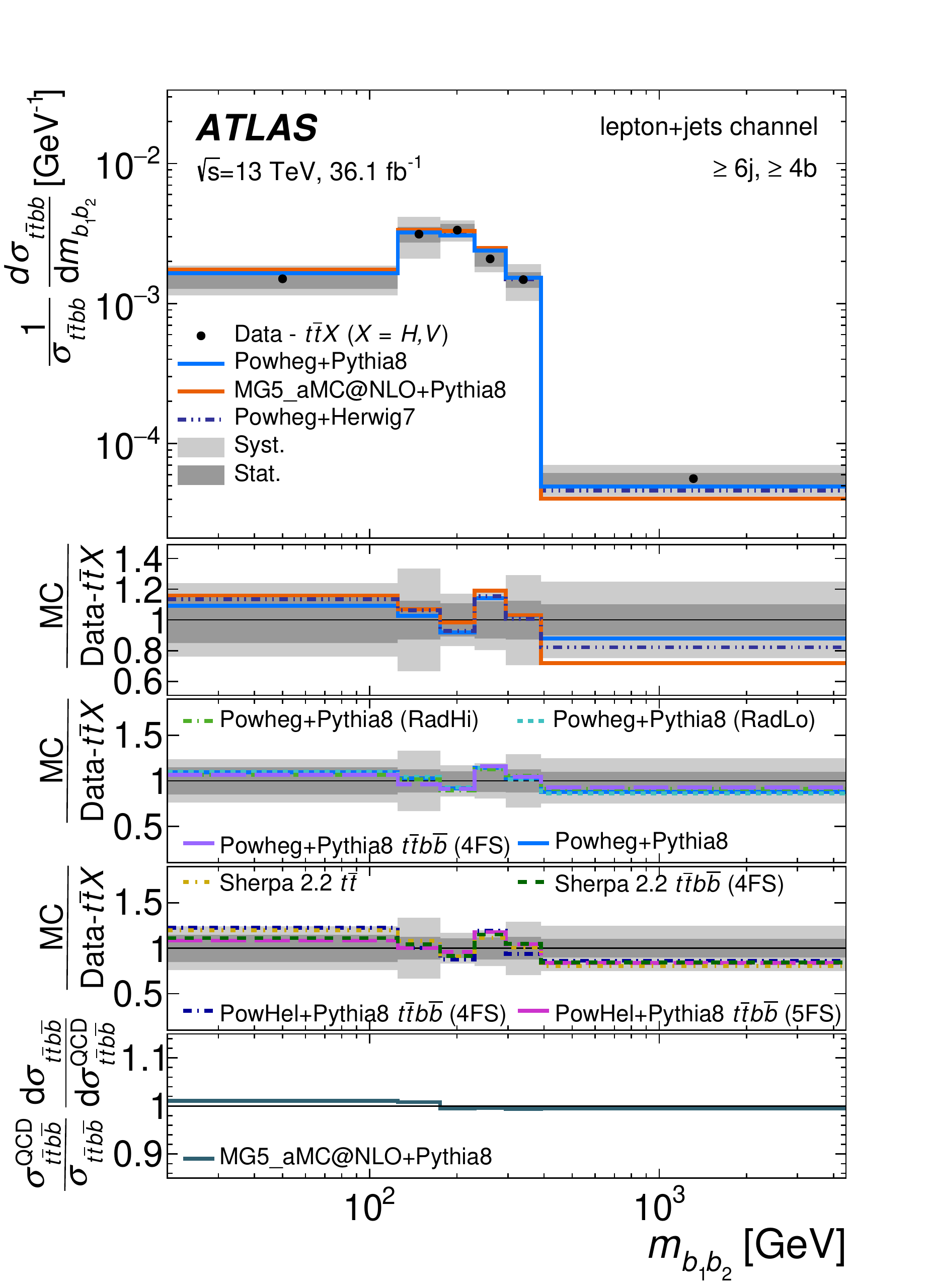}
  }
  \subcaptionbox{\label{fig:ptb1b2_ljets}}{
    \includegraphics[width=0.42\textwidth]{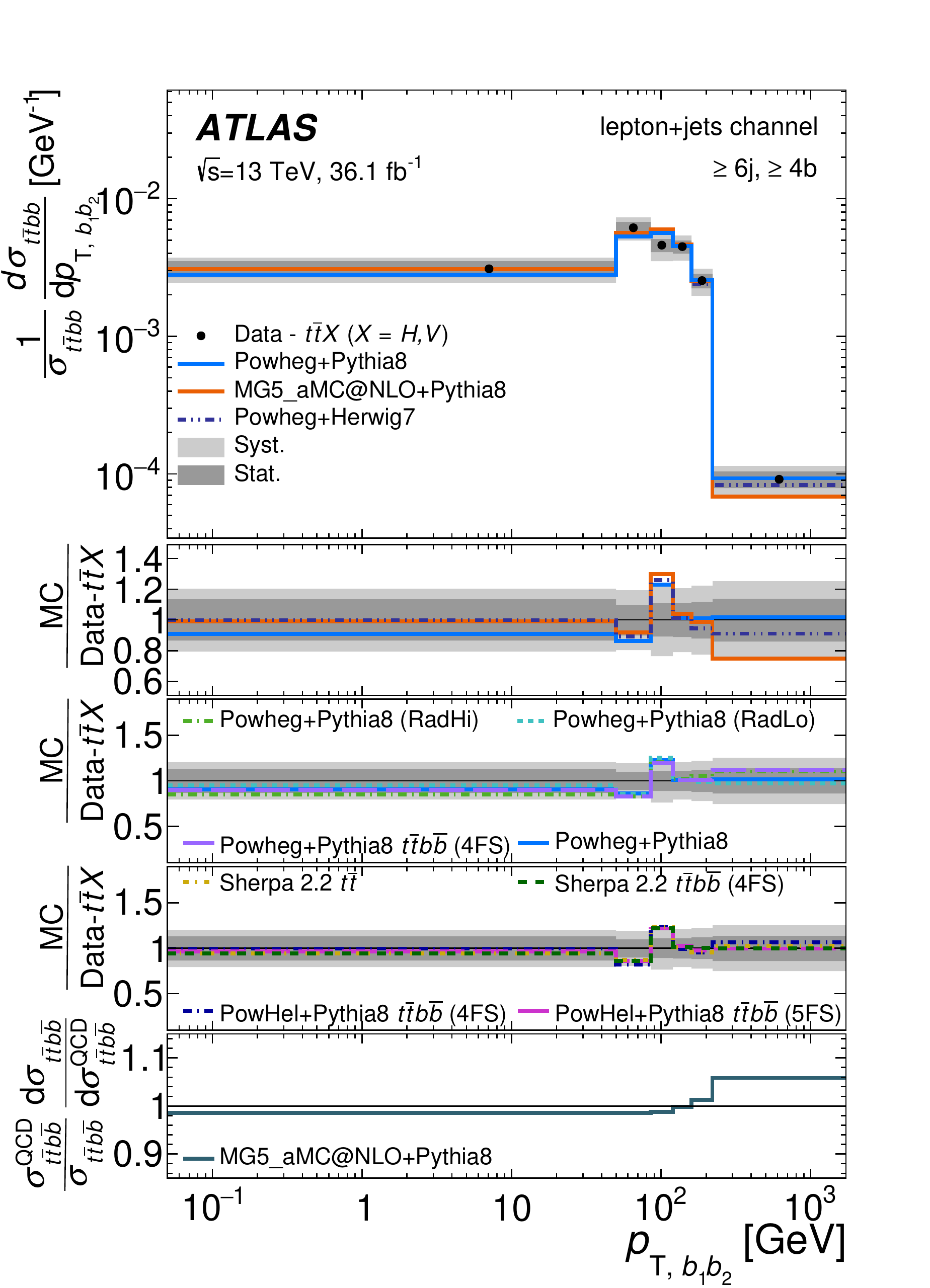}
  }
  \subcaptionbox{\label{fig:drb1b2_ljets}}{
    \includegraphics[width=0.42\textwidth]{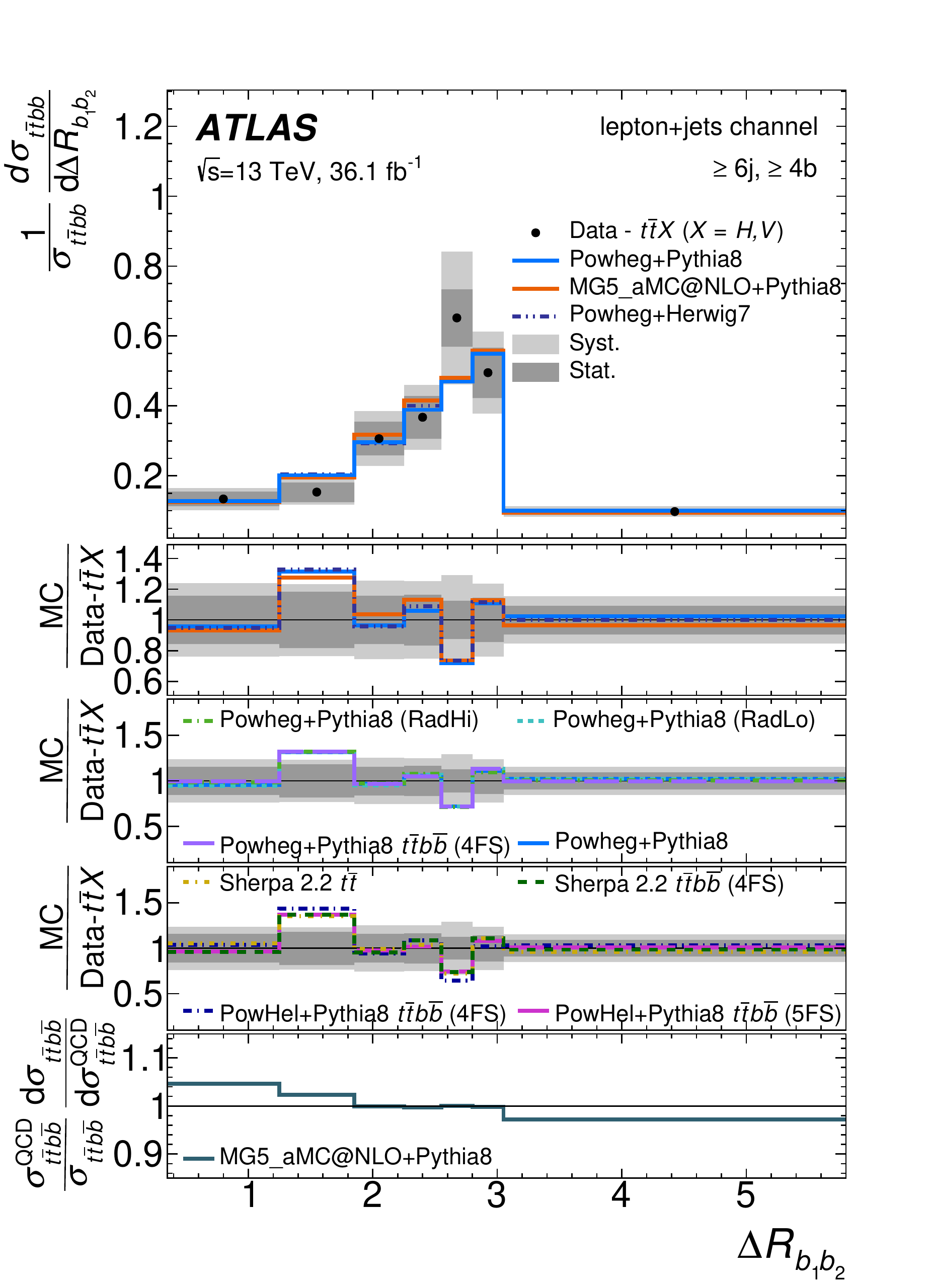}
  }
  \caption{Relative differential cross-sections as a function of
    \protect\subref{fig:mb1b2_ljets} $m_{b_1b_2}$,
    \protect\subref{fig:ptb1b2_ljets} $p_{\mathrm{T},b_1b_2}$, and
    \protect\subref{fig:drb1b2_ljets} $\Delta R_{b_1,b_2}$ of the two
    highest-\pt $b$-jets in events with at least four $b$-jets in the \ljets
    channel compared with various MC generators. The \ttH and \ttV contributions are subtracted from data. 
    Four ratio panels are shown, the first three of which show the ratios of
    various predictions to data. The last panel shows the ratio of predictions of
    normalised differential cross-sections from \amcnlopyeight including
    (numerator) and not including (denominator) the contributions from \ttV and
    \ttH production. Uncertainty bands represent the statistical and total systematic uncertainties as described in Section~\ref{sec:systematics}.  Events with observable values outside the axis range are not included in the plot.}
  \label{fig:res_dRptljets}
\end{figure}

 
Figures~\ref{fig:res_dRpt} and~\ref{fig:res_dRptljets} show the distribution of
the mass, the angular distance $\Delta R$ and \pt of the $b_1 b_2$ system built
from the two highest-\pt \bjets. The \pt of the $b_1 b_2$ system is measured
with a precision of 10\%--15\% over the full range in the $e\mu$ channel and
with an uncertainty of 20\%--25\% in the \ljets channel. It is well
described by the different MC predictions, which vary significantly less than the
experimental uncertainty.
The distributions of the $\Delta R$ between the two $b$-jets and the invariant mass
of the $b_1 b_2$ pair are measured
with slightly higher uncertainties and also show little variation between the
different predictions. Good agreement between the data and the models is
confirmed by the $p$-values listed in Table~\ref{tab:comb_bb_leading}.

\begin{table}
  \caption{Values of $\chi^2$ per degree of freedom and $p$-values between the
    unfolded normalised cross-sections and the various predictions for the mass,
    \pt and $\Delta R$ of the leading two $b$-jets in the $e\mu$ and
    \ljets channels. The number of degrees of freedom is equal to the
    number of bins in the measured distribution minus one.}
  \small
  \centering
  \input{tables/combined_tables_ttXsubtracted/bb_leading}
  \label{tab:comb_bb_leading}
\end{table}

\begin{figure}
  \centering
  \subcaptionbox{\label{fig:mbb_emu}}{
    \includegraphics[width=0.42\textwidth]{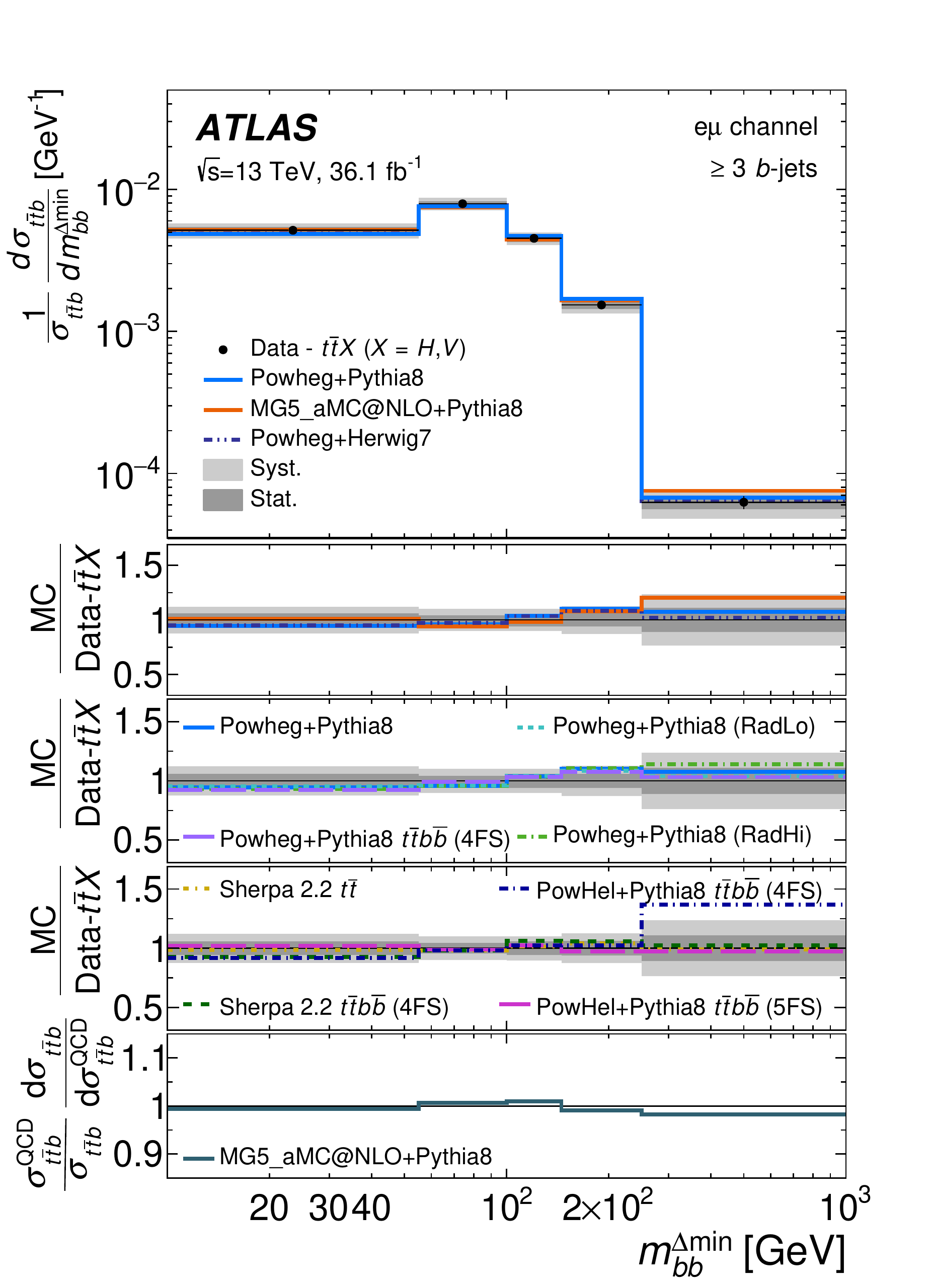}
  }
  \subcaptionbox{\label{fig:ptbb_emu}}{
    \includegraphics[width=0.42\textwidth]{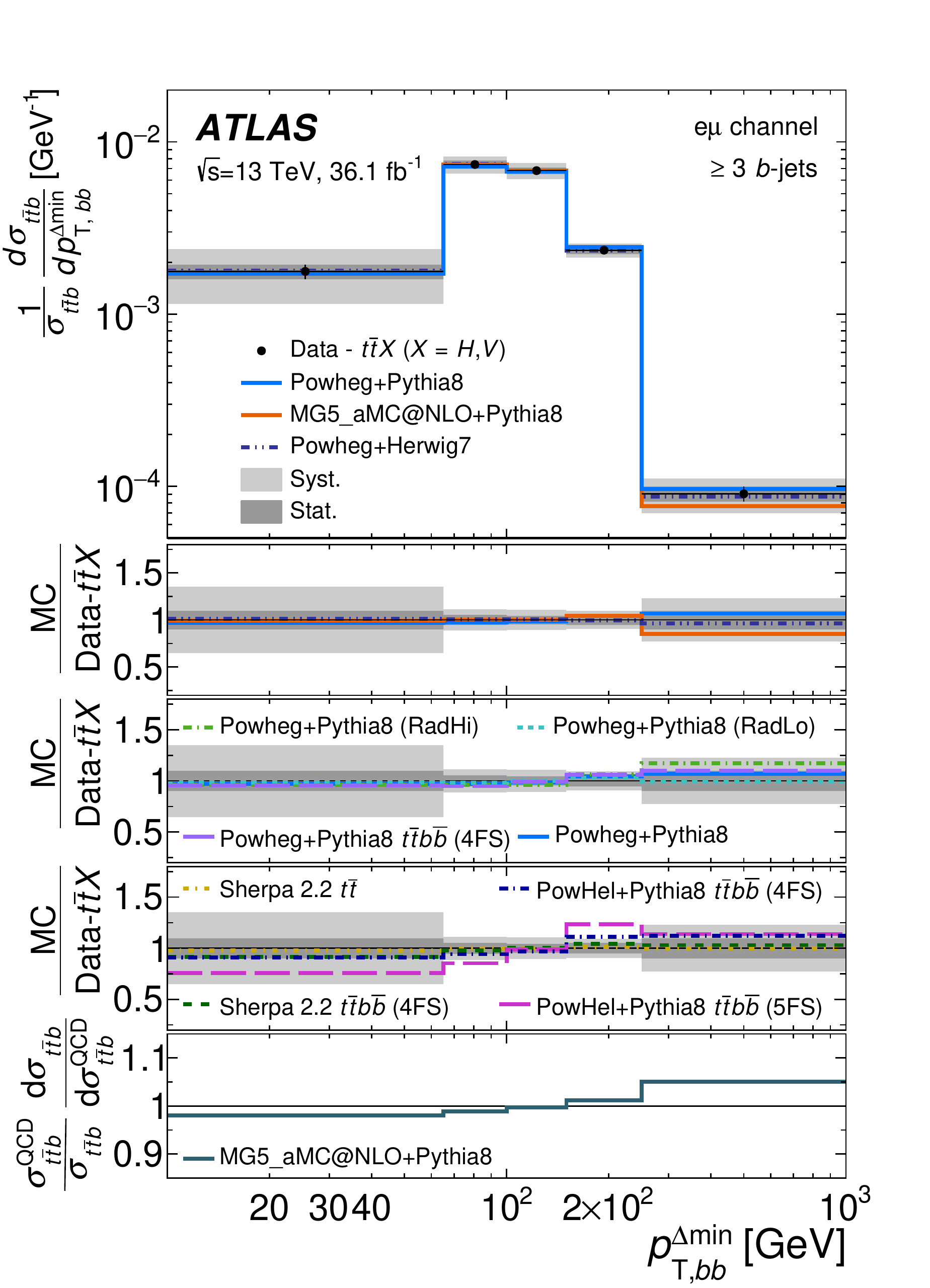}
  }
  \subcaptionbox{\label{fig:drbb_emu}}{
    \includegraphics[width=0.42\textwidth]{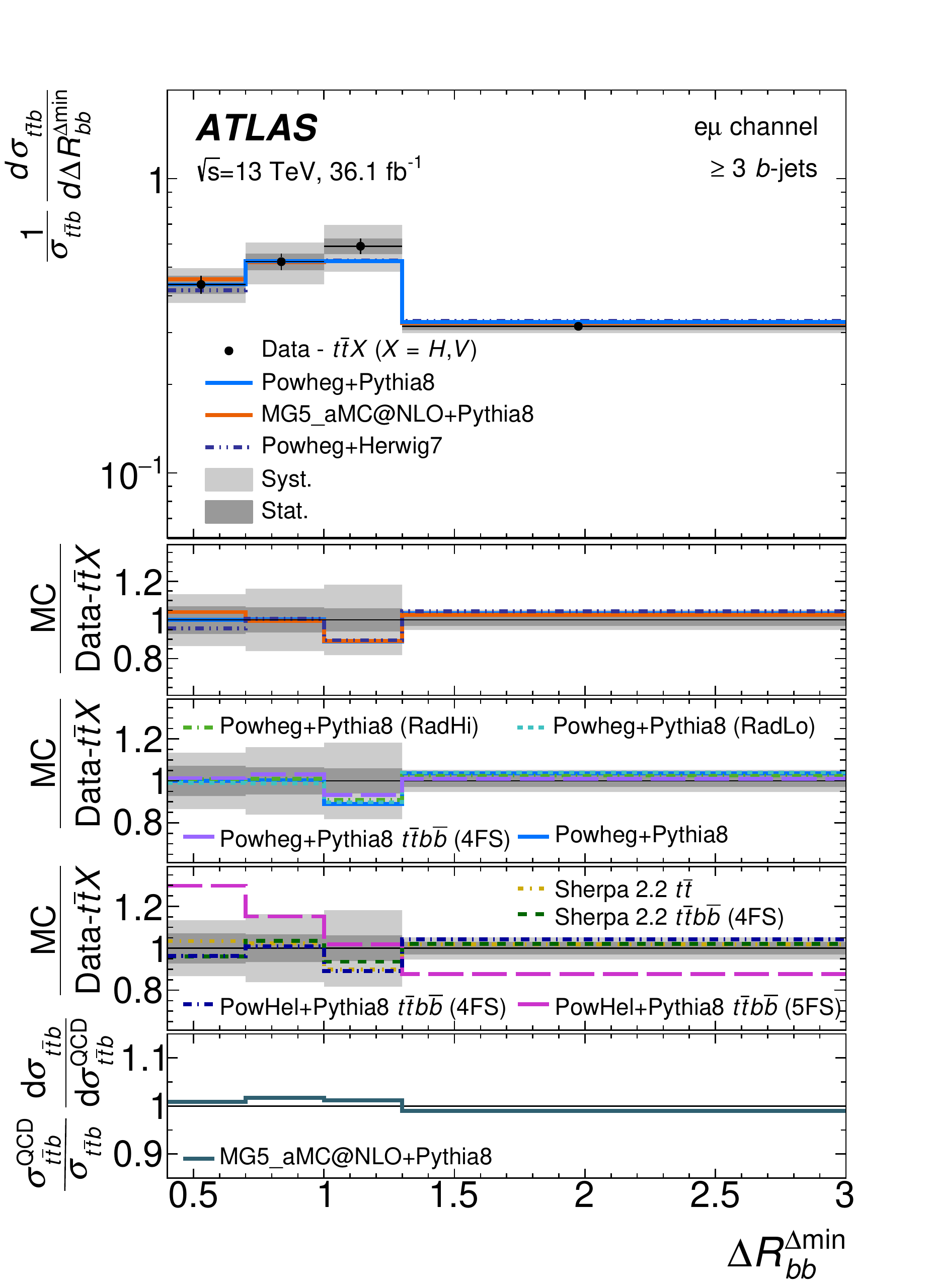}
  }
  \caption{Relative differential cross-sections as a function of
    \protect\subref{fig:mbb_emu} $m_{bb}^{\Delta \mathrm{min}}$,
    \protect\subref{fig:ptbb_emu} $p_{\mathrm{T},bb}^{\Delta \mathrm{min}}$ and
    \protect\subref{fig:drbb_emu} $\Delta R_{bb}^{\Delta \mathrm{min}}$ of two closest
    $b$-jets in $\Delta R$ in events with at least three $b$-jets in the $e\mu$
    channel compared with various MC generators. The \ttH and \ttV contributions are subtracted from data. 
    Four ratio panels are shown, the first three of which show the ratios of
    various predictions to data. The last panel shows the ratio of predictions of
    normalised differential cross-sections from \amcnlopyeight including
    (numerator) and not including (denominator) the contributions from \ttV and
    \ttH production. Uncertainty bands represent the statistical and total systematic uncertainties as described in Section~\ref{sec:systematics}.  Events with observable values outside the axis range are not included in the plot.}
  \label{fig:res_dRmin}
\end{figure}
\begin{figure}
  \centering
  \subcaptionbox{\label{fig:mbb_ljets}}{
    \includegraphics[width=0.42\textwidth]{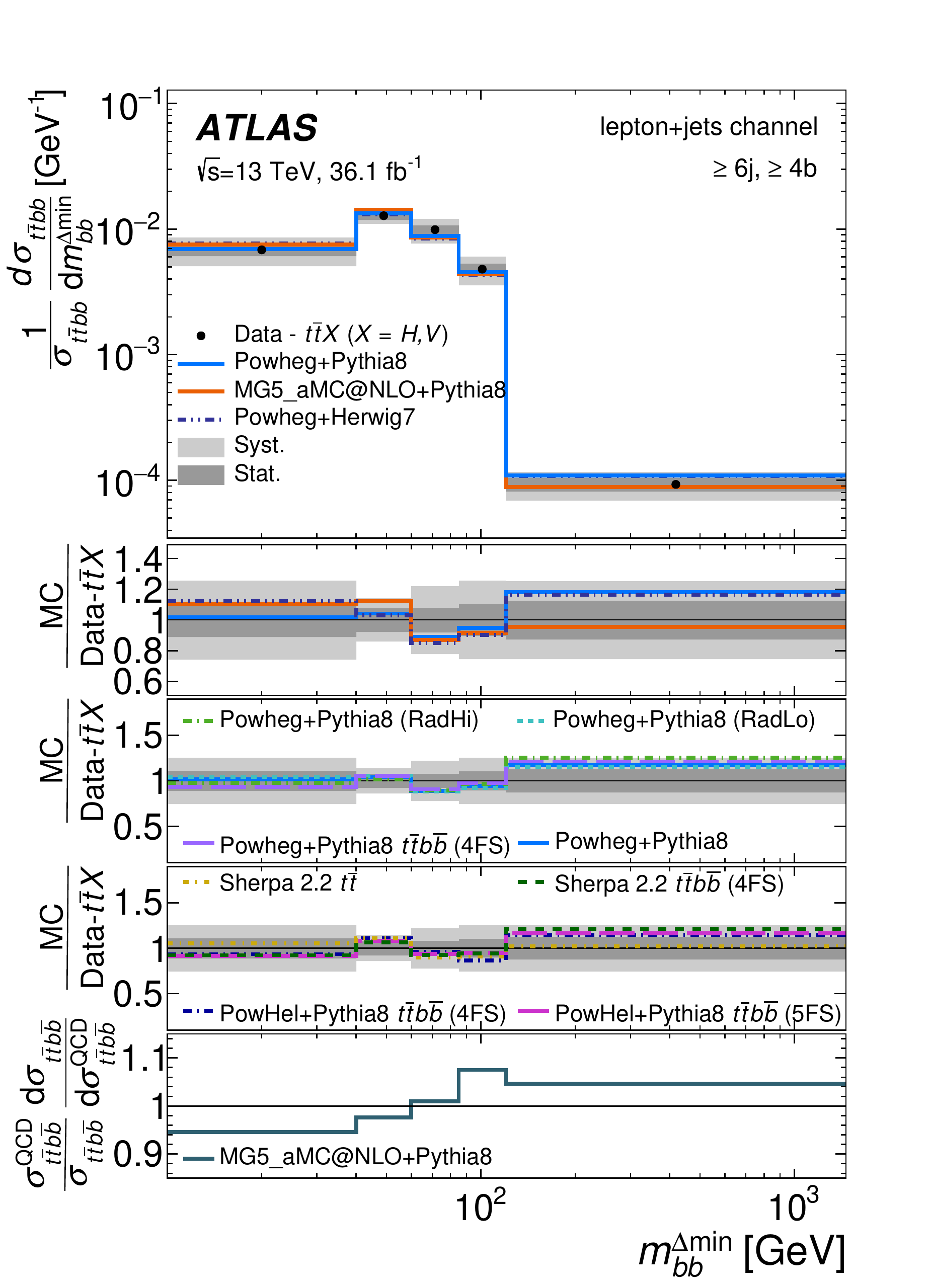}
  }
  \subcaptionbox{\label{fig:ptbb_ljets}}{
    \includegraphics[width=0.42\textwidth]{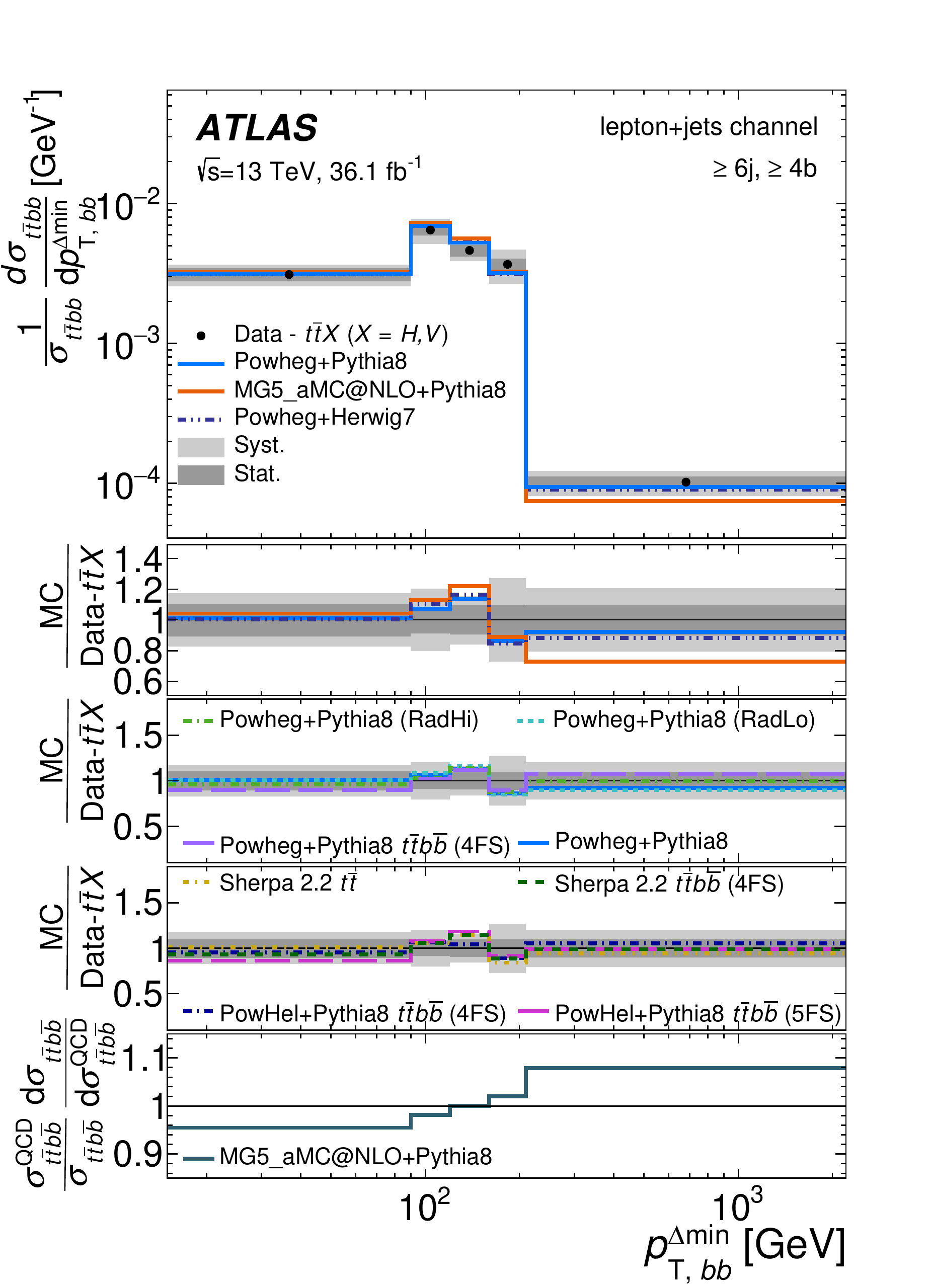}
  }
  \subcaptionbox{\label{fig:drbb_ljets}}{
    \includegraphics[width=0.42\textwidth]{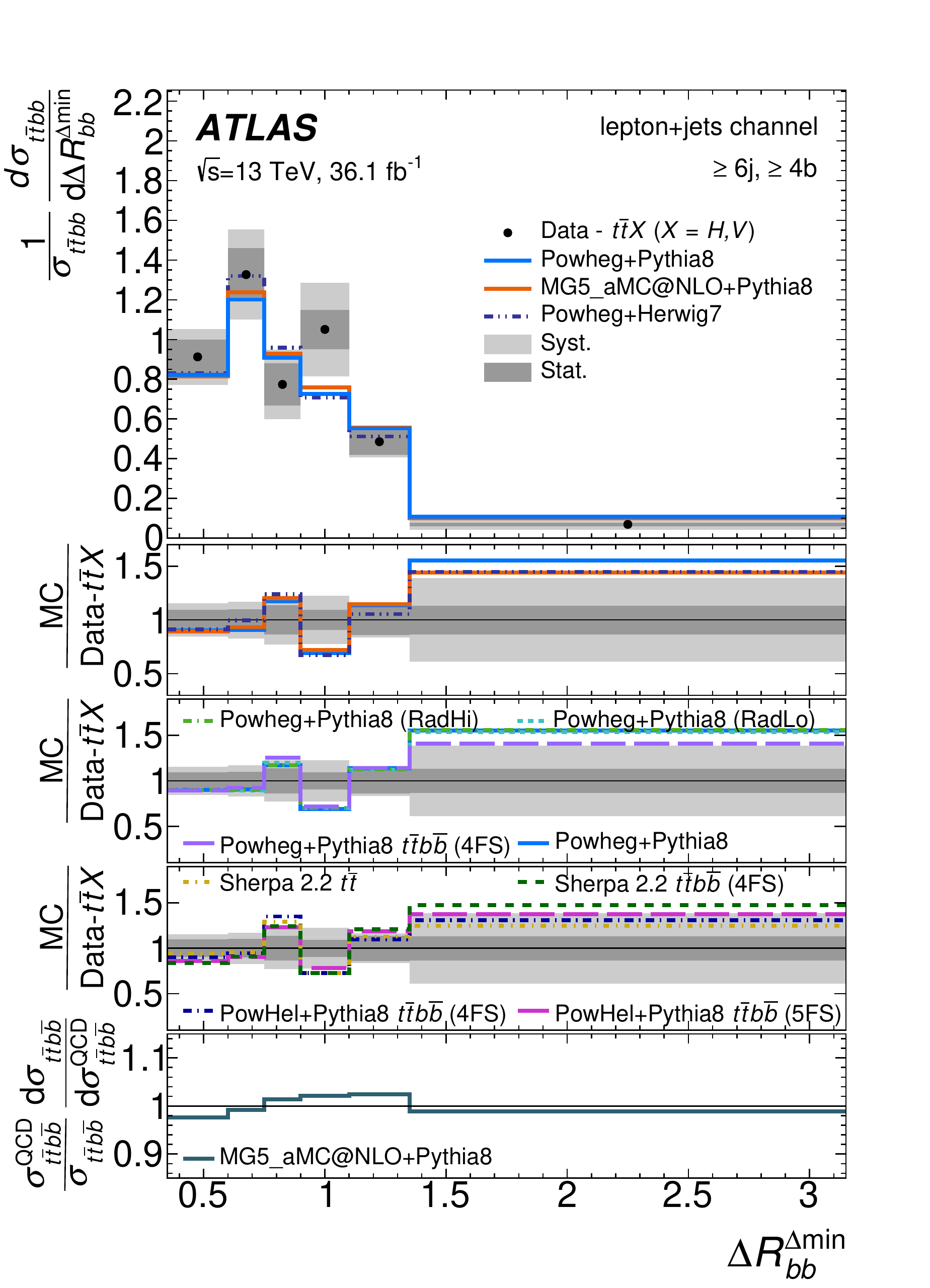}
  }
  \caption{Relative differential cross-sections as a function of
    \protect\subref{fig:mbb_ljets} $m_{bb}^{\Delta \mathrm{min}}$,
    \protect\subref{fig:ptbb_ljets} $p_{\mathrm{T},bb}^{\Delta \mathrm{min}}$ and
    \protect\subref{fig:drbb_ljets} $\Delta R_{bb}^{\Delta \mathrm{min}}$ of two closest
    $b$-jets in $\Delta R$ in events with at least four $b$-jets in the \ljets
    channel compared with various MC generators. The \ttH and \ttV contributions are subtracted from data. 
    Four ratio panels are shown: the first three  show the ratios of
    various predictions to data. The last panel shows the ratio of predictions of
    normalised differential cross-sections from \amcnlopyeight including
    (numerator) and not including (denominator) the contributions from \ttV and
    \ttH production. Uncertainty bands represent the statistical and total systematic uncertainties as described in Section~\ref{sec:systematics}.  Events with observable values outside the axis range are not included in the plot.}
  \label{fig:res_dRminljets}
\end{figure}

Figures~\ref{fig:res_dRmin} and~\ref{fig:res_dRminljets} show the same
observables but reconstructed from the pair of two closest \bjets in the event,
i.e. those with the smallest $\Delta R$, denoted by $m^{\Delta \mathrm{min}}_{bb}$,
$p^{\Delta  \mathrm{min}} _{\mathrm{T},bb}$, and $\Delta R^{\Delta \mathrm{min}} (b,b)$.
The experimental uncertainties are similar to those using the \bjet pair with
the highest \pt. However, the model variations are larger and
\powhelpyeight~(5FS) does not describe the data with $\geq 3$\bjets in the $e\mu$ channel well.


\begin{table}
  \caption{Values of $\chi^2$ per degree of freedom and $p$-values between the
    unfolded normalised cross-sections and the various predictions for the mass,
    \pt and $\Delta R$ of the closest two $b$-jets in the $e\mu$ and \ljets
    channels. The number of degrees of freedom is equal to the number of bins in
    the measured distribution minus one.}
  \small
  \centering
  \input{tables/combined_tables_ttXsubtracted/bb_closest}
  \label{tab:comb_bb_closest}
\end{table}

%% file: tables/fid_xsec_results.tex
\sisetup{round-mode=figures, round-precision=2}
  \begin{tabular}{l
    S[table-format=4.1, table-number-alignment=right]
    S[table-format=4.1, table-number-alignment=right]
    S[table-format=4.1, table-number-alignment=right]
    S[table-format=4.1, table-number-alignment=right]
    S[table-format=4.1, table-number-alignment=right]
    S[table-format=4.1, table-number-alignment=right]
    S[table-format=4.1, table-number-alignment=right]
    S[table-format=4.1, table-number-alignment=right]
}
  \toprule
  & \multicolumn{4}{c}{$e\mu$ [fb]} & \multicolumn{4}{c}{\ljets [fb]} \\
  & \multicolumn{2}{c}{$\geq 3b$} & \multicolumn{2}{c}{$\geq 4b$} & \multicolumn{2}{c}{$\geq 5j, \geq 3b$} & \multicolumn{2}{c}{$\geq 6j, \geq 4b$} \\
  \cmidrule(r){1-1}\cmidrule(lr){2-3} \cmidrule(lr){4-5} \cmidrule(lr){6-7} \cmidrule(lr){8-9}
  \multirow{3}{*}{Measured} & \multicolumn{2}{l}{181} & \multicolumn{2}{l}{27}  & \multicolumn{2}{l}{2450} & \multicolumn{2}{l}{359} \\
  & \multicolumn{1}{r}{$\pm$} & \multicolumn{1}{r}{$5 \mathrm{~(stat)}$} & \multicolumn{1}{r}{$\pm$} & \multicolumn{1}{r}{$3 \mathrm{~(stat)}$} & \multicolumn{1}{r}{$\pm$} & \multicolumn{1}{r}{$40 \mathrm{~(stat)}$} & \multicolumn{1}{r}{$\pm$} & \multicolumn{1}{r}{$11 \mathrm{~(stat)}$} \\
  & \multicolumn{1}{r}{$\pm$} & \multicolumn{1}{r}{$24 \mathrm{~(syst)}$} & \multicolumn{1}{r}{$\pm$} & \multicolumn{1}{r}{$7 \mathrm{~(syst)}$} & \multicolumn{1}{r}{$\pm$} & \multicolumn{1}{r}{$690 \mathrm{~(syst)}$} & \multicolumn{1}{r}{$ \pm$} & \multicolumn{1}{r}{$61 \mathrm{~(syst)}$} \\
  \midrule
  $t\bar{t}X (X=H,V)$ MC & \multicolumn{1}{r}{$\hspace{3.5mm}4$} &  & \multicolumn{1}{r}{$\hspace{2.0mm}2$} & & \multicolumn{1}{r}{$\hspace{3.6mm}80$} &  & \multicolumn{1}{r}{$\hspace{1.6mm}28$} \\
  Measured $-~t\bar{t}X$ & \multicolumn{2}{l}{177} & \multicolumn{2}{l}{25} & \multicolumn{2}{l}{2370} & \multicolumn{2}{l}{331} \\
  \midrule
  \SHERPAttbb (4FS) & \multicolumn{2}{l}{$103 \pm 30$} & \multicolumn{2}{l}{$17.3 \pm 4.2$} & \multicolumn{2}{l}{$1600 \pm 530$} & \multicolumn{2}{l}{$270 \pm 70$} \\
  \ppyeightttbb (4FS) & \multicolumn{2}{l}{104} & \multicolumn{2}{l}{16.5} & \multicolumn{2}{l}{1520} & \multicolumn{2}{l}{260} \\
  \powhelpyeight (5FS) & \multicolumn{2}{l}{152} & \multicolumn{2}{l}{18.7} & \multicolumn{2}{l}{1360} & \multicolumn{2}{l}{290} \\
  \powhelpyeight (4FS) & \multicolumn{2}{l}{105} & \multicolumn{2}{l}{18.2} & \multicolumn{2}{l}{1690} & \multicolumn{2}{l}{300} \\
  \bottomrule
\end{tabular}

%% file: tables/uncert_fidxsec_table.tex
\begin{table}[ht!]
  \caption{Main systematic uncertainties in percentage for particle-level
    measurement of inclusive cross-sections in $\ge$ 3 $b$ and $\ge$ 4 $b$ phase
    space.}
  \centering
\sisetup{table-number-alignment=right}
  \begin{tabular}{l
S[table-format=2.1, table-number-alignment=right]
S[table-format=2.1, table-number-alignment=right]
S[table-format=2.1, table-number-alignment=right]
S[table-format=2.1, table-number-alignment=right]
rrrr}
    \toprule
    Source  & \multicolumn{4}{c}{Fiducial cross-section phase space} \\
            & \multicolumn{2}{c}{$e\mu$}  & \multicolumn{2}{c}{\ljets} \\
    \cmidrule(r){2-3}\cmidrule(l){4-5}
                              & {$\ge 3 b$}   & {$\ge 4b$}    & {$\ge 5 j, \ge 3 b$}   & {$\ge 6 j, \ge 4 b$}   \\
                              & {unc. [$\%$]} & {unc. [$\%$]} & {unc. [$\%$]} & {unc. [$\%$]} \\
    \midrule 
    Data statistics           & 2.7         & 9.0         & 1.7         & 3.0         \\
    \midrule 
    Luminosity                & 2.1         & 2.1         & 2.3         & 2.3         \\
    Jet                       & 2.6         & 4.3         & 3.6         & 7.2         \\
    $b$-tagging               & 4.5         & 5.2         & 17          & 8.6         \\
    Lepton                    & 0.9         & 0.8         & 0.8         & 0.9         \\
    Pile-up                    & 2.1         & 3.5         & 1.6         & 1.3         \\
    $t\bar{t}c$ fit variation & 5.9         & 11          & \multicolumn{1}{r}{-}           & \multicolumn{1}{r}{-}           \\
    Non-\ttbar bkg            & 0.8         & 2.0         & 1.7         & 1.8         \\
    \midrule 
    Detector+background total syst.      & 8.5         & 14          & 18          & 12          \\
    \midrule 
    Parton shower             & 9.0         & 6.5         & 12          & 6.3         \\
    Generator                 & 0.2         & 18          & 16          & 8.7         \\
    ISR/FSR                   & 4.0         & 3.9         & 6.2         & 2.9         \\
    PDF                       & 0.6         & 0.4         & 0.3         & 0.1         \\
    $t\bar{t}V / t\bar{t}H$   & 0.7         & 1.4         & 2.2         & 0.3         \\
    MC sample statistics      & 1.8         & 5.3         & 1.2         & 4.3         \\
    \midrule 
    \ttbar modelling total syst.     & 10          & 20          & 21           & 12         \\
    \bottomrule 
    \toprule 
    Total syst.               & 13          & 24          & 28          & 17          \\
    Total                     & 13          & 26          & 28          & 17          \\
    \bottomrule 
  \end{tabular}
  \label{tab:xsec_unc}
\end{table}

%% file: tables/combined_tables_ttXsubtracted/nbjets.tex
\sisetup{table-number-alignment=left}
\begin{tabular}{l
S[table-format=2.3, table-number-alignment=left]
r
r
S[table-format=2.2, table-number-alignment=right]
}
\toprule
  Generators     & \multicolumn{2}{c}{$N_{b\mathrm{\textrm{-}jets}}: [2, 3, \ge 4 b]$} & \multicolumn{2}{c}{$N_{b\mathrm{\textrm{-}jets}}: [3, \ge 4 b]$} \\

                 &  {$\chi^2$ / NDF} &  {$p$-value} & {$\chi^2$ / NDF} & {$p$-value} \\
  \midrule
  \textbf{\bm{$e\mu$} channel} \\
  \ppyeight      & 18.1 / 2       & $<0.01$   & $<0.01$ / 1    & 1.0      \\
  \amcnlopyeight & 14.1 / 2       & $<0.01$   & 0.05 / 1    & 0.83      \\
  \SHERPAtt      & 0.85 / 2       & 0.65      & 0.06 / 1       & 0.80      \\
  \SHERPAttbb (4FS)   &  {-}              & -         & 0.37 / 1       & 0.54      \\
  \powhelpyeight (5FS) &  {-}              & -         & 0.33 / 1       & 0.56      \\
  \powhelpyeight (4FS) &  {-}              & -         & 0.76 / 1  &  0.38      \\
  \phwseven      & 39.4 / 2       & $<0.01$   & 0.26 / 1       & 0.61      \\
  \ppyeightttbb (4FS)  &  {-}       & -   & 0.28 / 1  &  0.60      \\
  \ppyeightradhi & 9.2 / 2       &  0.01   & 0.08 / 1       & 0.77      \\
  \ppyeightradlo & 27.0 / 2       & $<0.01$   & 0.01 / 1       & 0.92      \\
  \bottomrule
\end{tabular}

%% file: tables/combined_tables_ttXsubtracted/ht.tex
\sisetup{table-number-alignment=left}
\begin{tabular}{l
r
S[table-format=2.2, table-number-alignment=right]
rrrrrrrrrrrrrrrrrrrrrrrrr}
  \toprule
  {}                        & \multicolumn{2}{c}{$H_{\mathrm{T}}$} & \multicolumn{2}{c}{$H^{\mathrm{had}}_{\mathrm{T}}$} \\
  {}                        & {$\chi^2$ / NDF}                       & {$p$-value} & {$\chi^2$ / NDF} & {$p$-value}              \\
  Generator                 &                                      &           &                &                        \\
  \midrule
  \textbf{\bm{$e\mu$} channel, \bm{$\geq 3~b$}-jets}                                            \\
  \ppyeight            & 0.95 / 4 & 0.92  & 2.68 / 3 & 0.44              \\
  \amcnlopyeight       & 3.71 / 4 & 0.45  & 3.72 / 3 & 0.29              \\
  \SHERPAtt            & 0.58 / 4 & 0.97  & 2.26 / 3 & 0.52              \\
  \SHERPAttbb (4FS)    & 0.35 / 4 & 0.99  & 0.40 / 3 & 0.94              \\
  \powhelpyeight (5FS) & 4.88 / 4 & 0.30  & 1.85 / 3 & 0.60              \\
  \powhelpyeight (4FS) & 1.39 / 4 & 0.85  & 3.33 / 3 & 0.32              \\
  \phwseven            & 0.26 / 4 & 0.99  & 2.28 / 3 & 0.52              \\
  \ppyeightttbb (4FS)  & 0.63 / 4 & 0.96  & 3.93 / 3 & 0.27              \\
  \ppyeightradhi       & 4.09 / 4 & 0.39  & 6.43 / 3 & 0.09              \\
  \ppyeightradlo       & 0.14 / 4 & 1.0 & 1.06 / 3 & 0.79              \\
  \midrule
  \textbf{lepton+jets channel, \bm{$\geq 6$} jets, \bm{$\geq 4~b$}-jets} \\
  \ppyeight            & 0.60 / 4 & 0.96  & 1.41 / 4 & 0.84              \\
  \amcnlopyeight       & 9.88 / 4 & 0.04  & 17.6 / 4 & $<0.01$           \\
  \SHERPAtt            & 0.72 / 4 & 0.95  & 1.38 / 4 & 0.85              \\
  \SHERPAttbb (4FS)    & 1.09 / 4 & 0.90  & 2.58 / 4 & 0.63              \\
  \powhelpyeight (5FS) & 0.81 / 4 & 0.94  & 1.40 / 4 & 0.84              \\
  \powhelpyeight (4FS) & 1.38 / 4 & 0.85  & 2.38 / 4 & 0.67              \\
  \phwseven            & 4.27 / 4 & 0.37  & 7.00 / 4 & 0.14              \\
  \ppyeightttbb (4FS)  & 0.72 / 4 & 0.95  & 1.71 / 4 & 0.79              \\
  \ppyeightradhi       & 0.94 / 4 & 0.92  & 0.96 / 4 & 0.92              \\
  \ppyeightradlo       & 1.15 / 4 & 0.89  & 2.57 / 4 & 0.63              \\
  \bottomrule
\end{tabular}

%% file: tables/combined_tables_ttXsubtracted/bjet_pt.tex
\sisetup{table-number-alignment=left}
\begin{tabular}{l
S[table-format=2.3, table-number-alignment=left]
rrrrrrrrrrrrrrrrrrrrrrrrrr}
  \toprule
  {} & \multicolumn{2}{c}{$p_{\mathrm{T}}^{b_1}$} & \multicolumn{2}{c}{$p_{\mathrm{T}}^{b_2}$} & \multicolumn{2}{c}{$p_{\mathrm{T}}^{b_3}$} & \multicolumn{2}{c}{$p_{\mathrm{T}}^{b_4}$} \\
  {} &         {$\chi^2$ / NDF} & {$p$-value} &       {$\chi^2$ / NDF} & {$p$-value} &      {$\chi^2$ / NDF} &  {$p$-value} &      {$\chi^2$ / NDF} & {$p$-value} \\
  Generator                          &                        &           &                        &           &                        &           &                        &           \\
  \midrule
  \textbf{\bm{$e\mu$} channel, \bm{$\geq 3~b$}-jets}                                            \\
  \ppyeight      & 2.09 / 4 & 0.72 & 0.50 / 3 & 0.92 & 0.09 / 2    & 0.95    & -        & -    \\
  \amcnlopyeight & 2.62 / 4 & 0.62 & 0.27 / 3 & 0.97 & 0.33 / 2    & 0.85    & -        & -    \\
  \SHERPAtt      & 0.98 / 4 & 0.91 & 0.67 / 3 & 0.88 & 0.02 / 2 &  0.99 & -        & -    \\
  \SHERPAttbb (4FS) & 3.52 / 4 & 0.47 & 0.68 / 3 & 0.88 & 0.21 / 2    & 0.90    & -        & -    \\
  \powhelpyeight (5FS) & 10.9 / 4  &  0.03 & 2.58 / 3  &  0.46 & 3.91 / 2  &  0.14    & -        & -    \\
  \powhelpyeight (4FS) & 6.21 / 4  &  0.18 & 1.96 / 3  &  0.58 & 1.30 / 2  &  0.52    & -        & -    \\
  \phwseven      & 1.16 / 4 & 0.89 & 1.02 / 3 & 0.80 & 0.02 / 2    & 0.99 & -        & -    \\
  \ppyeightttbb (4FS) & 2.62 / 4  &  0.62 & 0.53 / 3  &  0.91 & 0.46 / 2  &  0.80    & -        & -    \\
  \ppyeightradhi & 2.71 / 4 & 0.61 & 0.56 / 3 & 0.91 & 0.26 / 2    & 0.88    & -        & -    \\
  \ppyeightradlo & 1.93 / 4 & 0.75 & 0.64 / 3 & 0.89 & 0.05 / 2    & 0.97    & -        & -    \\
  \midrule
  \textbf{lepton+jets channel, \bm{$\geq 6$} jets, \bm{$\geq 4~b$}-jets}                \\
  \ppyeight            & 2.09 / 4 & 0.72 & 2.98 / 3 & 0.40 & 1.42 / 3 & 0.70 & 0.20 / 2 & 0.90 \\
  \amcnlopyeight       & 5.20 / 4 & 0.27 & 5.31 / 3 & 0.15 & 1.87 / 3 & 0.60 & 0.08 / 2 & 0.96 \\
  \SHERPAtt            & 2.01 / 4 & 0.73 & 2.46 / 3 & 0.48 & 1.75 / 3 & 0.63 & 0.15 / 2 & 0.93 \\
  \SHERPAttbb (4FS)    & 2.04 / 4 & 0.73 & 2.82 / 3 & 0.42 & 1.23 / 3 & 0.75 & 0.52 / 2 & 0.77 \\
  \powhelpyeight (5FS) & 2.07 / 4 & 0.72 & 3.65 / 3 & 0.30 & 1.73 / 3 & 0.63 & 0.85 / 2 & 0.65 \\
  \powhelpyeight (4FS) & 2.52 / 4 & 0.64 & 2.37 / 3 & 0.50 & 2.41 / 3 & 0.49 & 0.18 / 2 & 0.91 \\
  \phwseven            & 2.58 / 4 & 0.63 & 3.50 / 3 & 0.32 & 1.30 / 3 & 0.73 & 0.26 / 2 & 0.88 \\
  \ppyeightttbb (4FS)  & 1.76 / 4 & 0.78 & 2.02 / 3 & 0.57 & 1.83 / 3 & 0.61 & 0.84 / 2 & 0.66 \\
  \ppyeightradhi       & 1.50 / 4 & 0.83 & 2.39 / 3 & 0.50 & 1.74 / 3 & 0.63 & 0.37 / 2 & 0.83 \\
  \ppyeightradlo       & 2.17 / 4 & 0.70 & 3.75 / 3 & 0.29 & 1.42 / 3 & 0.70 & 0.17 / 2 & 0.92 \\
  \bottomrule
\end{tabular}

%% file: tables/combined_tables_ttXsubtracted/bb_leading.tex
\begin{tabular}{lrrrrrrrrrrrrrrrrrrrrrrrrrr}
  \toprule
  {} & \multicolumn{2}{c}{$m_{b_1b_2}$} & \multicolumn{2}{c}{$p_{\mathrm{T}, b_1b_2}$} & \multicolumn{2}{c}{$\Delta R_{b_1b_2}$} \\
  {} & $\chi^2$ / NDF & $p$-value &           $\chi^2$ / NDF & $p$-value &      $\chi^2$ / NDF & $p$-value \\
  Generator                          &                &           &                          &           &                     &           \\
  \midrule
  \textbf{\bm{$e\mu$} channel, \bm{$\geq 3~b$}-jets}                         \\
  \ppyeight            & 1.55 / 4 & 0.82 & 1.74 / 3 & 0.63 & 0.70 / 4 & 0.95 \\
  \amcnlopyeight       & 1.73 / 4 & 0.79 & 1.08 / 3 & 0.78 & 3.73 / 4 & 0.44 \\
  \SHERPAtt            & 0.25 / 4 & 0.99 & 0.64 / 3 & 0.89 & 0.99 / 4 & 0.91 \\
  \SHERPAttbb (4FS)    & 2.88 / 4 & 0.58 & 0.76 / 3 & 0.86 & 2.88 / 4 & 0.58 \\
  \powhelpyeight (5FS) & 3.74 / 4 & 0.44 & 4.75 / 3 & 0.19 & 4.70 / 4 & 0.32 \\
  \powhelpyeight (4FS) & 1.35 / 4 & 0.85 & 2.90 / 3 & 0.41 & 0.86 / 4 & 0.93 \\
  \phwseven            & 0.48 / 4 & 0.98 & 0.42 / 3 & 0.94 & 0.97 / 4 & 0.91 \\
  \ppyeightttbb (4FS)  & 1.89 / 4 & 0.76 & 0.79 / 3 & 0.85 & 0.68 / 4 & 0.95 \\
  \ppyeightradhi       & 3.77 / 4 & 0.44 & 3.49 / 3 & 0.32 & 0.50 / 4 & 0.97 \\
  \ppyeightradlo       & 1.04 / 4 & 0.90 & 0.95 / 3 & 0.81 & 1.01 / 4 & 0.91 \\
  \midrule
  \textbf{lepton+jets channel, \bm{$\geq 6$} jets, \bm{$\geq 4~b$}-jets}     \\
  \ppyeight            & 1.82 / 5 & 0.87 & 1.66 / 5 & 0.89 & 2.48 / 6 & 0.87 \\
  \amcnlopyeight       & 4.11 / 5 & 0.53 & 4.63 / 5 & 0.46 & 2.90 / 6 & 0.82 \\
  \SHERPAtt            & 2.84 / 5 & 0.72 & 1.79 / 5 & 0.88 & 3.40 / 6 & 0.76 \\
  \SHERPAttbb (4FS)    & 2.40 / 5 & 0.79 & 1.76 / 5 & 0.88 & 3.37 / 6 & 0.76 \\
  \powhelpyeight (5FS) & 2.39 / 5 & 0.79 & 1.85 / 5 & 0.87 & 2.94 / 6 & 0.82 \\
  \powhelpyeight (4FS) & 3.71 / 5 & 0.59 & 2.49 / 5 & 0.78 & 4.79 / 6 & 0.57 \\
  \phwseven            & 2.46 / 5 & 0.78 & 2.60 / 5 & 0.76 & 2.80 / 6 & 0.83 \\
  \ppyeightttbb (4FS)  & 1.88 / 5 & 0.87 & 1.51 / 5 & 0.91 & 2.79 / 6 & 0.83 \\
  \ppyeightradhi       & 1.68 / 5 & 0.89 & 1.67 / 5 & 0.89 & 2.72 / 6 & 0.84 \\
  \ppyeightradlo       & 1.89 / 5 & 0.86 & 2.35 / 5 & 0.80 & 2.63 / 6 & 0.85 \\
  \bottomrule
\end{tabular}

%% file: tables/combined_tables_ttXsubtracted/bb_closest.tex
\sisetup{table-number-alignment=left}
\begin{tabular}{l
r
S[table-format=2.2, table-number-alignment=right]
r
S[table-format=2.2, table-number-alignment=right]
rrrrrrrrrrrrrrrrrrrrrrrr}
  \toprule
  {} & \multicolumn{2}{c}{$m_{bb}^{\Delta\mathrm{min}}$} & \multicolumn{2}{c}{$p_{\mathrm{T}, bb}^{\Delta\mathrm{min}}$} & \multicolumn{2}{c}{$\Delta \mathrm{R}^{\Delta\mathrm{min}}_{bb}$} \\
  {} &                {$\chi^2$ / NDF} & {$p$-value} &    {$\chi^2$ / NDF} & {$p$-value} &  {$\chi^2$ / NDF} & {$p$-value} \\
  Generator                          &                               &           &                                           &           &                                               &           \\
  \midrule
  \textbf{\bm{$e\mu$} channel, \bm{$\geq 3~b$}-jets}                                            \\
  \ppyeight            & 1.37 / 4 & 0.85  & 0.42 / 4 & 0.98    & 0.78 / 3 & 0.86     \\
  \amcnlopyeight       & 3.67 / 4 & 0.45  & 2.50 / 4 & 0.65    & 1.22 / 3 & 0.75     \\
  \SHERPAtt            & 0.17 / 4 & 1.0 & 0.06 / 4 & 1.0   & 0.99 / 3 & 0.80     \\
  \SHERPAttbb (4FS)    & 1.36 / 4 & 0.85  & 0.52 / 4 & 0.97    & 0.21 / 3 & 0.98     \\
  \powhelpyeight (5FS) & 0.18 / 4 & 1.0   & 12.7 / 4 & 0.01    & 27.9 / 3 & $< 0.01$ \\
  \powhelpyeight (4FS) & 4.29 / 4 & 0.37  & 2.36 / 4 & 0.67    & 0.81 / 3 & 0.85     \\
  \phwseven            & 0.87 / 4 & 0.93  & 0.06 / 4 & 1.0   & 0.95 / 3 & 0.81     \\
  \ppyeightttbb (4FS)  & 1.12 / 4 & 0.89  & 1.00 / 4 & 0.91    & 0.30 / 3 & 0.96     \\
  \ppyeightradhi       & 1.94 / 4 & 0.75  & 1.31 / 4 & 0.86    & 0.51 / 3 & 0.92     \\
  \ppyeightradlo       & 0.99 / 4 & 0.91  & 0.28 / 4 & 0.99    & 0.86 / 3 & 0.84     \\
  \midrule
  \textbf{lepton+jets channel, \bm{$\geq 6$} jets, \bm{$\geq 4~b$}-jets}             \\
  \ppyeight            & 0.86 / 4 & 0.93  & 0.99 / 4 & 0.91    & 3.22 / 5 & 0.67     \\
  \amcnlopyeight       & 1.01 / 4 & 0.91  & 4.33 / 4 & 0.36    & 3.19 / 5 & 0.67     \\
  \SHERPAtt            & 0.66 / 4 & 0.96  & 1.21 / 4 & 0.88    & 4.98 / 5 & 0.42     \\
  \SHERPAttbb (4FS)    & 1.44 / 4 & 0.84  & 0.89 / 4 & 0.93    & 4.07 / 5 & 0.54     \\
  \powhelpyeight (5FS) & 1.08 / 4 & 0.90  & 1.61 / 4 & 0.81    & 3.14 / 5 & 0.68     \\
  \powhelpyeight (4FS) & 1.93 / 4 & 0.75  & 0.30 / 4 & 1.0 & 5.43 / 5 & 0.37     \\
  \phwseven            & 1.32 / 4 & 0.86  & 1.47 / 4 & 0.83    & 4.53 / 5 & 0.48     \\
  \ppyeightttbb (4FS)  & 1.05 / 4 & 0.90  & 0.82 / 4 & 0.94    & 3.87 / 5 & 0.57     \\
  \ppyeightradhi       & 1.51 / 4 & 0.83  & 0.95 / 4 & 0.92    & 2.98 / 5 & 0.70     \\
  \ppyeightradlo       & 0.77 / 4 & 0.94  & 1.51 / 4 & 0.83    & 3.25 / 5 & 0.66     \\
  \bottomrule
\end{tabular}

%% file: sections/summary.tex
\section{Summary}
\label{sec:summary}
Measurements of inclusive and normalised differential cross-sections of pairs of
top-quarks in association with heavy-flavour jets in 13~\TeV\ $pp$ collisions are
presented using a data sample of \lumi\ collected by the ATLAS detector at the
LHC. The results are shown in both the $e\mu$ and \ljets channels within fiducial
phase spaces. The background coming from \ttbar production in association with
additional light-flavour and charm-quark jets is evaluated using a fit to a
binned $b$-tagging discriminant. The data after background subtraction are
unfolded to particle level to correct for detector and acceptance effects.
The fiducial cross-sections are measured for $\geq 3b$ and $\geq 4b$ phase
spaces in the $e\mu$ channel, and for $\geq 5j$, $\geq 3b$ and $\geq 6j$,
$\geq 4b$ phase spaces in the \ljets channel. The two cross-section measurements 
with the smallest uncertainties, 13\% and 17\%, are those for $\geq 3b$ in the 
$e\mu$ channel and $\geq 6j$, $\geq 4b$ in the \ljets channel, respectively.
The measured cross-sections, after subtracting estimated contributions from \ttH
and \ttV, are compared with various \ttbb predictions and are found to be higher
than predicted but compatible within the uncertainties.

The normalised fiducial differential cross-sections are presented as a function of
several relevant kinematic variables and global event properties. In general,
the different observables are measured with a precision of 10\% in most of the
phase space, rising to 30\% at the edge of the phase space for some of the
observables. The observables are well described by most MC predictions in both
channels. However, it is worth noting that in all the predictions where additional
\bjets are dominantly produced by the parton shower, they predict too few events with
more \bjets than those produced in top decays. Only \SHERPAtt describes the full \bjet
multiplicity spectrum, and in events with $\geq 3$ \bjets it yields the best agreement 
with data in most of the observables. \powhelpyeight~(5FS) shows poor agreement
in some of the observables in events with $\geq 3$ \bjets in the $e\mu$ channel.
The differential kinematic distributions are equally well described by
predictions that have additional \bjet production that is generated by the 
parton shower calculation and by predictions with additional $b$-quarks in the matrix element.

%% file: atlaslatex/acknowledgements/Acknowledgements.tex

We thank CERN for the very successful operation of the LHC, as well as the
support staff from our institutions without whom ATLAS could not be
operated efficiently.

We acknowledge the support of ANPCyT, Argentina; YerPhI, Armenia; ARC, Australia; BMWFW and FWF, Austria; ANAS, Azerbaijan; SSTC, Belarus; CNPq and FAPESP, Brazil; NSERC, NRC and CFI, Canada; CERN; CONICYT, Chile; CAS, MOST and NSFC, China; COLCIENCIAS, Colombia; MSMT CR, MPO CR and VSC CR, Czech Republic; DNRF and DNSRC, Denmark; IN2P3-CNRS, CEA-DRF/IRFU, France; SRNSFG, Georgia; BMBF, HGF, and MPG, Germany; GSRT, Greece; RGC, Hong Kong SAR, China; ISF and Benoziyo Center, Israel; INFN, Italy; MEXT and JSPS, Japan; CNRST, Morocco; NWO, Netherlands; RCN, Norway; MNiSW and NCN, Poland; FCT, Portugal; MNE/IFA, Romania; MES of Russia and NRC KI, Russian Federation; JINR; MESTD, Serbia; MSSR, Slovakia; ARRS and MIZ\v{S}, Slovenia; DST/NRF, South Africa; MINECO, Spain; SRC and Wallenberg Foundation, Sweden; SERI, SNSF and Cantons of Bern and Geneva, Switzerland; MOST, Taiwan; TAEK, Turkey; STFC, United Kingdom; DOE and NSF, United States of America. In addition, individual groups and members have received support from BCKDF, CANARIE, CRC and Compute Canada, Canada; COST, ERC, ERDF, Horizon 2020, and Marie Sk{\l}odowska-Curie Actions, European Union; Investissements d' Avenir Labex and Idex, ANR, France; DFG and AvH Foundation, Germany; Herakleitos, Thales and Aristeia programmes co-financed by EU-ESF and the Greek NSRF, Greece; BSF-NSF and GIF, Israel; CERCA Programme Generalitat de Catalunya, Spain; The Royal Society and Leverhulme Trust, United Kingdom. 

The crucial computing support from all WLCG partners is acknowledged gratefully, in particular from CERN, the ATLAS Tier-1 facilities at TRIUMF (Canada), NDGF (Denmark, Norway, Sweden), CC-IN2P3 (France), KIT/GridKA (Germany), INFN-CNAF (Italy), NL-T1 (Netherlands), PIC (Spain), ASGC (Taiwan), RAL (UK) and BNL (USA), the Tier-2 facilities worldwide and large non-WLCG resource providers. Major contributors of computing resources are listed in Ref.~\cite{ATL-GEN-PUB-2016-002}.

%% file: atlas_authlist.tex
 
\begin{flushleft}
{\Large The ATLAS Collaboration}

\bigskip

M.~Aaboud$^\textrm{\scriptsize 34d}$,    
G.~Aad$^\textrm{\scriptsize 99}$,    
B.~Abbott$^\textrm{\scriptsize 125}$,    
O.~Abdinov$^\textrm{\scriptsize 13,*}$,    
B.~Abeloos$^\textrm{\scriptsize 129}$,    
D.K.~Abhayasinghe$^\textrm{\scriptsize 91}$,    
S.H.~Abidi$^\textrm{\scriptsize 164}$,    
O.S.~AbouZeid$^\textrm{\scriptsize 39}$,    
N.L.~Abraham$^\textrm{\scriptsize 153}$,    
H.~Abramowicz$^\textrm{\scriptsize 158}$
% ...
\bigskip
\\

% ...
$^{13}$Institute of Physics, Azerbaijan Academy of Sciences, Baku; Azerbaijan.\\
% ...
$^{34}$$^{(a)}$Facult\'e des Sciences Ain Chock, R\'eseau Universitaire de Physique des Hautes Energies - Universit\'e Hassan II, Casablanca;$^{(b)}$Centre National de l'Energie des Sciences Techniques Nucleaires (CNESTEN), Rabat;$^{(c)}$Facult\'e des Sciences Semlalia, Universit\'e Cadi Ayyad, LPHEA-Marrakech;$^{(d)}$Facult\'e des Sciences, Universit\'e Mohamed Premier and LPTPM, Oujda;$^{(e)}$Facult\'e des sciences, Universit\'e Mohammed V, Rabat; Morocco.\\
% ...
$^{39}$Niels Bohr Institute, University of Copenhagen, Copenhagen; Denmark.\\
% ...
$^{91}$Department of Physics, Royal Holloway University of London, Egham; United Kingdom.\\
% ...
$^{99}$CPPM, Aix-Marseille Universit\'e, CNRS/IN2P3, Marseille; France.\\
% ...
$^{125}$Homer L. Dodge Department of Physics and Astronomy, University of Oklahoma, Norman OK; United States of America.\\
% ...
$^{129}$LAL, Universit\'e Paris-Sud, CNRS/IN2P3, Universit\'e Paris-Saclay, Orsay; France.\\
% ...
$^{153}$Department of Physics and Astronomy, University of Sussex, Brighton; United Kingdom.\\
% ...
$^{158}$Raymond and Beverly Sackler School of Physics and Astronomy, Tel Aviv University, Tel Aviv; Israel.\\
% ...
$^{164}$Department of Physics, University of Toronto, Toronto ON; Canada.\\
% ...
$^{a}$ Also at Borough of Manhattan Community College, City University of New York, NY; United States of America.\\
$^{b}$ Also at California State University, East Bay; United States of America.\\
$^{c}$ Also at Centre for High Performance Computing, CSIR Campus, Rosebank, Cape Town; South Africa.\\
$^{d}$ Also at CERN, Geneva; Switzerland.\\
$^{e}$ Also at CPPM, Aix-Marseille Universit\'e, CNRS/IN2P3, Marseille; France.\\
% ...
$^{*}$ Deceased

\end{flushleft}
